\documentclass[12pt]{article}
\usepackage[top=1.in, left=0.7in, bottom=1.in, right=0.7in]{geometry}                		
\usepackage{graphicx}				
								
\usepackage{pdflscape}
\usepackage{amsmath}
\usepackage{appendix}
\usepackage{pdflscape}
\usepackage{booktabs,caption}
\usepackage[flushleft]{threeparttable}
\usepackage{amssymb}
\usepackage{ntheorem}

\usepackage[natbibapa]{apacite}
\usepackage{rotating, graphicx}
\usepackage{array} 
\usepackage{dcolumn}
\usepackage{adjustbox}
\usepackage{booktabs} 
\usepackage{hyperref}
\hypersetup{colorlinks,linkcolor={blue},citecolor={blue},urlcolor={blue}} 
\usepackage[dvipsnames]{xcolor}
\usepackage{comment}

\usepackage{ragged2e}
\usepackage[onehalfspacing]{setspace}
   
\begin{document}
\linespread{1.25}


\title{\Large \bf Discrimination in the Venture Capital Industry:\\ Evidence from Field Experiments}

\author{Ye Zhang \footnote{Stockholm School of Economics Finance Department. Address: Drottninggatan 98, 111 60 Stockholm, Sweden. Email: Ye.Zhang@hhs.se Telephone number: (+46)720323662}}
\date{\today}
\maketitle

\thispagestyle{empty}
\begin{abstract}

\noindent This paper examines discrimination by early-stage investors based on startup founders’ gender and race using two complementary field experiments with real U.S. venture capitalists. Results show the following. (i) Discrimination varies depending on the context. Investors implicitly discriminate against female and Asian founders when evaluating attractive startups, but they favor female and Asian founders when evaluating struggling startups. This helps to reconcile the contradictory results in the extant literature and confirms the theoretical predictions of ``discrimination reversion” and ``pro-cyclical discrimination” phenomena. (ii) Among multiple coexisting sources of discrimination identified, statistical discrimination and implicit discrimination are important reasons for investors’ “anti-minority” behaviors. A consistent estimator is developed to measure the polarization of investors’ discrimination behaviors and their separate driving forces. (iii) Homophily exists when investors provide anonymous encouragement to startups in a non-investment setting. (iv) There was temporary, stronger discrimination against Asian founders during the COVID-19 outbreak. \\
\end{abstract}




\newpage
\setcounter{page}{1}
\hypertarget{sec:intro}{\section{Introduction}}\label{sec:intro}
The persistent gender gap and racial disparities (\cite{bertrand2010dynamics}, \cite{bertrand2001gender}, \cite{chetty2020race}) at the top of the earnings distribution often raise the question of whether explicit or implicit discrimination creates a glass ceiling for women and nonwhites in modern U.S. society (\cite{bertrand2005implicit}).\footnote{\cite{hegde2021race} document the ``glass ceiling" phenomenon by analyzing detailed administrative records for the universe of patent examiners at the U.S. Patent and Trademark Office. They find that minority examiners are less likely to be promoted to the most senior grades.} Therefore, a significant amount of attention and debate about discrimination concentrate on top-level, high-skilled labor markets and relevant financial markets.\footnote{``\href{https://www.forbes.com/sites/shereeatcheson/2021/05/13/having-a-glass-ceiling-to-break-through-is-privilege-heres-why/?sh=7633257520d3}{Having A Glass Ceiling To Break Through Is Privilege. Here’s Why.}" May 13, 2021, Forbes} Despite the existence of a large body of academic literature on empirically testing discrimination in various markets,\footnote{For a recent summary, see \cite{lang2020race}, \cite{lang2012racial}, and \cite{bertrand2017field}.} most prior work mainly focuses on the entry-level and low-skilled labor market and product market (\cite{bertrand2004emily}, \cite{giuliano2009manager}, \cite{kessler_incentivized_2019}, \cite{list2004nature}). For the few papers studying discrimination in the high-skilled labor market, such as the entrepreneurial finance market, 
previous literature has provided conflicting empirical findings due to a lack of natural experimental settings.\par

\vspace{2mm}
In this paper, I re-examine discrimination issues in the U.S. entrepreneurial finance market, which generates huge amounts of wealth and produces many business leaders. Through two complementary field experiments with real U.S. venture capitalists (VC), this paper first identifies the existence of gender and racial discrimination by early-stage investors as well as the nature of this discrimination. Moreover, the paper also investigates how the direction and magnitude of the discovered discrimination vary across the distribution of startups' attractiveness and in different market conditions.\footnote{Based on \cite{gompers_diversity_2017}, 87\% of U.S. venture capitalists are white, and investors may have an unconscious bias against minority founders. Given the uniqueness of the entrepreneurial financing setting, this paper mainly studies racial discrimination faced by Asians, who are the largest minority group in the U.S. entrepreneurial community. ``Asians" in this paper primarily stand for ``East Asian" groups who originate mainly from China, Korea, Vietnam, etc. According to \cite{gompers_diversity_2017}, Asians account for 18\% of new U.S. venture capitalists and 15\% of new entrepreneurs entering the market. Studying discrimination against African Americans and other under-represented minorities is an important question. However, this paper's experimental design would need to be adjusted for future researchers to study these important questions.} The documented distributional effect and dynamic changes of discrimination not only empirically confirm the theoretical predictions of the ``discrimination reversion" and ``pro-cyclical discrimination" phenomena in \cite{morgan2009diversity}, but also enable the reconciliation of the contradictory results of previous empirical papers. Since the VC industry plays an important role in fostering innovative and successful companies (\cite{bernstein2016impact}), rigorously identifying such discrimination and its nature is of critical importance not only for maintaining social fairness (\cite{fang2011theories}) and assessing the efficiency of capital allocation  (\cite{bertrand2020gender}), but also for explaining the persistent gender and racial gap at the top level.\par

\vspace{2mm}
While the stark gender funding gap and the less favorable treatment received by nonwhite founders in the fundraising process has been well-documented (\cite{ewens_are_2020}, \cite{guzman2019gender}, \cite{henderson2015credit}, \cite{hebert2020gender}),\footnote{\cite{gompers2017diversity} demonstrate that from 1990-2016, women have made up less than 10\% of the entrepreneurial and venture capital labor pool despite an increase in female labor market participation during this period. Based on \cite{gornall_gender_2020}, venture capitalists only invested 1 dollar in startups with female founding teams for every 35 dollars invested in startups with male founding teams in 2017. Also, \cite{guzman2019gender} document that female-led ventures are 63 percent less likely than male-led ventures to have obtained external funding (i.e., venture capital) from 1995-2001, even though women and men are equally likely to achieve exit outcomes through IPOs or acquisitions.} causally identifying discrimination and its nature in this setting poses new challenges for the commonly used empirical methods.\footnote{According to \cite{list2004nature}, empirically testing for marketplace discrimination has taken two quite distinct paths in economic research: regression-based methods and field experiments. For more discussion on field experiments, see \cite{list2010field}, \cite{list2007field}, \cite{levitt2009field}, \cite{list2011so}, etc.} Moreover, extant literature provides seemingly conflicting results. On the one hand, papers exploiting regression-based methods (\cite{ewens_are_2020}, \cite{guzman2019gender}, \cite{henderson2015credit}) often show that women-led and minority-led startups are associated with a lower likelihood of raising external funding, suggesting the existence of gender and racial discrimination. These non-experimental studies potentially suffer from omitted variable bias due to a lack of exogenous variations.\footnote{Since most venture capital funds and startup companies are private firms, some startups' unique comparative advantages are usually only observable to investors rather than to researchers. Also, most databases only observe the match outcomes between investors and startups, making it difficult to separate investors' preferences from founders' preferences.} However, natural experimental settings are rarely available to solve various endogeneity concerns. On the other hand, \cite{gornall_gender_2020} implement the first correspondence test, a widely used field experimental method, in the U.S. VC industry. They find that early-stage investors are biased \textit{towards} female and Asian founders because surprisingly, investors reply more frequently to fictitious cold call, pitch emails sent by female and Asian names compared to male and white names. The major concern with this field experiment is that sending cold call emails is not a mainstream fundraising method used by high-quality startups. Hence, results mainly capture how investors evaluate struggling startup teams and might not be generalizable to the situation of evaluating top startups. Also, the ``low-response-rate" problem is more severe in this setting compared with other markets,\footnote{According to \cite{gornall_gender_2020}, the response rate to cold call pitch emails with attractive startup characteristics is about 6.5\% in 2018, an economic boom period. This paper shows that the email response rate to pitch emails with quality variations (i.e., including both attractive and average startup characteristics) is roughly 1\% in 2021, an economic recession period. The difference in email response rates in these two papers comes from both the pro-cyclical nature of early-stage VC investment and the additional quality variations introduced in this study. \cite{bertrand2017field} discuss other standard limitations of the correspondence test method.} making it difficult for researchers to introduce variations in startup quality and test the nature of discrimination.\footnote{Disentangling the nature of discrimination requires researchers to separate various belief-based mechanisms (i.e., ``statistical discrimination”) (\cite{bertrand2017field}, \cite{altonji1999race}, \cite{phelps1972statistical}, \cite{arrow1973theory}) from different taste-based mechanisms (i.e., ``animus") (\cite{becker2010economics}). This disentanglement is difficult to accomplish in discrimination literature (\cite{gneezy2012toward}) despite its importance in both welfare analysis and policy-making (\cite{bohren2019dynamics}, \cite{neumark_detecting_2012}).}\par

\vspace{2mm}
To overcome the identification challenges mentioned above and reconcile the mixed results in the literature, this study implements two complementary field experiments in the U.S. entrepreneurial finance market, referred to as Experiment A and Experiment B in this paper. Experiment A combines two preference elicitation techniques: an ``incentivized resume rating (IRR)" experiment (\cite{kessler_incentivized_2019}) and a ``donation game" (\cite{carpenter2008altruistic}). To implement it, I collaborate with several accelerators and build a ``Nano-Search Financing Tool," a machine learning matching tool that facilitates VCs' deal sourcing process. In the first part of this matching tool, investors need to evaluate multiple dimensions of randomly generated startup profiles. Investors know the profiles are hypothetical, but truthful evaluations help the algorithm to better match investors with their preferred startups. Hence, the IRR experiment uses real investment opportunities to reveal VCs' investment preferences and directly identify the nature of discrimination.\footnote{To increase the sample size, some randomly selected investors also receive a ``monetary incentive" following \cite{armona2019home} so that the more accurate investors' evaluation results are, the larger the monetary award the lottery winners will receive. Although the ``monetary incentive" elicits slightly different sources of investors' beliefs compared to the ``matching incentive", Section \ref{sec:irr} justifies its usage by showing that investors' profitability judgements elicited by these two incentives do not differ quantitatively.} Besides investment, VCs also add value to startups by providing non-investment support. To test whether discrimination also exists in a non-investment setting, in the second part of this matching tool, investors can anonymously donate a portion of a provided unexpected \$15 Amazon Gift Card to randomly displayed startup teams. The researcher will use the donated money to purchase small gifts for the corresponding startup teams in the collaborative incubators and provide founders with anonymous encouragement from investors during the COVID-19 pandemic recession.\par

\vspace{2mm}
Experiment A yields the following findings. First, although Experiment A does not discover aggregate-level discrimination against minority founders, it does identify significant \textit{implicit} discrimination against female founders. After investors become fatigued from evaluating multiple profiles, their ratings of women-led startups begin to decline. At the end of the IRR experiment, women-led startups are 16.67\% less likely to be contacted by investors and also receive 19.4\% less funding compared to similar men-led startups. Second, results from quantile regressions demonstrate a distributional effect in the initial contact stage, suggesting that the direction and magnitude of implicit discrimination can vary across the spectrum of startups' attractiveness. Significant and strong implicit gender, racial, and age discrimination mainly exists when investors evaluate top startups, especially in the tech sector. Hence, results that capture the fundraising situation of struggling startups are not generalizable to that of attractive startups. Third, as predicted by \cite{morgan2009diversity}, implicit discrimination is pro-cyclical. Discrimination becomes more severe when investors' internal thresholds increase and their startup investment criteria become more selective. However, discrimination might reverse when investors are sufficiently unselective. Fourth, homophily exists in a non-investment setting.\footnote{``Homophily effect" refers to the tendency for people to seek out or be attracted to those who are similar to themselves.} Compared to male founders, male investors on average donate \$3.20 less to female founders and are also 77\% less likely to donate all money to female founders. Similar homophily exists within the white male group and within the other group.\par

\vspace{2mm}
Experiment A shows that some investors exhibit implicit discrimination against female and Asian founders. However, there are also some impact funds that support minority founders in the sample. So, how divided is the investment community in terms of their attitude towards minority founders, and what separates us? To answer this question in Experiment A, I develop a consistent ``decision-based heterogeneous effect" estimator by using the ``leave-one-out" technique and exploiting both the rich data and the exogeneous ``within-individual" randomization of the IRR experiment.\footnote{Junlong Feng has provided crucial help and discussions for developing this estimator.} Standard heterogeneous effects rely on participants' pre-determined demographic information. However, since each investor evaluates multiple randomized profiles, researchers can identify ``individual-level" preferences and classify the participants into an ``anti-minority" group and a ``pro-minority" group based on their indicated decisions. The estimator finds that investors' profitability ratings can explain investors' heterogeneous contact decisions more than other mechanisms. For example, investors who prefer not to contact female founders expect women-led startups to have 16.40 percentile ranks lower potential financial returns than men-led startups. However, investors who prefer to contact female founders expect women-led startups to have 7.93 percentile ranks higher potential financial returns than men-led startups. However, investors' availability ratings are not significantly different between the ``pro-minority" groups and ``anti-minority" groups. Similar ``decision-based heterogeneous effects" also exist when investigating racial discrimination and agism. Therefore, holding different beliefs in the profitability of minority-led startups and majority-led startups is an important reason for the polarization in investors' attitudes towards minority founders.\par

\vspace{2mm}
Although Experiment A identifies that discrimination mainly exists when investors evaluate highly-rated startups, it does not have enough statistical power to investigate how investors evaluate struggling startups. Hence, the paper follows up with Experiment B, a redesigned correspondence test that enables researchers to identify the nature of discrimination in the cold call, pitch email setting. During the COVID-19 outbreak (03-04/2020), I sent hypothetical pitch emails to more than 17,000 venture capitalists with randomized founder names indicative of gender and race, randomized founder educational backgrounds, and randomized startup project characteristics displayed in both the emails' \textit{subject lines} and in the emails' contents. By utilizing new email tracking technology, I can monitor detailed information acquisition behavior for each investor, including email opening behavior, time spent on pitch emails, click rate on the inserted startup’s website, the contents in email replies and the email response rate.\footnote{This experimental design and behavioral measurements generated enough experimental power to survive in the harsh experimental environment of the pandemic, when early-stage investors dramatically slowed down their investment pace (\cite{howell_financial_2020}).} Since most high-quality startups would prefer a ``warm" fundraising method, Experiment B mainly studies how investors evaluate struggling startups.\footnote{Before using this experimental design, the researcher originally tried to collaborate with real startup teams and send their randomized truthful information to investors. However, it was extremely hard to recruit enough real startups to cover major industries that VCs invest in. Startup founders are usually worried that the success rate of raising funding through cold call pitch emails is extremely low. Moreover, this ``cold" method might make a bad impression on VCs, indicating that the startup has failed with other warmer fundraising methods.}\par

\vspace{2mm}

Experiment B's results confirm that investors are slightly biased \textit{towards} female and Asian founders in the pitch email setting. Using minority names generally increases the email opening rate by roughly 1\% compared to using majority names. After further testing the nature of bias, the paper finds that the bias towards female founders is likely driven by taste-based mechanisms. The positive ``female name" effect is much larger for impact funds, and revealing additional quality signals does not shrink the gender gap. However, the bias towards Asians is likely driven by belief-based mechanisms because revealing additional quality signals reduces the racial gap. As Experiment B was mainly implemented between 03/2020-04/2020, it also finds a temporary discrimination against Asian founders during the COVID-19 outbreak. Investors spent 24\% less time on pitch emails sent by Asian names compared to white names in March 2020. However, this discrimination quickly reversed starting in April 2020. Hence, discrimination behaviors might also be temporarily affected by big societal events.\par

\vspace{2mm}
The contribution of this paper is both empirical and methodological. Empirically, combining results from Experiment A and Experiment B, this paper shows that in the pre-screening stage, investors discriminate \textit{against} minority groups when evaluating attractive startups and are biased \textit{towards} minority groups when evaluating struggling startups. This distributional effect not only provides an explanation to reconcile contradictory results in the literature, but also confirms the ``discrimination reversion" phenomenon that has been theoretically predicted by several fundamental discrimination theory papers (\cite{morgan2009diversity}, \cite{phelps1972statistical}, \cite{aigner1977statistical}, \cite{lundberg1983private}) but has never been empirically tested before.\footnote{Another possible explanation to reconcile the literature is that investors do not exhibit implicit discrimination when replying to pitch emails, which is also consistent with the results of \cite{gornall_gender_2020}. However, Experiment B finds that investors check emails at different times during the day. Considering that implicit discrimination is already apparent when investors have a relatively light cognitive workload in Experiment A, it is very likely that implicit discrimination also affects investors' behaviors when they check emails throughout the day. This makes this explanation less plausible.} Experiment A also confirms the ``pro-cyclical discrimination" phenomenon predicted by the model in \cite{morgan2009diversity}. Furthermore, it documents the existence of homophily in a non-investment setting and a temporary, stronger discrimination against Asian founders during the COVID-19 outbreak. Therefore, this paper empirically contributes to both discrimination literature and entrepreneurial finance literature.\par

\vspace{2mm}
Methodologically, the main contribution of this study is to provide a framework to identify discrimination and its nature in a financial and high-skilled labor market after solving multiple challenges faced by commonly used methods through a series of redesigned field experiments. Moreover, the developed ``decision-based heterogeneous effect" estimator, which uses rich data and ``within-individual" level randomization in the IRR experiment, helps to measure how divided the society is and what drives the polarization of people's attitudes towards minority groups.\par

\vspace{2mm}
To the best of my knowledge, this is also the first paper to implement the correspondence test and the IRR experiment in the same context and compare their results. The IRR experimental paradigm, which is an incentivized elicitation technique invented by \cite{kessler_incentivized_2019}, is motivated by the desire to provide a more ethical experimental design that can substitute for the standard correspondence test involving deception.\footnote{Special thanks go to Corinne Low for her insightful discussions clarifying the following important nature of the IRR experiment. Following the widely accepted Becker-Degroot-Marschak elicitation techniques of willingness to pay, the IRR experiment provides an incentive structure for eliciting true preferences and provides within-individual level exogenous variations. Also, ``the primary context of the IRR experiment is usually non-experimental, and subjects’ motivation for participating in the study is mainly to receive the commercial benefits". Unlike a ``survey," an IRR experiment requires much more social resources in order to reveal true preferences.} By comparing the results from these two experimental methods, this paper shows that in a high-skilled labor market, results from a correspondence test that mainly captures a ``cold" setting may not match results from an IRR experiment that captures a ``warm" setting, suggesting that neither may be generalizable without caveats. The rich data collected by the IRR experiment enable researchers to identify subtle mechanisms, test novel heterogeneous effects, distributional effects, and dynamic changes in different market conditions. Notwithstanding these impressive merits, increasing use of the IRR experimental method in the future may gradually decrease its ability to detect discrimination due to the potential consent form effect.\footnote{After the IRR experimental design and the corresponding data analysis methods become more well-known among the public, experimental participants might carefully control their behaviors even in the second half the study. I leave addressing these limitations to future research.}\par

\vspace{2mm}
This paper is organized as follows. Section \ref{sec:data} discusses the construction of the individual-level VC investor database. Section \ref{sec:irr} presents the design of Experiment A and analyzes investors' evaluations of startup profiles and donation behaviors. Section \ref{sec:correpondencetest} describes the design of Experiment B and analyzes investors' information acquisition behaviors. Section \ref{sec:discussion} provides a comparison of these two experiments, the link to discrimination theories, and related policy implications. Section \ref{sec:conclusion} concludes.\par

\hypertarget{sec:data}{\section{Data}}\label{sec:data}
To implement these field experiments, I have constructed a cross-sectional, individual-level global venture capitalist database, which contains updated demographic information and contact information for 17,882 investors before 02/2020.\footnote{All investors' email addresses had been verified by a testing email before the experiments began.} Since all experiments are implemented in English, only investors from English-speaking areas are included. This database combines multiple commercial databases and manually collected data. Detailed database descriptions and the key variable construction process are provided in Online Appendix \ref{sec:data_construction}.\par

\vspace{2mm}

Despite the granular information provided by this database, it is important to realize the following three limitations. First, this database contains systematically more investors from the U.S. as well as more senior VCs due to data availability online and the data collection method used by data companies.\footnote{Most of the commercial databases used here are provided by U.S. data companies and collected by English speakers with the exception of Zdatabase, which is the most comprehensive and timely database covering VC and PE activities in China.} Hence, it may not be representative of the true geographical distribution of all venture capitalists in the world. Second, because of the high turnover rate within the VC industry, the contact information and status of these investors need to be updated frequently before use. Third, except for the key variables like gender, seniority, and location, other demographic variables are only available for relatively famous investors whose biographies are more readily available online. \par
\vspace{2mm}

The Summary Statistics of the 17,882 investors’ demographic information is provided in Table \ref{investor_summary}. Panel A reports the location distribution of these investors, showing that US-based investors account for 84.91\% of this set of investors.\footnote{Maps of investors' global geographical distribution and U.S. geographical distribution are provided in Online Appendix Figure \ref{map_global} and \ref{map_US}.} Panel B shows that most investors are interested in the Information Technology industry. Other important preferred industries include Healthcare, Consumers, and Energy. Panel C summarizes investors' background information. On average, female investors account for 24\% of total investors. This is consistent with the NVCA/Deloitte survey results showing that women accounted for 21\% of investment professionals in the U.S. VC industry in 2018 due to recent progress in increasing diversity.\footnote{See \href{https://www2.deloitte.com/content/dam/Deloitte/us/Documents/audit/us-audit-egc-nvca-human-capital-suvey-2018.pdf.} {``NVCA–Deloitte
Human Capital Survey 2019."} \cite{gompers2014gender} also show that women are under-represented among senior investment professionals in the VC industry.} Senior investors, who are partners, presidents, C-level managers, or vice presidents and above, account for 84\% of total investors in the sample. Most investors are institutional investors, and angel investors only account for 11\% of the sample. Moreover, only 2\% of all investors work in not-for-profit impact funds.\footnote{Pitchbook classifies VC funds into not-for-profit funds and for-profit funds together with the description of their investment preferences. If I use indicative keywords in the fund descriptions to classify the VC funds following \cite{barber2020impact}, this percentage increases to 6\%-8\% depending on the keyword selection method.} \par
\hypertarget{sec:irr}{\section{Experiment A's Design and Results}}\label{sec:irr}
The goal of designing Experiment A as presented here is to elicit investors' preferences for startups with a stronger incentive (i.e., real investment opportunities). Since venture capitalists add value to startups by providing both funding (i.e., investment) and non-monetary support,\footnote{A large body of literature has investigated the non-monetary support of venture capitalists, see \cite{lerner2022venture}, \cite{bernstein2016impact}, etc.} Experiment A combines the following two preference elicitation techniques: i) the IRR experiment, designed to test discrimination and belief-based mechanisms in an investment setting, and ii) the dictator game, designed to test discrimination and potential taste-based mechanisms in a non-investment setting in which investors provide anonymous encouragement to startups during economic hardship.\par

\subsection{Experimental Design} \label{irr:design}

\textit{\textbf{A. Recruitment Process and Sample Investors}}\par

Experiment A was implemented from 03/2020 - 09/2020 using only online recruitment methods. I sent invitation emails together with instruction posters to the 15,000+ U.S. venture capitalists who also participated in Experiment B during the same period.\footnote{During the COVID-19 pandemic, Columbia IRB paused all field work which involves person-to-person activities due to COVID-19. The recruitment email templates and the instruction poster templates are provided in Online Appendix \ref{sec:appendix_irr} Figure \ref{fig:recruitment1}, Figure \ref{fig:recruitment2}, Figure \ref{fig:poster1}, and Figure \ref{fig:poster2}.} This ensures comparability across these experiments. To recruit real venture capitalists and create an experimental setting that closely mimics the real world, I have partnered with several real incubators and we have built a machine learning, algorithm-based matching tool called the ``Nano-Search Financing Tool". Developing these kinds of data-driven matching tools has become popular in the VC industry. Incubators and VC funds, such as Techstars, Social+ Capital, and Citylight Capital, have worked extensively on developing machine learning algorithms to help evaluate investments, seek deals, and complement face-to-face multiple stage investment strategies. Similarly, our tool aims to match investors with startups in the collaborating incubators and mainly captures investors' preferences in the pre-selection stage (i.e., ``ex ante screening" in \cite{cornell1996culture}).\par

\vspace{2mm}
In total, 69 real U.S. investors from 68 different VC companies participated in Experiment A, which provides 1,216 startup profile evaluation results.\footnote{At the beginning of the study, each investor evaluated 32 profiles, and 6 investors finished the 32-profile version of the evaluation task. However, to recruit more investors, later participants only needed to evaluate 16 profiles. Also, one investor participated in the experiment twice for two different VC funds. Results are similar after removing the first 6 investors.} The sample size is comparable with \cite{kessler_incentivized_2019}. Given that the response rate is roughly 0.5\%, Table \ref{irr_investor_summary} compares the observable characteristics of recruited investors and the venture capitalists recorded in Pitchbook Database to check for sample selection bias during the recruitment process. Panel A shows that the recruited investors' sectors of interest are representative and cover all major industries focused on by VC investors. Panel B shows that 67.1\% of the recruited investors are early-stage investors, who are interested in the Seed stage. Panel C shows that while the sample investors are representative in terms of gender, 42\% of investors belong to minority groups (i.e., Asian, Hispanic, African, etc.). This is higher than the percentage of minority investors in the U.S. Furthermore, 86\% of recruited investors are in senior positions, as their contact information is more readily available in existing databases. \par

\vspace{2mm}
The major concern is whether Experiment A only recruits investors who systematically discriminate more against women and Asians. Since each experimental participant's identity and affiliated VC company is observable, Panel E further compares the investment histories of the 69 recruited investors’ affiliated VC companies and those in Pitchbook. Results show that recruited investors are more likely to come from large and active VC companies. As the experiment was implemented during the pandemic recession, most early-stage VC investors shifted to ``survival mode" and paused new investments (\cite{howell_financial_2020}). Hence, only those large and active VC companies were still attracted by the potential investment opportunities offered by Experiment A. Importantly, recruited investors' affiliated VC companies do not invest less in women-led startups compared to other investors in Pitchbook. Some might also speculate that recruited investors may not care about the incentives provided by Experiment A. Since \cite{camerer1999effects} shows that experimental subjects behave in a more pro-social way in weakly incentivized experimental settings, this implies that Experiment A might underestimate investors' discrimination against women and Asians.\par

\vspace{2mm}
\textit{\textbf{B. Survey Tool Structure}}\par
\vspace{2mm}
The survey tool mainly contains the following two sections. After reading the consent form, investors first enter the profile evaluation section (i.e., the IRR experiment section), in which they need to evaluate 16 randomly generated startup profiles and answer standard background questions. The second section is the donation section (i.e., the dictator experiment section), in which investors decide how much of an unexpected \$15 Amazon Gift Card they want to donate to randomly displayed startup teams. Online Appendix Figure \ref{fig:irr_flow_chart} provides a flowchart for Experiment A.\par
\vspace{2mm}

To help participants understand how the incentive works, I also provide an instruction page before the first profile evaluation section. Although investors know that all the startup profiles are hypothetical, this instruction page emphasizes that ``the more accurately they reveal their preferences, the better outcomes the matching algorithm will generate (and the higher financial return that the lottery winner will obtain)." Moreover, since most VC investors only invest in startups in their industries and stages of interest (i.e., ``qualify/disqualify" test), I ask all the participants to assume that the generated startups they will be evaluating are in their industries and stages of interest. The matching tool will collect each investor’s
preferred industry and stage after the profile evaluation section.\footnote{One can also design separate startup profiles with more customized business models for investors from different industries. I did not do this because the market changes very quickly in the entrepreneurial community, especially during the COVID-19 period. Startups' business models created in the design stage are likely to be invalid during the recruitment stage. Moreover, to obtain insights from investors in diverse industries, Experiment A should provide general information that accommodate as many participants with diverse backgrounds as possible.\par}\par

\vspace{2mm}


\textbf{\textit{B.1 Profile Evaluation Section (IRR Experiment)}}\par
\vspace{2mm}

Following the factorial experimental design, Experiment A randomizes multiple startup characteristics simultaneously and independently in each created startup profile. The researcher first creates a set of startup team and project characteristics. Then the backend Javascript code randomly draws different characteristics and combines them together to dynamically create a hypothetical startup when each investor evaluates a new startup profile.\footnote{Sometimes the random combination may generate unusual cases, such as a startup with 50+ employees that still does not generate profits (see Amazon’s history). However, such cases account for a small percentage of total generated cases.} To test implicit discrimination, Experiment A also deliberately introduces a short break after investors evaluate the first eight startup profiles. This break presents a page indicating the investor's progress and encouraging them to finish all the evaluations. All the randomization of different startup components is provided in Table \ref{irr_randomization}.\footnote{The detailed construction process of the startup characteristics is provided in Online Appendix \ref{sec:appendix_irr}.}\par

\vspace{2mm}
To make profiles more realistic, Experiment A implements the following two designs. First, the information provided about startups follows a ``Crunchbase" format and is usually publicly available on LinkedIn, Crunchbase, or AngelList.\footnote{\hyperlink{https://about.crunchbase.com/products/crunchbase-pro/?utm_source=google&utm_medium=cpc&utm_campaign=Brand-Pro-NAM&utm_content=Conversion&gclid=CjwKCAjw8MD7BRArEiwAGZsrBeRWmEli5Q15EhRc2odUSojcPyVZUZ4RUauMt0Ff-n0j3yr_8julXRoCd34QAvD_BwE}{Crunchbase} is a commercial platform that provides public information about startups mainly in the U.S.} Investors, like Plug and Play Tech Centers, sometimes go to these public platforms to seek relevant startups that fit their portfolios. The current design mimics this startup-seeking setting. Second, descriptions of startup founders' experiences are all extracted from real startup founders' biographies.\par

\vspace{2mm}
\textbf{\emph{Manipulating Gender and Race.}} --- To indicate the gender and race of the startup founder, Experiment A randomly assigns each hypothetical startup team member a first name highly indicative of gender (male or female) and a last name highly indicative of race (Asian or white). To make such information more salient, all the members in the same startup team are assigned names of the same gender and race. Moreover, Experiment A also includes the founder’s name in the evaluation questions and uses indicative words like ``she/her/his/him/he". Similar to other components, the combination of first names and last names is dynamically implemented by Qualtrics.\footnote{Considering our collaborative incubators and startups have relatively more Asian founders and female founders, the ratio of female and male startup founders are both 50\% to maximize the experimental power. A similar ratio is used for Asian founders and white founders.} The detailed name selection process and the list of full names are provided in Online Appendix \ref{sec:appendix_irr} Table \ref{appendix_irr_full_name}.\par
\vspace{2mm}

\textbf{\emph{Manipulating Age}} --- The age of the startup founder is indicated by the graduation year from their college or graduate school rather than being listed directly. If a team has two co-founders, their age falls in the same range, which belongs to either the older group (who graduated before 2005) or the younger group (who graduated after 2005). I assume founders graduate from college at the age of 23,\footnote{It is suspicious to list age directly in a startup's profile because it is not common practice. Hence, using the graduation year as a proxy achieves more realism. Using 22 gives similar results. However, some investors may assume that founders graduate from graduate school rather than from an undergraduate program at these universities; hence, 23 is used.} so the approximated age is calculated by the formula: age $= 2020 - $graduation year$ + 23$.\par

\vspace{2mm}
\textbf{\textit{Evaluation Questions}}\par

\vspace{2mm}
The evaluation questions include three mechanism questions and two decision questions. Considering that most venture capitalists are well-educated and market savvy, I use probability or percentile ranking questions rather than Likert scale questions.\footnote{Using probability and percentile ranking questions has two advantages. First, these questions are more objective compared to the Likert scale. Second, the wide range from 1 to 100 provides richer evaluation results and additional statistical power. \cite{brock2020discriminatory} also use probability questions to replace Likert Scale questions when they recruit real Turkish bankers to evaluate different loan profiles.} This design provides continuous outcome variables, which allows researchers to implement infra-marginal analysis or distributional analysis that explore how investor preferences vary across the distribution of startups' attractiveness. Screenshots of evaluation questions are provided in Online Appendix \ref{sec:appendix_irr} Figure \ref{fig:Q1}, and Figure \ref{fig:Q5}.\par

\vspace{2mm}
\textbf{\textit{(Belief-based) Mechanism Questions}} --- The mechanism questions are designed to test the following three standard, belief-based discrimination mechanisms. First, being a minority can be indicative of a startup’s future financial returns (i.e., the first moment). To test this mechanism, investors need to evaluate the percentile rank of each startup's probability of generating higher financial returns compared to the startups they have previously invested in (i.e., ``profitability" evaluations $Q_1$). Second, given that the entrepreneurial financing process is essentially a two-sided matching process, investors also need to evaluate the probability that a startup will accept their investment rather than other investors' (i.e., ``availability" evaluations $Q_2$). Third, investors may use the founder's group membership as an indicator of a startup's risk (i.e., the second moment). Hence, investors also evaluate the risk percentile rank of each startup profile (i.e., ``risk" evaluations $Q_5$). $Q_5$ was not in the original design but was added later based on investors' feedback for the purpose of checking robustness. Therefore, it only applies to a small number of investors receiving the matching incentive.\footnote{$Q_5$ is placed after all the other evaluation questions to minimize its impact on the originally designed questions} Due to the small sample size of $Q_5$, risk-related results are not reported.\par

\vspace{2mm}
\textbf{\emph{Decision Questions}} --- The two decision questions are designed to examine both the investor's contact interest ratings ($Q_3$) and intended investment interest ratings ($Q_4$). Specifically, $Q_4$ asks for the relative investment magnitude rather than the absolute investment magnitude as different investors have different ranges of targeted investment amounts. In order to accommodate more investors, I have tried to make $Q_4$ as standardized as possible.\par

\vspace{2mm}
\textbf{\textit{Background Questions}}\par
\vspace{2mm}
After the investor evaluates all the profiles, Experiment A also collects standard background information about each participant to determine how representative the sample investors are and analyze potential heterogeneous effects based on predetermined investor characteristics.\footnote{All the background questions are placed after the evaluation section to avoid priming subjects to think about any particular characteristics that the research project aims to test.} Such background information includes investors' preferred industries, stages, special investment philosophies, gender, race, and other standard demographic information.\par

\vspace{2mm}
\textbf{\textit{B.2 Donation Section (Dictator Game)}}\par

Experiment A also inserts a donation section at the end of the survey tool to investigate whether investors treat minority groups and majority groups differently when providing non-investment support. Investors are informed that they will receive an \textit{unexpected} \$15 Amazon Gift Card to thank them for participating in this experiment.\footnote{To avoid polluting the incentive structure in Experiment A, the compensation with the \$15 Amazon Gift Card is mentioned only at the very end of the survey tool rather than in the consent form.} However, they can decide whether to donate a portion of the provided \$15 to randomly displayed startup founders. For instance, if the investor donates \$3, she/he will receive a \$12 Amazon Gift Card. The researcher will use the donated money to purchase a small gift for the corresponding type of startup founders in our collaborative incubators and give them anonymous encouragement from investors during the pandemic recession.\footnote{I provide a small gift rather than cash to founders because a small gift is usually more associated with warm encouragement, while giving a small amount of cash can sometimes be considered insulting.}\par

\vspace{2mm}

Experiment A orthogonally randomizes the gender and race of displayed startup founders by changing the pictures displayed and the wordings used in the description. The options investors may randomly be provided with include donating to the ``Women’s Startup Club” (mainly white female founders), ``Asian Women’s Startup Club” (mainly Asian female founders), ``Asian Startup Club” (mainly Asian male founders), or just ``our Startup Club” (mainly white male founders). To make the information of gender and race more salient, I also add a picture containing four startup founders of the same gender and race so that experimental participants understand what type of founders they are donating to.\footnote{The concern with using pictures in the experiment is that the appearance or other messages delivered by the pictures cannot be fully controlled. To ameliorate this issue, I use four founders’ pictures combined together to send the signal of gender and race. All the pictures are obtained from a public library (i.e., Wikimedia Commons, Freeimages, etc.) with no copyright problems.}  All individuals in the pictures are smiling and professionally dressed to make sure that they are as much on equal footing as possible. The detailed donation question and an example founder's picture are provided in Online Appendix \ref{sec:appendix_irr} Figure \ref{donation_founder}.\par

\vspace{2mm}

\textbf{\emph{Limitations and Justifications}} --- The donation game does not capture discrimination in the investment process. Instead, it provides insights into whether discrimination exists when investors provide non-investment support, which is also crucial for startups' success. Some may be concerned that investors may not care about \$15, suggesting that this sub-experiment does not have high enough stakes. According to \cite{camerer1999effects}, this implies that the donation game underestimates the level of discrimination due to both extra noise and the ``presentation effect", which is a typical limitation of weakly incentivized experiments. \par

\vspace{2mm}


\textbf{\textit{C. Incentive Structure}} \par

\vspace{2mm}
As a preference elicitation technique, one key point of the IRR experimental design is its incentive structure. The following incentives are designed not only to increase the stakes of Experiment A and impose costs for making inefficient and inaccurate evaluations, but also to bring real value to all the experimental participants.\par
\vspace{2mm}

\textbf{\textit{Matching Incentive}} --- For some randomly selected investors who receive the recruitment email (Version 1), I only provide the following “matching incentive” following \cite{kessler_incentivized_2019}. Basically, after each investor evaluates all the startup profiles, a machine learning algorithm is used to identify matching startups from the collaborative incubators. If the matched startups are also interested in the investor’s investment philosophy, they will contact the investor for a potential collaboration opportunity. The matching algorithm uses all of the investors' evaluation answers to identify their investment preferences. Therefore, all five evaluation questions are incentivized by this incentive. A description of the algorithm is provided in the consent form.\footnote{This ``matching incentive" has the following merits. First, researchers can apply it to any other two-sided matching markets. Second, it can incentivize all the evaluation questions, unlike the monetary incentive. Third, if the designed matching algorithm can improve the matching efficiency, such an incentive can bring real value to both sides of the matching market. Despite the merits mentioned above, implementing this incentive often requires researchers to have certain social resources and connections.}\par

\vspace{2mm}

\textbf{\emph{Monetary Incentive}} --- To increase the sample size, I provide both the ``matching incentive" and an extra “monetary incentive”, as used by \cite{armona2019home}, to the remaining randomly selected investors who received the recruitment email (Version 2). This ``monetary incentive” is essentially a lottery in which 2 experimental participants will be randomly selected to receive \$500 each plus an extra monetary return closely related to their evaluations of each startup’s quality. The more accurate their evaluations are of each startup’s profitability, the bigger the financial return they will obtain as a lottery winner.\footnote{For example, consider if Peter Smith, one hypothetical experimental participant, is chosen as the lucky draw winner. In his survey, he indicates that on average, male teams are more likely to generate higher financial returns. In that case, the researcher can construct a portfolio containing more real startups with male teams. After one year, based on the financial performance of real startups in Pitchbook, this portfolio containing more startups with male teams generates a 10\% return. Thus, Peter Smith receives \$500 + \$500*10\% = \$550 as his finalized monetary compensation one year after he participates in the survey. \$500*10\%=\$50 is the ``extra monetary return.” The historical return of the VC industry is roughly between -15\% and +15\%, which means that the range of expected monetary compensation is roughly between \$425 and \$575.} However, this is only used to incentivize the ``profitability evaluation question" (i.e., ``Q1"). The evaluation results will be determined based on the Pitchbook data published in the 12 months following completion of the recruitment process. I separately informed these two lottery winners that they would receive the award at the end of July 2020. The evaluation algorithm is provided in the consent form (Version 2).\footnote{This ``monetary incentive" has both merits and limitations. First, it mimics the real investing process in which investors have a certain amount of principal and need to evaluate different startups accurately to generate maximum return. Second, it does not require many social resources. Third, researchers can apply it to more general situations besides a two-sided matching market. However, the current version can only incentivize the evaluation of startups' profitability (i.e., $Q_1$) to avoid distorting participants’ evaluations on other questions. If the collaboration likelihood (i.e., $Q_2$) is added to the financial return algorithm, then all the participants may claim that the best startups would be willing to collaborate with them even if that is not true. Similarly, if contact interest ratings (i.e., $Q_3$) and investment interest ratings (i.e., $Q_4$) are added to the financial return algorithm, participants may be motivated to distort their true evaluations in order to maximize their financial return as both $Q_3$ and $Q_4$ can be affected by $Q_2$.}\par

\vspace{2mm}

\textbf{\textit{Justification}} --- One concern with adding the ``monetary incentive” is the possibility of attracting participants who do not value the matching incentive, which results in extra noise. Another concern is that the ``monetary incentive" essentially elicits each subject's judgement of how the market evaluates each startup's profitability, which might be different from the subject's own judgement of each startup's profitability as incentivized by the ``matching incentive".\footnote{Although ``monetary incentive" and ``matching incentive" elicit two slightly different sources of beliefs in a startup's profitability, these beliefs are all individual-level investors' beliefs of a startup's profitability that help to prove the existence of ``statistical discrimination" or belief-driven discrimination. In this experimental setting, these two quality judgements do not differ quantitatively.} To address these concerns, I have compared the evaluation results of investors who receive only the ``matching incentive” and those who receive both incentives. The comparison results are provided in Online Appendix \ref{sec:appendix_irr} Table \ref{irr:incentive_comparison}, showing that these two incentive structures do not cause systematically different evaluations. The interaction terms between the incentive structure and a startup's gender and race are not significant. Moreover, this experiment discovers multiple highly significant startup team and project characteristics that are crucial for investors' investment interest ratings as shown in \cite{zhang2020venture}. This means that investors understand the incentives and evaluate all the questions carefully.\footnote{Researchers can also separately ask subjects questions that can test their understanding of the incentive (\cite{casaburi2018time}).}\par

\vspace{2mm}

\subsection{Experimental Results}\label{irr:result}

I denote an investor $i$’s evaluation of a startup profile $j$ on evaluation question $k$ as $Y_{ij}^{(k)}$ and estimate the following regression.  Formally,
\begin{eqnarray}
    Y_{ij}^{(k)}= X_{ij}\beta^{(k)} +\alpha_i+\epsilon_{ij}^{(k)}
\end{eqnarray}
$X_{ij}$ represents any founder's demographic information, like gender and race. $\alpha_i$ are investor fixed effects that account for different average ratings across investors. Since each type of startup characteristic is randomized orthogonally and independently, the coefficient $\beta^{(k)}$ has a causal interpretation. Standard errors are clustered at the investor level.\par
\vspace{2mm}

Online Appendix Table \ref{Group-level IRR ATE} reports corresponding regression results testing group-level ATE of a startup founder's gender (Panel A), race (Panel B), and age (Panel C) using the total 1,216 profile evaluation results. ``Female Founder", ``Asian Founder", and ``Older Founder" are indicators that equal one if the startup founder has a female first name, Asian last name, and graduated before 2005, respectively. They are equal to zero, otherwise. The dependent variables are investors' evaluations of startups' profitability ($Q_1$), availability ($Q_2$), contact interest ratings ($Q_3$), and investment interest ratings ($Q_4$). After adjusting p-values to account for multiple hypothesis testing, the study does not find any discrimination evidence in aggregate.\footnote{The study also does not find investors spend significantly different amounts of time on evaluating majority founders' and minority founders' profiles. This rules out attention discrimination in the profile screening process (see Online Appendix \ref{sec:appendix_irr} Table \ref{table_attention_irr}).}\par
\vspace{2mm}

The lack of aggregate-level significant evidence occurs in almost all papers that use the IRR experimental method to detect socially sensitive preferences (\cite{kessler_incentivized_2019}, \cite{zhang2021impact}). This null result may potentially be due to the following reasons. First, investors understand that they are participating in a research project and hence might behave in a more friendly manner to minorities due to the Hawthorne effect. Second, as there are more minorities among recruited investors, sample selection bias might also contribute to the null result. Third, heterogeneous effects and implicit discrimination might exist as shown in \cite{kessler_incentivized_2019}. Fourth, if discrimination mainly exists at the top level, the aggregate-level ATE cannot capture it. All these reasons call for more thorough data analysis. As shown below, the study finds that discrimination against minorities mainly concentrates on top startups and exists in the second half of the study when investors become more fatigued or more familiar with the evaluation process.\par

\vspace{2mm}
\textbf{A. (IRR Experiment) Discrimination in the Second Half of Study}\par
\vspace{2mm}
\textbf{A.1 (Average Treatment Effect) \textit{Implicit} Discrimination Against Female Founders.}\par
\vspace{2mm}
Given the aggregate-level null result, the study further investigates whether \textit{implicit} gender and racial discrimination exists. ``Implicit discrimination” refers to the attitudes or stereotypes that affect investors' decisions in an unconscious manner. According to \cite{bertrand_implicit_2005} and \cite{cunningham2015implicit}, implicit discrimination can significantly affect people's behaviors when the task involves ambiguity, time pressure, evaluating mixed attributes, or causes a higher cognitive load and inattentiveness.\par

\vspace{2mm}

The task of screening startups in the pre-selection stage satisfies these criteria. First, the startup selection process usually involves considerable ambiguity since there are no clear standards for evaluating the attractiveness of each potential deal. Second, many investors need to evaluate a large number of startups quickly before narrowing down their potential investment targets. Considering that the level of fatigue introduced in this experiment is not high, any detected ``implicit discrimination" against minorities may play a more important role in the real-world investment process. \par

\vspace{2mm}
\cite{kessler_incentivized_2019} creates two methods of testing implicit discrimination using the IRR experiment: 1) comparing evaluations of the first half of the profiles and the second half of the profiles \emph{within} each block; and 2) comparing evaluations of the first half of the study and evaluations of the second half of the study. The logic behind both methods is that when investors become more rushed or fatigued, their implicit preferences are more likely to emerge and affect their judgments. This discrimination due to the ``fatigue effect” has been classified as implicit discrimination in the IRR experiment. Since only the second method has been pre-registered for this experiment, results of implicit discrimination in this paper rely on the comparison of evaluations in the first half of the study and those in the second half of the study.\footnote{The pre-registered analysis plan for this experiment (AEARCTR-0004982) is restricted from access by the public, but is available upon request. The method of testing implicit discrimination is pre-registered on Page 5 of ``Analysis Plan Version 1.pdf". }\par
\vspace{2mm}

Table \ref{implicit bias gender and race} displays aggregate-level \textit{implicit} discrimination results based on a founder's gender (Panel A) and race (Panel B). It shows whether investors' ratings of minority founders decline after the inserted short break by using the second method proposed by \cite{kessler_incentivized_2019}. ``Second Half of Study" is an indicator variable for profiles shown among the last half of profiles viewed by the investor. In Column (1), the dependent variable is investors' response time, defined as the number of seconds before each page submission and winsorized at the $95^{th}$ percentile (59.23 seconds on average). Each regression adds investor fixed effects and clusters standard errors at the investor level.\par

\vspace{2mm}
Results show that female founders receive significantly lower ratings in the second half of the study compared to those in the first half of the study. Column (1) shows that investors indeed spent 27 seconds fewer evaluating each profile after the inserted break, suggesting the potential existence of more fatigue and rushed evaluations at the end of the study. In Panel A Columns (2)-(5), all the interaction terms of ``Female Founder" and ``Second Half of Study" are negative. Specifically, the p-value of the interaction term in Column (5) is 0.02 and its Holm-Bonferroni p-value is 0.06 after accounting for multiple hypothesis testing. This suggests that investors' investment interest ratings (i.e., $Q_4$) of women-led startups decline by 10.3\% compared with similar men-led startups when entering the second half of the study. The marginally significant interaction term in Column (1) suggests that statistical discrimination might play a role as investors' profitability ratings (i.e., $Q_1$) decline by 4.26\%. Although the interaction terms of ``Asian Founder" and ``Second Half of Study" are not significant at the aggregate level, Section \textcolor{blue}{A.2} later shows that significant belief-driven discrimination against Asians with economically large magnitude does exist at the top level in the second half of the study. \par

\begin{figure}
    \centering
    \includegraphics[scale=0.42]{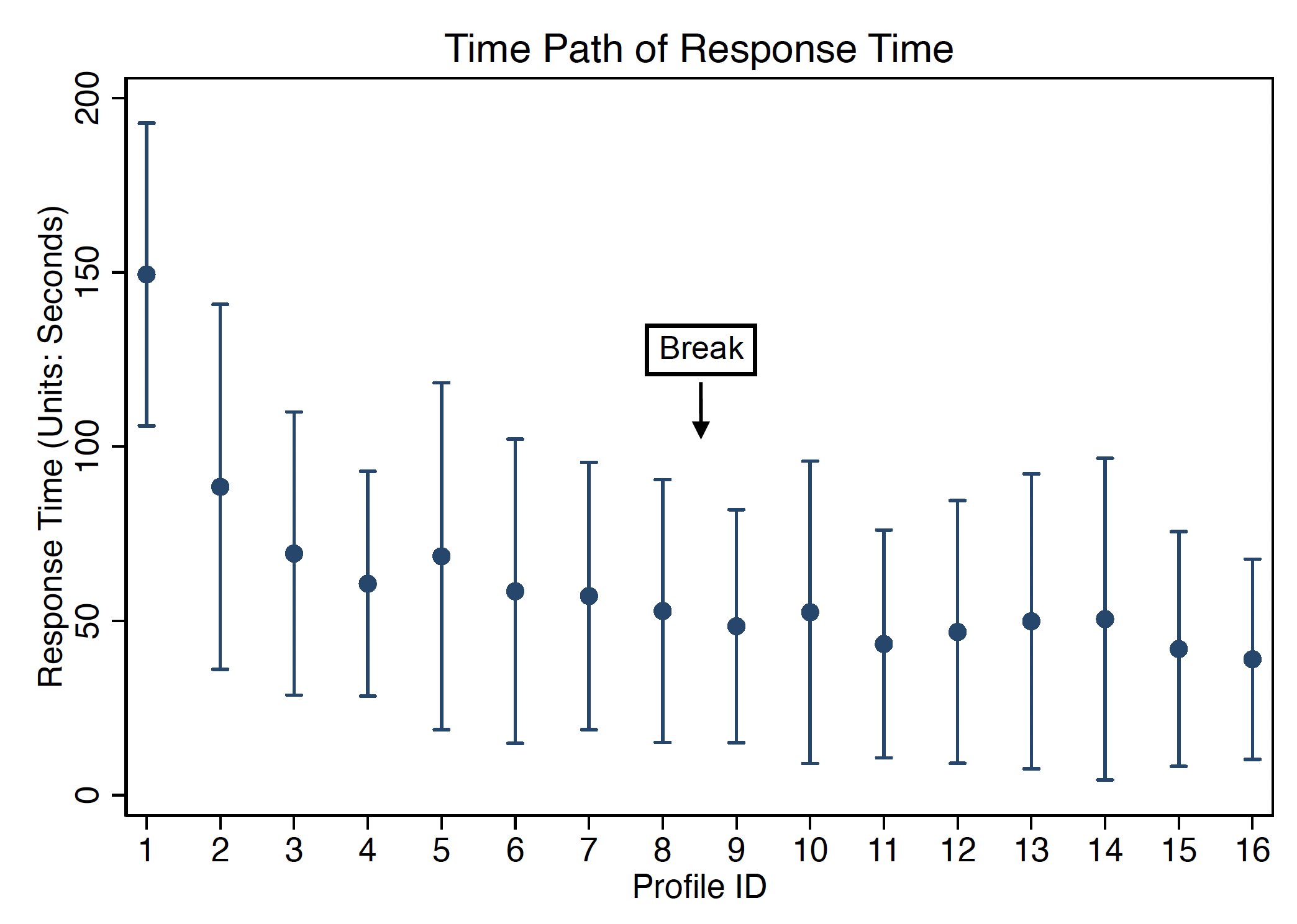}
    \caption{Time Path of Response Time}
    \label{fig:irr_timepath}
    \caption*{\footnotesize \emph{Notes.} This figure demonstrates the time-path of investors' response time as the study progresses to the end. The x-axis is the profile ID, which indicates the order of profiles displayed to each investor. The y-axis reports the mean and standard deviation of investors' response time measured in seconds.}
\end{figure}

\begin{figure}
    \begin{center}
    \includegraphics[scale=1.30]{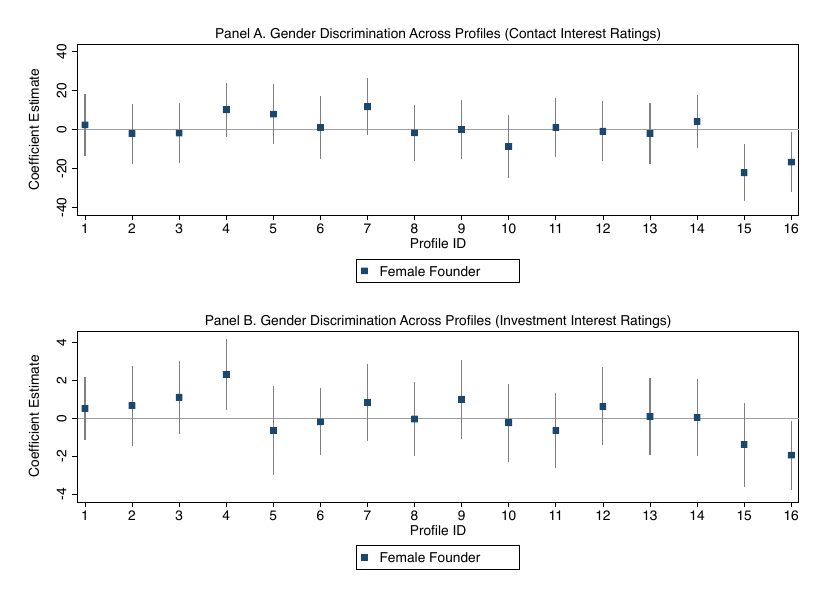}
    \caption{Gender Discrimination Across Profiles}
    \label{fig:gender_dynamic}
 \caption*{\footnotesize \emph{Notes.} This figure illustrates the impact of startup founders' gender on investors' evaluations across profiles. The x-axis is the profile ID, which indicates the order of profiles displayed to each investor. The y-axis reports the coefficient of ``Female Founder" and the 95\% confidence interval using the following regression $Q_{ij}^{k}=\alpha_j+\beta_j \text{Female Founder}_{ij}+\epsilon_{ij}$ for each profile ID $j$. Robust standard errors are used in these cross-sectional regressions. Panel A focuses on investors' contact interest ratings (i.e., $Q_3$). Panel B focuses on investors' investment interest ratings (i.e., $Q_4$).}
  \end{center}
\end{figure}

\vspace{2mm}
To further consolidate the result of implicit gender discrimination in Table \ref{implicit bias gender and race}, Figure \ref{fig:irr_timepath} and Figure \ref{fig:gender_dynamic} demonstrate the time-path of investors' response time and how investors' evaluations vary across profiles as the study progresses to the end, respectively. In Figure \ref{fig:irr_timepath}, there is a clear pattern that investors' response time gradually declines as the evaluation task progresses, confirming that attention is costly \cite{bartos_attention_2016}. Importantly, response time does not temporarily increase after the break and becomes smallest at the end of the study.\par

\vspace{2mm}
Figure \ref{fig:gender_dynamic} shows that investors' gender discrimination based on contact interest ratings (i.e., $Q_3$) and investment interest ratings (i.e., $Q_4$) mainly concentrate at the end of the study in terms of both significance and magnitude. The x-axis is the profile ID and the y-axis reports the coefficient of ``Female Founder" and the 95\% confidence interval using the following regression $Q_{ij}^{k}=\alpha_j+\beta_j \text{Female Founder}_{ij}+\epsilon_{ij}$ for each profile ID $j$. Panels A and B focus on $Q_3$ and $Q_4$ respectively. In Panel A, the coefficient of ``Female Founder" is -22.10\% for Profile 15 with p-value equal to 0.004, and becomes -16.67\% for Profile 16, the final profile evaluated, with p-value equal to 0.04. This suggests that women-led startups are roughly 17\% to 22\% less likely to be contacted by investors at the end of the study. Similarly, in Panel B, the coefficient of ``Female Founder" is -1.94 for Profile 16, which is statistically significant at the 5\% level. This shows that women-led startups receive 19.4\% less investment compared to similar men-led startups if the profile is displayed at the end of the evaluation section. Both the significance and economic magnitude of the discovered gender discrimination are strong when the evaluation task is close to the end, which is consistent with the implicit discrimination hypothesis (i.e., the ``fatigue effect"). To further pin down the implicit discrimination channel, I will rule out other alternative interpretations below. \par
\vspace{2mm}

\begin{figure}
    \begin{center}
    \includegraphics[scale=1.30]{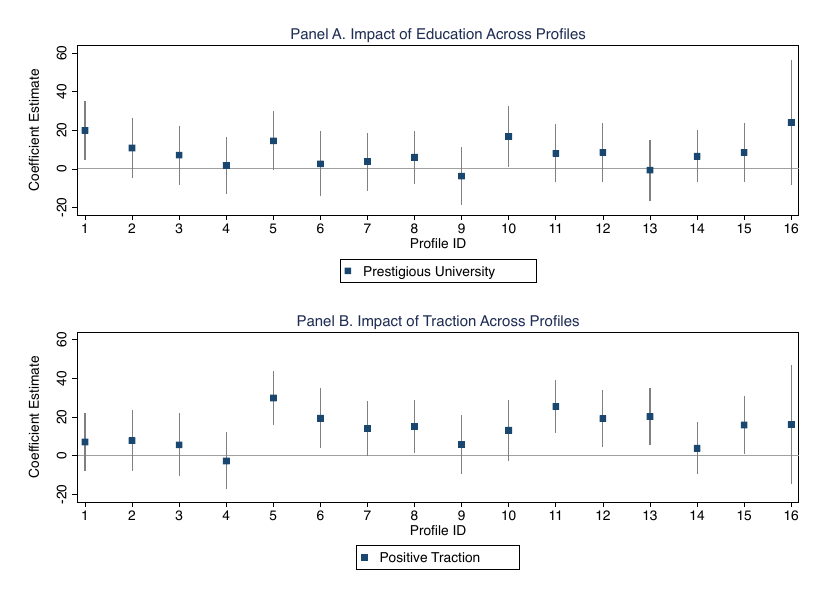}
    \caption{Evaluations of Education and Traction Across Profiles}
    \label{fig:startup_dynamic}
 \caption*{\footnotesize \emph{Notes.} This figure illustrates the impact of founders' educational backgrounds and startups' traction on investors' contact interest ratings across profiles. The x-axis is the profile ID, which indicates the order of profiles displayed to each investor. The y-axis reports the coefficient of ``Startup Characteristic" and the 95\% confidence interval using the following regression $Q_{ij}^{k}=\alpha_j+\beta_j \text{Startup Characteristic}_{ij}+\epsilon_{ij}$ for each profile ID $j$. Robust standard errors are used in these cross-sectional regressions. In Panel A, the ``Startup Characteristic" is an indicator that equals one if the startup founder graduated from a prestigious university, and equals zero if the founder graduated from a less prestigious university. In Panel B, the ``Startup Characteristic" is an indicator that equals one if the startup generates positive revenue and equals zero otherwise.}
 \end{center}
\end{figure}

\textbf{\textit{Rule Out Learning Effect}} One alternative interpretation of the findings in Table \ref{implicit bias gender and race} and Figure \ref{fig:gender_dynamic} is a ``learning effect". This indicates a worse situation where investors \emph{consciously} discriminate against women more when they are more familiar with the evaluation task. If the ``learning effect" dominates, investors' evaluations of other startup characteristics should be more accurate and less noisy as the IRR experiment progresses to the end. If the ``fatigue effect" dominates, investors' evaluations should demonstrate more noise at the end. Online Appendix Figure \ref{fig:startup_dynamic} demonstrates how investors evaluate founders' educational backgrounds and startups' traction across different profiles.\footnote{These two characteristics have been well documented to possibly affect investors' decisions in the entrepreneurial finance literature (\cite{bernstein_attracting_2017}), \cite{kaplan2009should}. For the impact of other startup characteristics on investors' investment, please read \cite{zhang2020venture}.} It shows that investors' evaluations of these non-sensitive startup characteristics have a larger variance for Profiles 15 and 16, which rules out the ``learning effect" and further supports the ``fatigue effect". Moreover, after removing the first few evaluations of each investor, the ``fatigue effect" still holds, indicating that ``learning" may not be the main reason for the observed results.\par
\vspace{2mm}

\textbf{\textit{Rule Out ``Balance-the-Sample" Hypothesis}} Another alternative interpretation for the discovered ``fatigue effect" is that investors may expect the overall population of entrepreneurs to follow the distribution in the real world, say ``80-20", rather than ``50-50" as used in this experiment. Thus, near the end of the evaluation process, investors may be tempted to skew the sample in the direction they expect the population to be, leading to biased experimental results.\footnote{Special thanks go to Peter DeMarzo who raised this brilliant point and suggested the corresponding testing method. ``50-50" has been chosen to maximize the experimental power. However, using this ratio may suggest to investors that the experiment's purpose is to test discrimination, making investors behave more friendly to minority groups, especially in the second half of the study where the balanced ratio is more obvious.} To test this hypothesis, I have checked whether investors' evaluation results are influenced by the ``mixed profiles" they have already evaluated in the first half of the study. However, as shown in Online Appendix \ref{sec:appendix_irr} Table \ref{irr:robustnesscheck}, investors who evaluated more minority founders in the first half of the experiment were not systematically tougher on minority groups in the second half of the experiment. Moreover, the ``50-50" ratio is also used for the randomization of founders' educational backgrounds and the startups' traction to maximize experimental power. If the ``Balance-the-Sample" hypothesis dominates, similar ``fatigue effect" should also exist for these two startup characteristics, which is not supported by Figure \ref{fig:startup_dynamic}. \par
\vspace{2mm}
 
It should be noted that the IRR experiment does not directly observe whether investors \emph{knowingly} discriminate against female founders. A rigorous interpretation of results above is that as the experiment goes on, investors get tired of the evaluation task and give lower ratings to women-led startups compared to similar men-led startups. For simplification purposes, discrimination that is associated with such ``fatigue effect" or occurs in the second half of the IRR experiment is called ``implicit discrimination" in this paper. Considering that the cognitive workload of evaluating these profiles is not large, the discovered ``implicit discrimination" may potentially play an important role in an intensive pre-screening process in the VC industry. \par

\vspace{2mm} 
\textbf{A.2 (Distributional Effect) Implicit Discrimination Concentrates on Top Startups.}\par
\vspace{2mm}
Several fundamental theory papers in discrimination have predicted an equilibrium situation in which evaluators discriminate against minority groups at relatively high rating scores and favor minority groups at very low rating scores (\cite{phelps1972statistical}, \cite{aigner1977statistical}, \cite{lundberg1983private}, \cite{morgan2009diversity}). The model in \cite{morgan2009diversity} also predicts that when evaluators are selective, this leads to minority underrepresentation in the workplace. However, when evaluators are sufficiently unselective, this leads to overrepresentation of minorities. These discrimination theories imply the existence of a distributional effect of discrimination across the spectrum of startups' attractiveness. Although this theoretical ``discrimination reversion" phenomenon has not been empirically tested before, it may potentially reconcile the contradictory results in empirical studies that test discrimination in the entrepreneurial financing setting.\par

\vspace{2mm}
Conflicting results in the literature have been generated by \cite{ewens_are_2020} and a classical correspondence test in \cite{gornall_gender_2020}. Both papers essentially study investors' intentions to \emph{contact} startups in the pre-screening stage. Compared to profitability ratings (i.e., $Q_1$) that only measure one dimension of startups, contact interest ratings (i.e., $Q_3$) provide more comprehensive evaluations of startups' quality. This summary measure of all information acquired by evaluators is what test scores or rating scores mean in discrimination theory papers (\cite{lundberg1983private}). Therefore, when testing the distributional effect of discrimination, contact interest ratings (i.e., $Q_3$) are the most appropriate measures for the purpose of reconciling the literature.\footnote{Online Appendix Table \ref{irr_investor_profile_gender_quantile_Q1} and \ref{irr_investor_profile_race_quantile_Q1} also report quantile regression results using $Q_1$.} \par
\vspace{2mm}

\textbf{(Gender)} Table \ref{irr_investor_profile_gender_quantile} investigates whether implicit gender discrimination varies across startups' received contact interest ratings (i.e., $Q_3$) using quantile regressions. The sample includes profile evaluations in the second half of the IRR experiment. Motivated by the literature discussing gender issues in science or STEM fields (\cite{carrell2010sex}, \cite{goldin2014grand}, and \cite{kessler_incentivized_2019}) and the crucial importance of the tech sector in the entrepreneurial community (see Table \ref{irr_investor_summary}), Panel A focuses on implicit gender discrimination among investors in the tech sector. Panel B uses all recruited investors' evaluation results. In each of Columns [1]–[9], the dependent variable is the $k$th percentile ($k\in{10,20,...,90}$) of the distribution of startups' received contact interest ratings (i.e., $Q_3$). In Column [10], the dependent variable is the average investors' contact interest ratings. Standard errors in parentheses are clustered at the investor level.\par

\vspace{2mm}
Results show that implicit gender discrimination against female founders is most prevalent for attractive startups. Panel A finds that in the tech sector, compared to similar men-led startups, investors are 10\% less likely to contact women-led startups whose contact interest ratings are in the 90th and 60th quantile of the attractiveness distribution. The negative coefficients of ``Female Founder" are statistically significant at the 5\% level. Panel B confirms this phenomenon by using the full sample. For startups whose contact interest ratings are at the 50th, 60th, and 80th percentile position, women-led startups are 10\%, 8\%, and 7\% less likely to be contacted by investors, respectively. Results are statistically significant at the 5\% level. However, in both Panels A and B, this gender discrimination disappears for women-led startups whose contact interest ratings are below the 40th percentile of the attractiveness distribution. Experiment A might not have enough statistical power to test investors' preferences towards women-led startups at very low rating scores, especially if the magnitude of this gender preference is small. Hence, this study uses Experiment B to capture the fundraising situation of struggling startups as shown in Section \ref{sec:correpondencetest}.\par
\vspace{2mm}

\textbf{(Race)} Table \ref{irr_investor_profile_race_quantile} tests similar distributional effect of implicit racial discrimination in the second half of the study using quantile regressions. Similar to implicit gender discrimination, discrimination against Asian founders is more salient among highly-rated startups. As shown in Panel A, investors in the tech sector are 9\% less likely to approach Asian founders compared to similar white founders if their startups are rated at the 90th percentile. Panel B finds that investors in the full sample give 10\% lower contact interest ratings to Asian founders whose startups are rated at the 80th percentile. This result is statistically significant at the 1\% level. However, in both Panels A and B, the coefficients of ``Asian Founder" become slightly positive for startups rated below the 20th percentile. Although this result is not statistically significant in the IRR experiment, it suggests the potential existence of weak preference towards Asian founders of startups with very low rating scores if the sample size is large enough (see Experiment B).\footnote{Some might have a concern that the distributional effect of implicit gender and racial discrimination is driven by investors who have extreme preferences. Online Appendix Table \ref{irr_investor_profile_outlier_quantile} shows that results are robust after removing evaluations of investors with extreme preferences.}

\par \vspace{2mm}

\textbf{(Age)} Although agism is not the main question of this paper, Online Appendix Table \ref{irr_investor_profile_age_quantile} finds that a ``discrimination reversion" phenomenon also exists when testing agism in the second half of the study. Panel A finds that tech investors are 8\% less likely to contact older founders compared to similar younger founders whose startups are rated at the 90th percentile. The coefficient of ``Older Founder" is statistically significant at the 10\% level. This is consistent with a large amount of anecdotal evidence and surveys that indicate wide-spread ageism during VCs' investment process, especially in the tech sector.\footnote{See Forbes ``The Biggest Bias In Tech That No One Talks About” (April 10th, 2019) by Maren Thomas Bannon, an early-stage technology venture capitalist.} However, for startups rated at the 20th and 40th percentile, investors are 10\% more likely to approach older founders. Results are statistically significant at the 10\% and 5\% level, respectively. Panel B finds that the coefficient of ``Older Founder" is -6\% for startups rated at the 80th percentile and +7\% for startups rated at the 40th percentile, which are both statistically significant at the 10\% level. \par

\vspace{2mm}
Quantile regression results above show that the magnitude and direction of evaluators' socially sensitive preferences, such as gender, race, and age discrimination, can vary across the distribution of candidates' rating scores. This provides one possible explanation for the contradictory results in the literature. On the one hand, papers exploiting regression-based methods often use observational data that record relatively mature and promising startups.\footnote{Standard databases often record startups that have successfully raised external funding. Also, startups whose information is posted on large fundraising platforms are relatively more mature and well-known than startups whose information is not recorded by these platforms.} Hence, they mainly capture the middle and right part of the startup attractiveness distribution where significant implicit discrimination exists. On the other hand, the correspondence test often focuses on the cold call pitch email setting. According to Crunchbase, ``Cold emails have a bad reputation in venture capital ... and can come off as begging".\footnote{See \hyperlink{https://about.crunchbase.com/blog/cold-email-to-a-venture-capitalist/}{``The Art of a Cold Email to a Venture Capitalist"} posted on Crunchbase, July 17, 2020.} Hence, high-rated startups generally do not choose this fundraising method and the correspondence test mainly captures the fundraising situations of struggling startups. Since different papers reveal investors' preferences at different positions on the startup distribution, the direction and magnitude of their results might be different. Importantly, results of correspondence tests might not be generalizable to highly-rated startups' fundraising situations.\par

\vspace{2mm}
\textbf{(Investment Interest Ratings)} There are several key differences between investors' investment interest ratings (i.e., ``Q4") and contact interest ratings (i.e., ``Q3"). First, contact interest ratings mainly capture whether a startup can raise funding from VC investors (i.e., extensive margin) while investment interest ratings mainly reflect how much funding an investor will invest in each startup (i.e., intensive margin). Extant literature and conflicting results mainly focus on the former subject, but scarce evidence exists for the latter. Second, investment interest ratings (i.e., ``Q4") are noisier than contact interest ratings (i.e., ``Q3") in the IRR experiment because this study does not provide any soft information about the startup founder, which is also crucial to investment decisions. Third, the investment intention is mainly affected by beliefs rather than tastes as suggested by \cite{zhang2020venture}. Therefore, the distributional effect measured by ``Q4" is not used to reconcile the literature. Instead, it tests whether discrimination also exists along the intensive margin.\footnote{The ``intensive margin" here refers to whether investors provide a smaller amount of funding to women-led startups and Asian-led startups.}\par

\vspace{2mm}
Table \ref{irr_investor_profile_Q4} shows the distributional effect of a startup founder's gender and race on investors' investment interest ratings in the second half of the study. Panels A and B focus on gender and racial discrimination, respectively. Panel A shows that investors provide 10\% less funding to women-led startups compared to similar men-led startups if these startups are rated between the 30th and 60th percentile based on $Q_4$. Hence, implicit gender discrimination along the intensive margin is more significant among mid-level startups. However, Panel B shows that significant implicit racial discrimination still concentrates among top-level startups. For startups rated at the 90th percentile, Asian-led startups receive 20\% less funding compared to similar White-led startups. This result is statistically significant at the 1\% level. However, for startups rated below the 20th percentile, this racial discrimination disappears. The negative coefficients of ``Female Founder" and ``Asian Founder" in Panels A and B indicate that minority groups receive double the penalty along both the extensive margin (see Table \ref{irr_investor_profile_gender_quantile} and Table \ref{irr_investor_profile_race_quantile}) and the intensive margin (see Table \ref{irr_investor_profile_Q4}).\par

\vspace{2mm}
\textbf{A.3 (Pro-cyclical Discrimination) Implicit Discrimination Varies with Market Conditions.}\par
\vspace{2mm}
 
\begin{figure}
    \centering
    \includegraphics[scale=0.45]{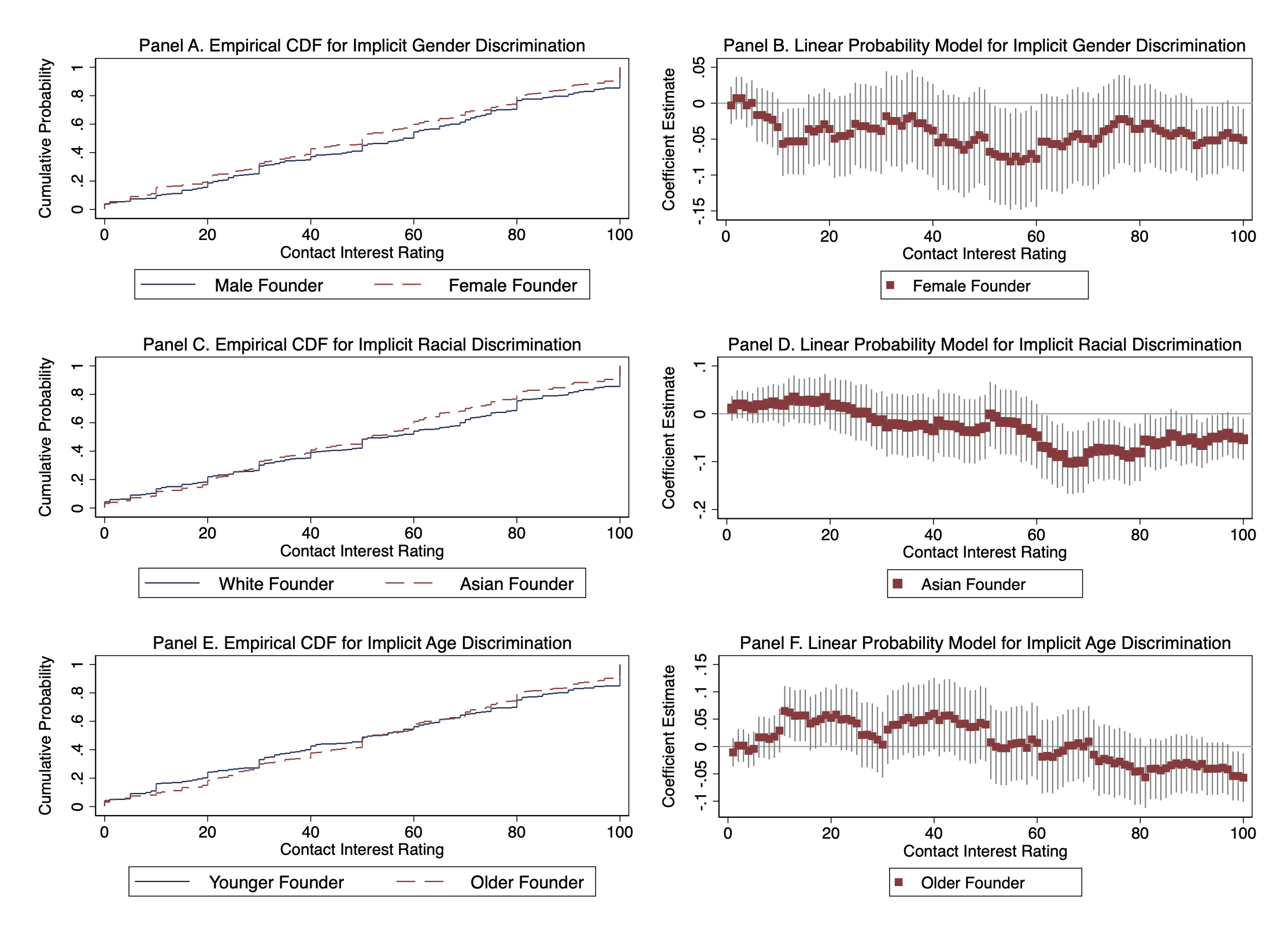}
    \caption{Investors' Implicit Gender, Racial, and Age Discrimination in Different Market Conditions}
    \label{fig:distribution_irr}
    \caption*{\footnotesize \emph{Notes.} This figure demonstrates how implicit gender, racial, and age discrimination varies across investors' internal thresholds measured by investors' contact interest ratings. The sample includes all recruited investors' evaluations in the second half of the IRR experiment. Panel A provides the empirical CDF for founder's gender on investors' contact interest rating (i.e., $Pr(\text{Contact Interest}\leq x|\text{Female Founder})$ and $Pr(\text{Contact Interest}\leq x|\text{Male Founder})$). Panel B provides the OLS coefficient estimates (i.e., $Pr(\text{Contact Interest}\leq x|\text{Female Founder})-Pr(\text{Contact Interest}\leq x|\text{Male Founder}))$ and the corresponding 90\% confidence level. Similarly, Panels C and E provide the empirical CDF for founder's race and age, respectively. Panels D and F provide the OLS coefficient estimates for founder's race and age, respectively.}
\end{figure}

Similar to \cite{kessler_incentivized_2019}, this subsection investigates how investors' implicit discrimination varies in different market conditions as measured by investors' internal thresholds. When the capital supply is abundant (limited) in the market during an economic boom (bust), investors are less (more) selective when investing in startups, and their internal thresholds become lower (higher). Figure \ref{fig:distribution_irr} illustrates these dynamic changes in investors' gender, racial, and age discrimination based on investors' contact interest ratings. The sample includes all recruited investors' evaluations in the second half of the IRR experiment. Panels A, C, and E provide the empirical cumulative density function (CDF) for a founder's gender, race, and age across investors' contact interest ratings, respectively. Panels B, D, and F provide the corresponding OLS coefficient estimates and their 90\% confidence intervals for a founder's gender, race, and age across investors' contact interest ratings, respectively.\footnote{Regressions used in Panels B, D, and F are the same as those used in \cite{kessler_incentivized_2019}: for each selected internal threshold $x$,
\begin{eqnarray*}
\text{Callback}_{ij}=\beta_0+\beta_1\text{Startup Characteristics}_{ij}+\epsilon_{ij}
\end{eqnarray*}
where $\text{Callback}_{ij}=1$ if $\text{Contact Interest Ratings}_{ij}>=x$ and $\text{Callback}_{ij}=0$ if $\text{Contact Interest Ratings}_{ij}<x$. The confidence intervals are calculated using robust standard errors. However, results are similar when clustering standard errors at the investor level.}\par

\vspace{2mm}
Figure \ref{fig:distribution_irr} shows that both the direction and magnitude of investors' implicit gender, racial, and age discrimination can vary across investors' internal thresholds. Panel A shows that the CDF for a female founder is generally to the left of the CDF for a male founder, which means that the CDF for a male founder first-order stochastically dominates the CDF for a female founder in most situations. Panel B shows that the coefficients of ``Female Founder" are negative for most selected investors' internal thresholds with the largest magnitude existing when the internal thresholds are around 60\% of contact interest ratings. Results in Panels A and B suggest the existence of widespread implicit gender discrimination against female founders in most market conditions. Unless investors contact almost all startups on the market, it is generally more difficult for female founders to approach investors compared to similar male founders. \par

\vspace{2mm}
Panels C and D show that implicit racial discrimination mainly exists when investors' internal thresholds are high. In Panel C, when contact interest ratings are above 30\%, the CDF for an Asian founder is to the left compared with the CDF for a white founder. The opposite is seen when contact interest ratings are below 30\%. Similarly, in Panel D, the coefficients of ``Asian Founder" gradually become negative as investors' internal thresholds increase. When internal thresholds are above 60\% of contact interest ratings, these negative coefficients become statistically significant. This indicates that Asian founders face more difficulties in raising funding when the capital supply is low. \par

\vspace{2mm}
Panels E and F display a more salient ``discrimination reversion" phenomenon of agism across investors' internal thresholds. In Panel E, the CDF for an older founder gradually moves to the left compared with the CDF for a younger founder as investors become more selective in funding startups. In Panel F, the coefficients of ``Older Founder" are significantly positive when contact interest ratings are around 20\% and are significantly negative when contact interest ratings are close to 100\%. Results indicate that implicit agism exists when investors only consider top startups.\par

\vspace{2mm}
These dynamic changes of implicit discrimination in various market conditions confirm the theoretical prediction of ``pro-cyclical discrimination" in \cite{morgan2009diversity}. Results are also fundamentally consistent with the distributional effect documented in the previous subsection. Both subsections show that implicit gender, racial, and age discrimination mainly exist when investors focus on top startups. However, when investors also consider startups with relatively low rating scores, discrimination against minority groups becomes smaller and might even reverse although the reversion is not statistically significant for gender and racial discrimination due to the small sample size of Experiment A.\par

\vspace{2mm}
\textbf{A.4 Decision-based Heterogeneous Effect Estimator and Polarization of Discrimination Attitudes}\par

\vspace{2mm}
In the real world, investors' preferences towards a founder's gender, race, and age are heterogeneous. Some favor minority founders, while some favor majority founders. In a divided society, different groups potentially make opposing decisions based on different motivations. For example, ``pro-minority" investors' decisions can be driven by taste if they simply want to support minorities rather than maximize financial returns. However, ``anti-minority" investors' investment decisions can be driven by belief if they perceive minority founders' startups to be less profitable. Understanding the separate driving forces of these two different groups' decisions has important policy implications. To test what separates investors and how divided the investment community is, this paper develops a consistent, ``\textit{decision-based}" heterogeneous effect estimator using the ``leave-one-out" technique.\par

\vspace{2mm}
This estimator exploits several unique features of the IRR experiment compared to the correspondence test. First, the IRR experiment collects both decision data, such as contact interest ratings, and mechanism data, such as profitability and availability ratings. However, the correspondence test generally does not collect mechanism data. Second, the IRR experiment provides continuous outcome variables, enabling researchers to measure the level of discrimination, while the outcome variables of the correspondence test are usually binary, which provides a limited amount of information.\par
\vspace{2mm}

The logic behind how this estimator works is very simple. The IRR experiment introduces within-individual level randomization and requires investors to reveal both their decisions and beliefs about multiple randomized startup profiles. When the number of profiles evaluated by each investor is large enough, researchers can identify ``individual-level" preferences based on evaluators' decisions. Hence, it is feasible to tell which investors discriminate against minorities and classify recruited investors into a ``pro-minority" group and an ``anti-minority" group based on their decisions. Researchers can then run separate pooled regressions within each group to investigate their mindset (i.e., mechanisms). While the estimator in \cite{kline2021reasonable} tests the fraction of evaluators who discriminate in the population using correspondence test data, the decision-based heterogeneous effect estimator focuses on the degree of polarization and links the heterogeneity of decisions and mechanisms.\par

\vspace{2mm}
Formally, assume there are I investors, and each evaluates K questions for J profiles, $i\in \{1,2,...,I\}, j\in \{1,2,...,J\}$, and $K=4, k\in \{1,2,3,4\}$ since each investor needs to evaluate $Q_1$ (profitability), $Q_2$ (availability), $Q_3$ (contact), and $Q_4$ (investment). The pooled regression to test group-level preferences is $Y_{ij}^{(k)}=X_{ij}\beta_i^{(k)}+\alpha_i+\epsilon_{ij}^{(k)}$. For simplicity, let's assume $X_{ij}$ contains only one gender indicator. This means that $X_{ij}=1$ if the founder's gender is female for the $j^{th}$ generated profile evaluated by investor $i$, and
$X_{ij}=0$ if otherwise.\par
\vspace{2mm}

Let's further assume that $\epsilon_{ij}^{(k)}=\eta_i^{(k)}+v_{ij}^{(k)}$, where $v_{ij}^{(k)}$ are independent and identically distributed random variables. $\eta_i^{(k)}$ stands for the investor fixed effect and will enter the constant term if researchers run the individual-level regressions. Since all startup profiles are dynamically and independently generated, under this residual structure, it is natural to have the following assumption without loss of generality: $\epsilon_{ij}^{(k)}\perp\epsilon_{ij'}^{(k)}$ if $j\neq j'$. However, $\epsilon_{ij}^{(k)}\not\perp\epsilon_{ij}^{(k')} $ if $k\neq k'$. This study classifies investors based on their contact decisions using $\beta_i^{(3)}$. It defines ``anti-minority" investors as those whose $\beta_i^{(3)}<0$ (i.e., investors who do not want to contact minority founders' startups), and it defines ``minority-friendly" investors as those whose $\beta_i^{(3)}>0$ (i.e., investors who prefer contacting minority founders' startups).\par

\vspace{2mm}
It is interesting to investigate $\beta_i^{(1)}$ (i.e., profitability ratings) and $\beta_i^{(2)}$ (i.e., availability ratings) in both the ``anti-minority" group and the ``minority-friendly" group to understand their separate driving mechanisms. For illustration purposes, I will focus on $\beta_i^{(1)}$ in the paper. However, the same logic applies to $\beta_i^{(2)}$. While the traditional heterogeneous effect relies on subjects' predetermined demographic information,\footnote{For example, researchers separately test the investment preferences of well-educated investors and uneducated investors.} the ``\textit{decision-based}" heterogeneous effect estimator classifies the investor pool into a ``pro-minority" group and an ``anti-minority" group based on their indicated decisions.\par

\vspace{2mm}
In the ideal case where $\beta_i^{(1)}$ is observable or predetermined (i.e., $\beta_i^{(1)}\perp\epsilon_{ij}^{(k)}$), researchers can divide I investors into 2 groups based on the sign of $\beta_i^{(3)}$ and run the following regression, which is similar to how they estimate traditional heterogeneous effects:
\begin{eqnarray*}
Y_{ij}^{(k)}=\gamma_1 1(\beta_i^{(1)}<0)X_{ij}+\gamma_2   1(\beta_i^{(1)}>0)X_{ij}+\alpha_i+\epsilon_{ij}^{(k)}
\end{eqnarray*}
Since $ 1(\beta_i^{(1)}<0)X_{ij}\perp \epsilon_{ij}^{(k)}$ and $1 (\beta_i^{(1)}>0)X_{ij}\perp \epsilon_{ij}^{(k)}$, there is no endogeneity problem.\par

\vspace{2mm}
However, since $\beta_i^{(1)}$ is not directly observable, the previous estimation method generates biased results due to the ``generated regressor problem". This is because if $\hat{\beta_i^{(1)}}=\frac{\sum_j X_{ij}Y_{ij}^{(1)}}{\sum_j X_{ij}^2}=\beta_i^{(1)}+\frac{\sum_j X_{ij}\epsilon_{ij}^{(1)}}{\sum_j X_{ij}^2}$, then $1(\hat{\beta_i^{(1)}}<0)X_{ij}= 1 (\beta_i^{(1)}+\frac{\sum_j X_{ij}\epsilon_{ij}^{(1)}}{\sum_j X_{ij}^2}<0)X_{ij}$, which $\not\perp\epsilon_{ij}^{(k)}$ since $\epsilon_{ij}^{(1)}\not\perp\epsilon_{ij}^{(k)}$. A similar problem applies to $1(\hat{\beta_i^{(1)}}>0)X_{ij}$. In this case, the estimation suffers from the endogeneity issue. To solve this ``generated regressor problem", the decision-based heterogeneous effect estimator uses the ``leave-one-out" technique and takes the following steps:\par
\vspace{2mm}

\textbf{Step 1:} for each i \& j, estimate $\beta_i^{(1)}$ leaving the $j^{th}$ observation out: $\hat{\beta_{ij}^{L(1)}}=\frac{\sum_{j'\neq j} X_{ij'}Y_{ij'}^{(1)}}{\sum_{j'\neq j} X_{ij'}^2}$ (when $|J|\rightarrow\infty, \hat{\beta_{ij}^{L(1)}}\overset{p}{\to} \beta_i^{(1)}$ for each j). Now we have $I\times J$ estimated $\hat{\beta_{ij}^{L(1)}}$ \par
\vspace{2mm}
\textbf{Step 2:} classify $I\times J$  $\hat{\beta_{ij}^{L(1)}}$ into two groups based on their signs. (This means that investor i can enter both the ``anti-minority" group and the ``minority-friendly" group in a finite sample. However, as $|J| \overset{p}{\to} \infty$, this situation will not happen)\par
\vspace{2mm}
\textbf{Step 3:} run the following pooled regressions

\begin{eqnarray*}
Y_{ij}^{(k)}=\gamma_1 1(\hat{\beta_{ij}^{L(1)}}<0)X_{ij}+\gamma_2  1(\hat{\beta_{ij}^{L(1)}}>0)X_{ij}+\alpha_i+\epsilon_{ij}^{(k)}
\end{eqnarray*}

Now, $\hat{\beta_{ij}^{L(1)}}\perp \epsilon_{ij}^{(k)}$ since $\hat{\beta_{ij}^{L(1)}}$ has left the $j^{th}$ term out and $\epsilon_{ij}^{(1)}$ does not enter $\hat{\beta_{ij}^{L(1)}}$, which breaks the connection with $\epsilon_{ij}^{(k)}$. Remember the assumption: $\epsilon_{ij}^{(k)}\perp\epsilon_{ij'}^{(k)} $ if $j\neq j'$, then $1(\hat{\beta_{ij}^{L(1)}}<0)X_{ij}\perp \epsilon_{ij}^{(k)}$. Hence, there is no longer an endogeneity problem using this estimation procedure.\par

\vspace{2mm}
Theoretically speaking, researchers can classify the group based on $\beta_i^{(k)}$ for $\forall k$. The interpretation of the results will change since the ``anti-minority" group and the ``minority-friendly" group are defined by different $\beta_i^{(k)}$. In the IRR experiment, the goal is to connect the heterogeneity of evaluators' revealed contact decisions and their ratings in mechanism questions. Hence, $\beta_i^{(3)}$ is used to classify the group. The estimator's feasibility relies on the carefully designed evaluation questions and an incentive structure that can incentivize all the questions.\par

\vspace{2mm}

Table \ref{2nd_irr_hetgender} shows the estimated decision-based heterogeneous effect for founders' gender, race, and age. The sample includes all recruited investors' evaluations in the second half of the study. Panels A, B, and C report the contact decision-based heterogeneous effect of a startup founder's gender, race, and age, respectively. All the coefficients and standard errors in the parentheses are calculated using the ``leave-one-out" estimation procedure and a bootstrap method due to the small sample size.\par

\vspace{2mm}
Results displayed in Table \ref{2nd_irr_hetgender} show that investors' profitability ratings can better explain investors' heterogeneous contact decisions than their availability ratings. Panel A shows that ``pro-women" (i.e., $\beta^3>0$) investors and ``anti-women" investors (i.e., $\beta^3<0$) have very different expectations for women-led startups' profitability. In Column (1), ``anti-women" investors perceive women-led startups to have 16.40 percentile ranks lower potential financial returns than similar men-led startups. On the contrary, ``pro-women" investors expect women-led startups to generate 7.93 percentile ranks higher potential financial returns than men-led startups. Both results are statistically significant at the 1\% level. Column (2) shows that investors' availability ratings are not significantly different for these two groups. A similar phenomenon can also be seen in Panels B and C for racial and age discrimination. These results suggest that expectations about minority-led startups' profitability rather than availability play a crucial role in driving investors' polarized decisions.\footnote{There might also exist other mechanisms that drive investors' divided decisions, such as taste. If taste is directly observed in an IRR experiment, the estimator can also test this mechanism.}\par

\vspace{2mm}
One interesting observation is that investors who give lower contact interest ratings to minority founders also invest less funding in minority founders' startups compared to similar majority founders' startups. Panel A shows that ``anti-women" investors on average invest 26.1\% less capital in women-led startups while ``pro-women" investors on average invest 10.8\% more capital in women-led startups. Similar results are also seen in Panels B and C. This shows that investors' discrimination behaviors are consistent along both the extensive margin and the intensive margin. Another interesting observation is that the division in investors' attitudes towards female founders is slightly larger than that towards Asian and older founders. The degree of implicit gender discrimination of the ``anti-women" group is -16.40 percentile ranks in Column (1), -21.81\% in Column (3), and -26.1\% in Column (4). These are higher than the degree of implicit racial or age discrimination of the ``anti-minority" group as shown in Panels B and C.\par
\vspace{2mm}

\textbf{Do women-led startups really perform worse than men-led startups?} To examine the accuracy of investors' beliefs, I compare the performance of women-led startups and men-led startups between 07/2020-07/2021 (i.e., during the one-year period after the experiment). Online Appendix \ref{sec:appendix_irr} Table \ref{compare_performance_gender} shows that conditional on 
being funded, women-led startups generally perform similarly to men-led startups in terms of the likelihood of raising new funding, going out of business, or successfully exiting through IPO or M\&A. However, women-led startups underperform in the IT industry and specifically are significantly associated with a higher likelihood of going out of business in this industry. This, to some extent, justifies investors' gender discrimination against women as investors need to screen out more women-led startups in the pre-selection stage to achieve a similar financial performance to investing in men-led startups.\footnote{\cite{barber2021explains} also show that research productivity falls more for women during the COVID-19 pandemic. I do not investigate the performance of Asian and older founders' startups because information about founders' race and age is not well-recorded on Pitchbook.} It should be noted that the under-performance of women-led startups in the IT industry may be a temporary phenomenon due to COVID-19. Also, since female founders' performance is endogenous and likely to be self-fulfilled due to investors' evaluation criteria, \cite{lundberg1983private} still view differential treatments of minority groups as discrimination even if minority groups are associated with worse performance.\par

\vspace{2mm}

\textbf{\textit{B. (Donation Game) Homophily Exists When Investors Provide Non-investment Support.}}\par

\vspace{2mm}
VCs adds value to startups by providing both capital and non-investment support. While the IRR experiment captures investors' preferences during the investment process, the donation game tests whether investors treat majority and minority groups differently when providing anonymous encouragement during economic hardship. Table \ref{irr_dictator} reports the regression results from the donation game in which investors' \textit{anonymous} donation decisions do not affect their investment opportunities or help improve investors' social image. The dependent variable of Columns (1) and (3) is the donated amount measured in dollars, ranging from \$0 to \$15. The dependent variable of Columns (3) and (4) is an indicator that equals one if the investor donates all \$15, and zero otherwise.\footnote{Investors who did not select a donation amount receive \$15; hence, their decisions are treated as ``donate \$0".} The ``Profitability Evaluations" are calculated based on $Q_1$, which is the coefficient $\beta_i$ of the regression $Q_{1ij}=\beta_0+\beta_i \text{Startup Characteristics}_{ij} +\epsilon_{ij}$ for each investor $i$. It stands for the causal effect of ``Startup Characteristics" on the investor $i$'s profitability ratings. ``Startup Characteristics" is ``Female Founder" in Panel A and ``Asian Founder" in Panel B. To increase the statistical power, Panel C defines the majority group as those who are white male and the minority group as everyone else. ``Startup Characteristics" is ``White Male Founder" in Panel C to determine whether homophily exists within the white male group. All regressions use robust standard errors reported in parentheses.\par

\vspace{2mm}
Panel A reports the existence of gender homophily. Column (1) shows that male investors on average donate \$3.20 less to female founders compared to male founders. This result is statistically significant at the 10\% level. On the other hand, female investors donate roughly \$4 more to female founders although the positive interaction term of ``Female Founder" and ``Female Investor" is not statistically significant. Column (3) shows that male investors are 77\% less likely to donate all money to female founders compared to male founders, which is statistically significant at the 5\% level. However, female investors are 68\% more likely to donate all money to female founders. The positive interaction term is statistically significant at the 10\% level.\footnote{Consistent with \cite{dellavigna2013importance}, the significantly negative coefficients of ``Female Investor" in Columns (1) and (3) show that men are usually more generous than women during the donation process. } Columns (2) and (4) show that the same results still hold even after controlling investors' profitability evaluations. This indicates that the reason male investors donate more to male founders is not because they perceive male founders to be more deserving due to their startups' higher profitability. Hence, other factors, such as taste, might play a role. While gender homophily has been documented in VCs' investment process (\cite{raina2019vcs}), this study shows that homophily also exists in a setting where investors provide anonymous non-investment support.\par

\vspace{2mm}
Although Panel B does not show the existence of significant racial homophily phenomenon, Panel C shows that strong homophily exists within the white male group and within the other group. Column (1) shows that white male investors donate \$2.97 more to white male founders compared to other founders. Investors who are not white male donate \$6.81 less to white male founders compared to other founders. Column (3) shows that white male investors are 82\% more likely to donate all money to white male founders compared to other founders. However, investors who are not white males are also less likely to donate all money to white male founders. Columns (2) and (4) show that this homophily is not driven by beliefs in the profitability of different founders' startups. In all columns, the coefficients of ``White Male Founder" and its interaction term with ``White Male Investor" are statistically significant at the 5\% level.\par

\subsection{Discussion}\label{irr:discussion}
Despite the rich results generated by Experiment A, it still has several limitations. Besides the well-documented sample selection bias and potential consent form effect, Experiment A does not have enough statistical power to test discrimination and its nature when investors evaluate ``low-type" startups.\footnote{Since the IRR experiment does not generate real economic outcomes, it would still be helpful to exploit ``natural experiments" in the future.} To fully understand whether investors can be biased towards minority groups with low rating scores as predicted by discrimination theories, I follow up with Experiment B, which uses a redesigned correspondence test to identify the nature of discrimination in the cold call, pitch email setting. This also allows for the comparison of the IRR experiment with the currently widely used correspondence test experiment, which is particularly helpful to elucidate the benefits and drawbacks of the IRR experimental design prior to its widespread adoption by the academic community.\par

\hypertarget{sec:correpondencetest}{\section{Experiment B's Design and Results}}\label{sec:correpondencetest}
Experiment B implements a redesigned correspondence test to study gender and racial discrimination when investors evaluate struggling startups. Unlike \cite{gornall_gender_2020}, this correspondence test can examine the nature of detected discrimination and utilizes a new email technology that tracks investors' detailed information acquisition behaviors. In addition to randomizing the startup founders' gender and race, this technology also enables researchers to introduce variations in startup quality when the email response rate is extremely low, such as during an economic recession.\footnote{To test the nature of discrimination, researchers need to randomize startup characteristics that affect investors' contact decisions. In Experiment B, I orthogonally randomize the startup founders' gender, race, educational backgrounds, and the projects' comparative advantages.}\par


\subsection{Experimental Design}\label{cor:design}

\textbf{\textit{A. Recruitment Process and Sample Investors}}\par
\vspace{2mm}
In Experiment B, the research team sent hypothetical cold call pitch emails to 17,000+ early-stage venture capitalists who are mainly from the U.S. and other English-speaking areas, as documented in Section \ref{sec:data}. Online Appendix \ref{sec:appendix_cor} Table \ref{Hypothetical Startup Summary Statistics} provides the industry distribution of the created hypothetical startups. In total, Experiment B prepared 67 startup ideas and more than 200 names to make sure that experimental results are not driven by any particular startup ideas or founder names. These startup ideas cover all major industries that venture capitalists are interested in, such as Information Technology, Healthcare, Consumers, Energy, etc.\par

\vspace{2mm}
Experiment B was implemented between 03/2020 - 04/2020 during the outbreak of COVID-19. Due to the concern of an economic recession, many early-stage venture capitalists paused new investments during this period (\cite{howell_financial_2020}). Hence, the email response rate in Experiment B is lower than that in \cite{gornall_gender_2020}, as their experiment was implemented during an economic boom. Another reason for the lower response rate in this paper is that Experiment B introduces quality variations while \cite{gornall_gender_2020} only use pitch emails with the most attractive startup characteristics. As the expression ``Chinese Virus” was widely used in 03/2020, Experiment B also accidentally captures temporary anti-Asian discrimination during this special period.\footnote{See \href{https://theconversation.com/donald-trumps-chinese-virus-the-politics-of-naming-136796}{``Donald Trump’s ‘Chinese virus’: the politics of naming"} and \href{https://www.forbes.com/sites/rachelsandler/2020/03/23/trump-abruptly-stops-calling-coronavirus-chinese-virus-at-daily-press-briefing/?sh=4f1cb5fa47ad}{``Trump Abruptly Stops Calling Coronavirus ‘Chinese Virus’ At Daily Press Briefing”}. Considering the unusualness of this period, I implemented another round of the correspondence test on the same pool of investors in 10/2020. However, the second-round experiment provides very noisy results as many investors have realized the existence of this experiment. Hence, this paper mainly shows results from the first round. Results of the second-round experiment are available upon request.}\par

\vspace{2mm}
\textbf{\textit{B. Randomization and Design}}\par
\vspace{2mm}
\textbf{\textit{Manipulating Identity of the Entrepreneur}} --- I assign four co-founders to each created startup team, which include a white female co-founder, a white male co-founder, an Asian female co-founder, and an Asian male co-founder.\footnote{Having co-founders for a startup is very common, especially for highly innovative and complicated companies. Based on Pitchbook data, startups with multiple co-founders account for 50\% of all startups.} Each co-founder has a randomly assigned first name and last name that signal their gender and race. To make sure that investors associate the names with the correct gender and race information, I have recruited 107 US-based Amazon Mechanical Turk users to assess the gender and race of these selected names to delete any ambiguous names. The name lists and the name generation process details are provided in Online Appendix \ref{sec:appendix_cor}.\par
\vspace{2mm}

\textbf{\textit{Manipulating the Startup's Quality}} --- I randomize the startup team's educational background and the project's comparative advantages in both the subject line and the contents of each email. For the educational background, the control group does not mention the founders’ educational backgrounds. However, the treatment group indicates that the startup team comes from a prestigious university in the U.S. in both the email's subject line and contents.\footnote{Prestigious universities used include Ivy League colleges, MIT, and Stanford. In the experiment implemented between 03/2020 and 04/2020, I also included Northwestern University, Caltech, Johns Hopkins University, the Juilliard School, and other top schools in the field related to the startup. For example, if the startup is related to music, I mention that the founding team members come from Columbia University and the Juilliard School.} Similarly, for the project characteristics, the control group does not mention any specific comparative advantages of the startup, while the treatment group mentions comparative advantages such as ``22\% MOM Growth Rate."\footnote{MOM is an abbreviated form of ``month over month" growth.} \par

\vspace{2mm}
\textbf{\textit{Pitch Email Design and Website Construction}} --- The pitch emails, covering various startup ideas written for this experiment, follow the template and structure provided by \cite{gornall_gender_2020} and ``good pitch email template" examples posted on Quora. The startup ideas are provided by my research team members, who are usually young startup founders or members of startup-related clubs at Columbia University and other Ivy League colleges. We only choose valid startup ideas with relatively good coverage of key industries after discussions with practitioners. Wix, a commercial website builder, is used to make the related startup websites which are in the under-construction stage. A pitch email example is provided in Online Appendix Figure \ref{cor:pitchemail}, and a website example is provided in Online Appendix Figure \ref{cor:website}.\par 
\vspace{2mm}

\textbf{\textit{Manipulating Access to Information}} --- The randomization of startups' characteristics is implemented in the following two stages. In the first stage, before investors open the pitch email, they will see the randomly assigned email sender's name, indicating the sender’s gender and race, and also the randomly generated email subject line, indicating whether the startup has a well-educated founding team and a project with an impressive advantage.\footnote{Although large companies may ask a secretary or investor relationship manager to contact investors, for early-stage startups, it is usually the startup's founding team members themselves who contact investors in order to show their sincerity.} In the second stage, after investors open the pitch email, they will decide how much attention to devote to reading the pitch email. In each email's contents, the co-founder’s name occurs multiple times (including in the introductory paragraph, email addresses, the email signature, and email senders' names) to make the gender and race information more salient. If the email's subject line mentions an Ivy League educational background or project advantages, there are extra sentences inserted to emphasize this information again in the email's contents while keeping the rest of the contents the same. After reading the email's contents, investors can decide whether to reply or forward the email to other related investors who may potentially also be interested in the same pitch email. All the technical details about sending a large number of emails and the preparation work involved are provided in Online Appendix \ref{sec:appendix_cor}.\par
\vspace{2mm}

\textbf{\textit{Email Behavior Measurements}} --- I track the following email behavior measurements: the email opening status and the corresponding time stamp, the email staying time measured in seconds, the sentiment of email replies analyzed through LIWC, the click rate of the inserted startup websites, the response rate, and contents of replies.\footnote{LIWC (Linguistic Inquiry and Word Count) is a text analysis program used for sentiment analysis.} Despite these rich behavior measurements, only email opening rate and email staying time generate enough statistical power to analyze investors' responses. All the other traditionally used behavior measurements do not survive in this recession period when the ``low-response-rate" problem is very severe. The detailed mechanisms of recording different email behaviors and whether such behavior measurements are used in previous literature are described in Online Appendix \ref{sec:appendix_cor} Table \ref{cor:tracebehavior}. A flow chart of Experiment B is provided in Figure \ref{fig:ct_flow_chart}.\par

\subsection{Results}\label{cor:analysis}

\subsubsection{\emph{Investors Favor Female and Asian Founders in the Pitch Email Setting.}}

Table \ref{table_open} Panel A summarizes investors' major information acquisition behaviors in the correspondence test. On average, the pitch email opening rate is 12.03\% and each investor spends roughly 24 seconds reading the cold call pitch email during the 03/2020-04/2020 testing period. However, both the startup website click rates and the email response rates are very low (roughly 1\%), indicating that early-stage investors are sensitive to business cycles as documented by \cite{howell_financial_2020}. Therefore, traditionally recorded investors' email behaviors, such as the email response rate, do not generate enough statistical power during the COVID-19 pandemic. All the experimental results in this paper rely on the new email behaviors recorded by the latest email tracking technology.\par

\vspace{2mm}
Table \ref{table_open} Panel B reports regression results of investors' email opening behaviors for randomized pitch emails. The dependent variable equals one when an investor opens the pitch email, and zero otherwise. Columns (1), (2), (3), and (5) use all the observations collected between 03/2020-04/2020. In Column (4), results are reported for the sub-sample where the startup team's educational background is from ``purely Ivy League colleges", Stanford, and MIT.\footnote{``Pure\_Ivy" represents cases like ``Team from Columbia University," while ``Mixed\_Ivy" represents cases like ``Team from Columbia University and Juilliard Music School." For some startups in the music or medical industry, I combine an Ivy League college with a university well-known for having a top program in that specific area for the treatment group.} All the regressions include startup fixed effects to control for any idiosyncratic characteristics of each startup pitch email, such as the business models, etc. Hence, I am comparing investors’ email opening rates within the same startup’s pitch email, and all the results are similar after including investor fixed effects. Following \cite{bernstein_attracting_2017}, all standard errors are clustered at the investor level to account for the correlated opening decisions across different pitch emails received by the same investor.\par

\vspace{2mm}
Results of Table \ref{table_open} Panel B show that on average, investors favor female, Asian, and well-educated founders during the experimental period. Column (1) shows that using a female first name in a pitch email raises the opening rate by 1\% compared to using a male first name. This result is statistically significant at the 1\% level. Column (2) shows that using an Asian last name in a pitch email raises the opening rate by 0.7\% after President Trump stops using the wording ``Chinese Virus” in 03/2021.\footnote{See \hyperlink{https://www.bloomberg.com/news/articles/2020-03-25/trump-says-he-ll-stop-using-chinese-virus-easing-blame-game}{Trump Says He’ll Stop Using the Term ‘Chinese Virus’}.} This difference is statistically significant at the 10\% level and represents a 6\% increase in opening rates compared with using a white last name. Similarly, Column (3) shows that mentioning a good educational background in the email's subject line increases the opening rate by 0.7\% compared with not mentioning education in the email's subject. This effect increases to 1.2\% for the sub-sample that mentions only a pure Ivy League educational background (i.e., ``Team from Columbia University” rather than ``Team from Columbia University and Juilliard Music School ”). However, Columns (4) and (5) show that mentioning the project's advantages in the email's subject line does not significantly increase the email opening rate, which is consistent with \cite{bernstein_attracting_2017}. Results provided by Table \ref{table_open} confirm the surprising results found in previous literature that investors are biased towards female and Asian founders in the pitch email setting.\par

\subsubsection{\emph{Temporary Anti-Asian Discrimination Exists During the COVID-19 Outbreak.}}

Table \ref{table_open} Panel C reports regression results of how startup characteristics affect investors' staying time on each email. The dependent variable is the time spent on each pitch email measured in seconds, which approximates how much attention each investor devotes to the email. In Columns (1) and (2), I include unopened emails and replace their email staying time with 0 seconds. Considering the potential truncation issue, I also report the sub-sample of opened emails in Column (3).\footnote{During the COVID-19 outbreak, no matter how biased investors were against Asian founders, the worst possible situation was that investors did not open the pitch emails sent by Asian last names, and hence the staying time is 0 seconds. This truncation issue at the 0 second mark will bias our results towards zero. Therefore, it is important to compare magnitudes of the race effect when the regression only focuses on opened emails.} Similar to Panel B, all the regressions include startup fixed effects, and standard errors in parentheses are clustered at the investor level.\par

\vspace{2mm}
Table \ref{table_open} Panel C shows that although investors generally spent more time on female and Asian founders' emails, there was temporary discrimination against Asian founders during the COVID-19 outbreak in 03/2020. Column (1) shows that using a female first name raises the time spent on a pitch email by 0.36s in 03/2020 and 0.12s in 04/2020. This magnitude is not large due to the truncation issue. Similarly, Columns (2) and (3) show that using an Asian last name raises the staying time by 0.38s in the full sample and 2.49s among opened emails in 04/2020, which accounts for a 10\% increase in the staying time. However, the significantly negative coefficients of the interaction term between ``Asian Founder" and ``March" indicate that using an Asian last name reduces the staying time by 0.28s in the full sample and 2.99s among opened emails in 03/2020. This accounts for a 12.5\% decrease in staying time. Results suggest that there is discrimination against Asian founders in 03/2020, and the direction of discrimination flips after 04/2020. Hence, the direction of discrimination also depends on timing and might be temporarily affected by big societal events, such as the COVID-19 outbreak.\par

\vspace{2mm}
It should be noted that the temporary discrimination against Asian founders in 03/2020 should be interpreted as the lower bound of the discrimination in the real world. As shown in Section \ref{irr:discussion}, investors exhibit more implicit discrimination against Asian founders when evaluating highly-rated startups compared to evaluating struggling startups. If discrimination against Asian founders even exists in the cold call, pitch email setting, this racial discrimination may be even worse in other mainstream fundraising settings. Moreover, the experimental setting of Experiment B is very noisy as many factors can affect investors' email behaviors. Hence, the racial discrimination must be salient enough in order to generate significant results. Both reasons suggest an even harsher fundraising environment for Asians during the COVID-19 outbreak. \par


\subsection{Mechanisms (Testing the Nature of Discrimination)}\label{cor:mechanism}

The mainstream discrimination theories studied with the correspondence test experiment can be classified into the following three types (\cite{bohren_dynamics_2019}): belief-based mechanisms, taste-based mechanisms, and amplifying mechanisms. Although results show that gender bias is likely driven by taste-based mechanisms and that the racial bias is mainly driven by belief-based mechanisms, multiple subtle mechanisms can coexist.\par

\subsubsection{Belief-Based Mechanisms}\label{cor:mec_belief}


\textbf{\emph{Expected Quality and Financial Returns (First Moment)}} Investors' bias towards female and Asian founders can stem from them foreseeing higher future returns from these founders' startups compared to other founders who also send cold call emails.\footnote{Investors may hold this belief because of previously documented facts (\cite{ewens_are_2020}), \cite{fairlie2010race}), the self-selection effect of minority founders (\cite{rosette2010agentic}, \cite{fernandez2009culture}, \cite{buttner1997women}, \cite{puri2013economic}, \cite{baron2001perceptions}, \cite{fryer2007belief}, \cite{bohren2019dynamics}, \cite{howell_networking_2019}, and \cite{kacperczyk2019illegitimacy}.), the lower negotiation power of minority founders (\cite{amatucci2004women}), more pleasant collaboration experiences (\cite{shane2012inventors}), etc.} Table \ref{table_interaction} tests this channel and finds that only the bias towards Asians is mainly driven by this belief. Column (4) shows that mentioning an Ivy League educational background reduces the bias towards Asian founders compared to white founders by 0.7\% in email opening rates. Column (5) shows that the interaction effect of using an Asian name and mentioning an Ivy League educational background increases to -3.2\% if I focus on the sub-sample of emails sent after 03/23/2020 that only mention ``pure" Ivy League colleges.\footnote{President Trump stopped using the phrase ``Chinese Virus” on 03/23/2020.} This interaction effect is statistically significant at the 1\% level. According to the discrimination model in \cite{ewens2014statistical}, the result that positive signal shrinks the racial gap is only consistent with a statistical discrimination hypothesis. Intuitively, if discrimination is driven by beliefs in productivity, more signals about the startup's quality will correct this belief and reduce the racial gap. Hence, investors favor Asians in the email setting because among struggling startups, Asian founders' startups are considered to be more profitable than similar white founders' startups.\footnote{This conclusion is also consistent with the positive coefficient of ``Asian Founder" in Panel B of Online Appendix Table \ref{irr_investor_profile_race_quantile_Q1}. Although the positive coefficient is not statistically significant, it suggests that investors' profitability evaluations of Asian founders can be positive if their startups' profitability is below the 10th percentile of the distribution.}\par

\vspace{2mm}
However, I do not find any suggestive evidence supporting belief-driven gender bias. According to Columns (1) and (3) of Table \ref{table_interaction}, the interaction terms of being a female founder and attending prestigious universities are insignificant. In Column (2), the interaction term even becomes positive, which is statistically significant at the 10\% level. Some might speculate that graduating from prestigious universities is not a valid positive signal for women-led startups, leading to the insignificant interaction terms. Online Appendix Table \ref{valid_signal} shows that good educational backgrounds improve investors' profitability evaluations across the whole distribution of both women-led startups and Asian-led startups. Hence, the ``invalid signal" hypothesis is not supported by Experiment A. Based on \cite{ewens2014statistical}, in certain situations, both belief-driven discrimination and taste-driven discrimination can generate the result that a positive signal widens the gender gap. Hence, this paper uses Neumark's model below to empirically test the relative variance of women-led startups and men-led startups before pinning down the dominant mechanism for investors' gender bias.\par

\vspace{2mm}

\textbf{\emph{Expected Variance of Different Groups (Second Moment)}}
According to the Heckman and Siegelman [HS] Critique (\cite{siegelman1993urban}; \cite{heckman1998detecting}), even in the ideal case in which both observed and unobserved group averages (i.e., first moment statistics) are identical, the correspondence test can generate spurious evidence of discrimination in either direction when the belief of unobserved productivity variance differs.\footnote{This is because a standard correspondence test only observes a nonlinear binary decision outcome (i.e., reply vs. no reply, etc.) and this outcome can be affected by higher moment statistics. Hence, the relative variance of unobservable characteristics of majority groups and minority groups also affects evaluators' callback rates.} \cite{neumark_detecting_2012} develops a model that can address this concern and recover an unbiased estimate of discrimination.\footnote{This model uses a Heteroscedastic Probit Model after imposing several parametric assumptions. I extend his model a little bit by adjusting his assumed monotonic hiring rules in Online Appendix Section \ref{sec:appendix_cor}. The full discussion and review of this model are provided in \cite{neumark_detecting_2012}.} Table \ref{probithet} shows that results are still robust after correcting for the source of bias from unobserved variance using Neumark's model. Column (1) demonstrates that using a female name significantly increases the email opening rate by 1\%, and I cannot reject the hypothesis that the variances between female and male founders are the same. However, the relative variance of female founders and male founders is smaller than 1, indicating that investors expect female founders to be more homogeneous. Columns (2) and (3) show that using an Asian last name still increases the email opening rate by 0.7\% and that the relative variance of Asian founders and white founders decreases from 1.12 in 03/2020 to 1.09 in 04/2020. This means that investors expect Asian-led startups to have more uncertainties than white-led startups during the COVID-19 outbreak. Fortunately, these uncertainties decrease starting in April. In a nutshell, the expected variance of different groups is not the main driver of the detected bias towards minority founders because results still hold after using Neumark's model.\par

\vspace{2mm}
\textbf{\emph{Strategic Channel}} The entrepreneurial financing process in the VC industry is a two-sided matching process (\cite{sorensen2007smart}). Theoretically speaking, investors may prefer minority founders if similar majority founders are ``over-qualified" and have weaker willingness to collaborate with them due to many outside options. Since Experiment B mainly captures a non-mainstream fundraising method that highly-rated startups generally do not use, investors are unlikely to reject a startup team because the founders are ``too good" or ``overqualified". Table \ref{table_open} Columns (4) and (5) also rule out this strategic channel by showing that mentioning an excellent educational background still significantly increases investors’ email opening rates. \par

\subsubsection{Taste-Based Mechanisms}\label{cor:mec_taste}


\textbf{\emph{Friendly Support}} Investors are likely to be biased towards minority founders because they want to support disadvantaged groups. For example, some impact funds or angel groups, such as 37 Angels, only invest in women-led startups. Table \ref{table_ESG} supports this hypothesis and finds that the bias towards women is much higher for impact funds compared to common funds. Although impact funds are also slightly more biased towards Asian founders, this result is not statistically significant.\footnote{Not-for-profit impact funds are defined using the primary investor type from Pitchbook.} Columns (1)-(3) show that using female names increases the email opening rate by 10.3\% for impact fund investors and only 1.1\% for profit-driven fund investors who do not have special ESG goals. The magnitude of this gender effect for impact funds is roughly 10 times that for common funds. Columns (4)-(6) find that impact funds open more emails sent with an Asian last name, although the magnitude of this racial effect for impact funds is only 2 times the effect for profit-driven funds, and results are not significant. Results above show that the bias towards female founders partially stems from friendly support from impact funds. However, this channel is not significant for Asian founders.\par

\vspace{2mm}
It should be noted that investors from profit-driven VC funds also favor female founders although the magnitude of their gender bias is much smaller than that of impact fund investors. \cite{ewens2014statistical} shows that only when the variance of women-led startups is larger than the variance of men-led startups (i.e., ``Case 2" in \cite{ewens2014statistical}), statistical discrimination predicts that a positive signal widens the gender gap. However, results from Neumark's model in Table \ref{probithet} shows that the variance of women-led startups is smaller than the variance of men-led startups (i.e., ``Case 3" in \cite{ewens2014statistical}). In this situation, only taste-based discrimination can generate a widened gender gap due to a positive signal. This is also consistent with the negative coefficient of ``Female Founder" in Panel B of Table \ref{irr_investor_profile_gender_quantile_Q1} when startups' profitability ratings are below the 10th percentile. Hence, the gender bias towards female founders from profit-driven investors is likely driven by taste. However, Experiment B cannot identify whether this taste-based mechanism comes from the friendly support channel, social image effect, or potential sexual attractiveness.\footnote{“\href{https://www.wired.com/story/female-founders-still-face-sexual-harassment-from-investors/}{Female Founders Still Face Sexual Harassment from Investors,}” October 15, 2018, shows that among respondents to the survey sent by Y Combinator, more than 20\% of women said they had been harassed.}\par

\vspace{2mm}

\textbf{\emph{Homophily}} Homophily means that people prefer groups that share similar backgrounds to themselves (\cite{egan2017harry}). Online Appendix Table \ref{table_homophily} shows that results do not support homophily within gender in Experiment B.\footnote{I do not test the homophily effect based on race because the racial information of investors is not provided in the data, and race prediction algorithms based on names are very noisy.} Columns (1) and (4) report that the difference in female and male investors' information acquisition behaviors are not significantly different from zero as shown by the interaction terms of ``Female Founder" and ``Female Investor". Columns (2) and (3) show that using female names increases the email opening rate by 0.8\% among female investors and 1.1\% among male investors. However, Columns (5) and (6) show that female investors spend more time on pitch emails sent by female names. Since results are not significant, homophily is not identified in Experiment B.\par

\subsubsection{Amplifying Mechanisms (Attention Discrimination \& Implicit Bias)}\label{cor:mec_amplify}
Mechanisms that can magnify both taste-based discrimination and belief-based discrimination may also exist. These amplifying mechanisms include attention discrimination and implicit bias. ``Attention discrimination” theory (\cite{bartos_attention_2016}) predicts that even if complete information about an individual is readily available, discrimination can still occur because investors may endogenously allocate their scarce attention to their preferred groups before they make their decisions. Considering that all the outcome variables (i.e., email opening rates, time spent on each pitch email) actually measure the attention investors devote to cold call pitch emails rather than finalized investment decisions, results naturally support the existence of the attention discrimination channel. ``Implicit bias” refers to the attitudes or stereotypes that affect investors' decisions in an unconscious manner. Unfortunately, Experiment B cannot test this channel.\par

\subsubsection{Alternative Mechanisms}\label{cor:mec_alternative}

\textbf{\emph{Uninformative Email Replies}} Investors may pretend to behave in a more friendly manner to minorities in an email setting. Hence, these email replies are not indicative of their true investment preferences. However, this hypothesis cannot explain the results found through measuring email opening rates and email reading time because these behaviors are usually not observed by the founders directly or used in previous correspondence tests. Therefore, I can rule out this mechanism.\par

\subsection{Merits and Limitations}\label{cor:limitation}
Experiment B has several merits due to the new email tracking technology. First, it improves the internal validity compared with the classical correspondence test design. By tracking multiple investors' detailed information acquisition behaviors, which were unobservable before, researchers can mitigate the ``low-response-rate" problem and introduce meaningful variations in startups' quality. This helps to identify discrimination and its nature when response rates are low.\par

\vspace{2mm}
However, Experiment B also suffers from several limitations. The most important limitation is that sending cold call pitch emails is not the mainstream fundraising method. Hence, it is only used to complement Experiment A by investigating the fundraising situation of relatively low-quality startups. Other standard limitations of correspondence tests include noisy experimental settings, only recording investors' behaviors in the initial contact stage, and the difficulty of implementation in European Union countries due to the data confidentiality rule.\footnote{Since correspondence tests usually do not provide a consent form to participants, it has been harder to implement this type of experiment in the EU area starting in 2018.\par Based on the attention discrimination theory by \cite{bartos_attention_2016}, investors benefit more from providing their limited attention to their preferred startup groups in a cherry-picking market (i.e., venture capital investment setting). Hence, if no further bias exists in later-round communication stages, the amount of attention measured is indicative of investors' internal preferences. Some research work (\cite{hu2020human}, \cite{kanze2018we}) has analyzed video data to study the later communication stage.}

\hypertarget{sec:discussion}{\section{Discussion}}\label{sec:discussion}
\subsection{Comparison of IRR Experiment and Correspondence Test}
The IRR experiment and correspondence test, both as experimental methods of testing discrimination, complement each other due to their different benefits and limitations. First, when studying high-skilled labor markets, the IRR experiment can provide a ``warm" setting if a correspondence test is not feasible or only captures a non-mainstream ``cold" setting as shown in this paper. Second, if a correspondence test can capture an appropriate natural experimental setting, a complementary IRR experiment can help the correspondence test to investigate coexisting mechanisms that are technically hard to identify otherwise. Third, while the correspondence test generally recruits a large number of subjects and does not suffer from the ``consent form” effect, the IRR experiment collects richer data that enables researchers to obtain novel empirical results. These results include discrimination's distributional effects, decision-based heterogeneous effects, and dynamic changes across evaluators' internal thresholds, which are usually impossible to obtain in a classical correspondence test. However, both methods cannot directly identify taste-driven discrimination and its sub-mechanisms. Hence, more experimental tools should be developed to test taste-driven preferences in a field setting.

\subsection{Link to Extant Theoretical Literature}
Although multiple discrimination theories have predicted a ``discrimination reversion" phenomenon within the framework of statistical discrimination, these theories mainly explain the findings related to the racial discrimination in this paper, which is affected by investors' beliefs.\footnote{See \cite{phelps1972statistical}, \cite{aigner1977statistical}, \cite{lundberg1983private}, and \cite{morgan2009diversity}.} However, the reversion of gender discrimination is partially caused by taste-driven preference towards female founders when investors evaluate struggling startups. Therefore, empirically speaking, distributional effects may follow other patterns depending on the strength and direction of taste-driven preferences.\par   

\subsection{Policy Implications}
Experimental results provide the following policy implications for handling discrimination in the entrepreneurial financing process. First, any actions that mitigate \textit{implicit} discrimination are helpful. For example, in fundraising activities like Startup Pitch Night, minority founders can take earlier time slots while investors are still focused. Second, since discrimination that exists among top startups is mainly driven by beliefs and attention discrimination might exist, investors should be encouraged to invite more minority founders to enter the communication stage and learn more details of these startups.\footnote{Special thanks to Paul Beaumont for pointing this out. Since most experimental results in this paper focus on the initial contact stage rather than the investment stage, it indicates that minority founders have fewer opportunities to elaborate on details of their startups because they are less likely to enter the communication stage. This is a form of ``attention discrimination" as investors collect less information about minority founders' companies.} Third, considering that this paper's empirical results echo several theoretical predictions in \cite{morgan2009diversity}, especially the ``discrimination reversion" and ``pro-cyclical discrimination" phenomena, I restate some of their insightful but counter-intuitive implications here. For example, high-level worker protections against dismissal and online interviews decrease diversity.\par

\hypertarget{sec:conclusion}{\section{Conclusion}}\label{sec:conclusion}
This paper mainly studies whether early-stage investors discriminate against female founders and Asian founders during the U.S. entrepreneurial financing process. Despite the importance of this question, the literature generates conflicting results. To identify discrimination and its nature in the VC industry, this paper implements complementary field experiments with real U.S. venture capitalists. Experiment A combines an IRR experiment and a donation game. Investors are invited to evaluate multiple hypothetical startup profiles using a machine learning matching tool so that they can find real matched startups from collaborating incubators. Investors can also use the tool to donate a small amount of money to randomly displayed startup teams and provide their anonymous encouragement during the COVID-19 recession. Experiment B exploits a redesigned correspondence test to study how investors evaluate struggling startups in a cold call, pitch email setting. With new email behavior tracking technologies, Experiment B compares investors' detailed information acquisition behaviors when evaluating pitch emails with randomized startups' information.\par

\vspace{2mm}
Results discover strong implicit discrimination against female and Asian startup founders. While the implicit discrimination against women and Asians mainly exists when investors evaluate attractive startups and become selective during the pre-screening process, investors are biased towards female and Asian founders when they evaluate unattractive startups and become sufficiently ``unselective". The discovered ``discrimination reversion" and ``pro-cyclical discrimination" phenomena confirm the prediction of discrimination theories and provide one explanation to reconcile the contradictory results in the literature. By developing a decision-based heterogeneous effect estimator, the paper finds that investors' beliefs about startups' profitability can better explain their heterogeneous treatments of minority groups than their beliefs about startups' availability. Besides discrimination in the investment setting, the paper also finds that homophily exists when venture capitalists provide non-investment support to startup founding teams. Lastly, the paper detects a temporary, stronger discrimination against Asian founders during the COVID-19 outbreak.\par

\vspace{2mm}
Overall, this paper contributes to experimental design and the debate about discrimination in a high-skilled labor market and a financial market setting. The discovered distributional effect emphasizes the importance of investigating discrimination at the top level. Researchers can test whether discrimination also exists in other parts of the entrepreneurial financing system and investigate its implications for equilibrium outcomes in the future. Since this paper discovers discrimination against Asians, the largest minority group in the U.S. entrepreneurial community, it suggests that future studies should test discrimination against other under-represented minority groups as well. In addition, studies exploiting any ``natural experiment” settings are helpful to implement welfare analysis.\par


\hypertarget{sec:acknowledgement}{\section*{Acknowledgement}}\label{sec:acknowledgement}
I would like to express my deepest appreciation to Jack Willis, Harrison Hong, and Wei Jiang for their guidance, support, and profound belief in the value of my work. I am also grateful to Donald Green, Sandra Black, Mark Dean, Alessandra Casella, Matthieu Gomez, Jose Scheinkman, Eric Verhoogen, Jushan Bai, Junlong Feng, Michael Best, Bentley MacLeod, Alexander Ljungqvist, Ulrike Malmendier, Bernard Salenie, Per Strömberg, Corinne Low, Shi Gu, Andrew Prat, Xavier Giroud, Johannes Stroebel, Olivier Toubia, and Patrick Bolton for their valuable comments. I thank the participants at the PhD Colloquiums at Columbia University and the investors who participated in these experiments. Special thanks go to Corinne Low, Colin Sullivan, and Judd Kessler for sharing their IRR Qualtrics code package. This project was supported by PER funding from the Columbia University Economics Department, by the Columbia University Eugene Lung Entrepreneurship Center Fellowship, and by the Columbia CELSS Seed Grant. The project is registered at AEA RCT Registry (AEARCTR-0004982) and approved by Columbia IRB. All errors are my own.\par

\clearpage
\bibliographystyle{aer}
\bibliography{reference.bib}


\begin{table}
 \caption{Summary Statistics for Investors}
\begin{center}
 \label{investor_summary}
\scalebox{0.85}{ 
\begin{tabular}{l m{5cm} m{2cm} c m{4cm} c c } 
\multicolumn{7}{c}{Panel A: Investor Location Distribution}\\
\hline
\multicolumn{2}{l}{Country}&N&\multicolumn{2}{l}{Percentage}&\multicolumn{2}{l}{Female Percentage}\\
\hline
\multicolumn{2}{l}{US}&15,184& \multicolumn{2}{c}{84.91\%}&\multicolumn{2}{c}{23.57\%}\\
\multicolumn{2}{l}{Canada}&647& \multicolumn{2}{c}{3.62\%}&\multicolumn{2}{c}{29.68\%}\\
\multicolumn{2}{l}{Israel}& 456 & \multicolumn{2}{c}{2.55\%}&\multicolumn{2}{c}{29.39\%}\\
\multicolumn{2}{l}{UK}&93& \multicolumn{2}{c}{0.52\%}&\multicolumn{2}{c}{22.58\%}\\
\multicolumn{2}{l}{India}&514& \multicolumn{2}{c}{2.87\%}&\multicolumn{2}{c}{18.87 \%}\\
\multicolumn{2}{l}{Singapore \& Hong Kong}&454& \multicolumn{2}{c}{2.54\%}&\multicolumn{2}{c}{21.59\%}\\
\multicolumn{2}{l}{Australia \& New Zealand}& 228  & \multicolumn{2}{c}{1.28\%}&\multicolumn{2}{c}{25.44\%}\\
\multicolumn{2}{l}{Others}& 306 & \multicolumn{2}{c}{1.71\%}&\multicolumn{2}{c}{21.57\%}\\
\hline
&&&&&&\\
\multicolumn{7}{c}{Panel B: Investor Industry Distribution}\\
\hline
\multicolumn{3}{l}{Industry}& \multicolumn{2}{c}{N}& \multicolumn{2}{c}{Percentage}\\
\hline
\multicolumn{3}{l}{Information Technology}& \multicolumn{2}{c}{13,628 }& \multicolumn{2}{c}{76.21\%}\\
\multicolumn{3}{l}{Healthcare}& \multicolumn{2}{c}{6,056 }& \multicolumn{2}{c}{33.87\%}\\
\multicolumn{3}{l}{Consumers}& \multicolumn{2}{c}{6,256}& \multicolumn{2}{c}{34.98\%}\\
\multicolumn{3}{l}{Energy}& \multicolumn{2}{c}{ 4,234 }& \multicolumn{2}{c}{23.68\%}\\
\multicolumn{3}{l}{Life Sciences}& \multicolumn{2}{c}{3,347}& \multicolumn{2}{c}{18.72\%}\\
\multicolumn{3}{l}{Finance}& \multicolumn{2}{c}{3,023}& \multicolumn{2}{c}{16.91\%}\\
\multicolumn{3}{l}{Media \& Entertainment}& \multicolumn{2}{c}{2,533}& \multicolumn{2}{c}{14.17\%}\\
\multicolumn{3}{l}{Agriculture \& Food}& \multicolumn{2}{c}{2,072}& \multicolumn{2}{c}{11.59\%}\\
\multicolumn{3}{l}{Transportation}& \multicolumn{2}{c}{1,743}& \multicolumn{2}{c}{9.75\%}\\
\multicolumn{3}{l}{Education}& \multicolumn{2}{c}{ 1,359}& \multicolumn{2}{c}{7.60\%}\\
\multicolumn{3}{l}{Clean Technology}& \multicolumn{2}{c}{1,201}& \multicolumn{2}{c}{6.72\%}\\
\multicolumn{3}{l}{Others}& \multicolumn{2}{c}{ 3,271 }& \multicolumn{2}{c}{18.29\%}\\
\hline
&&&&&&\\
\multicolumn{7}{c}{Panel C: Investor Characteristics}\\
\hline
\multicolumn{3}{l}{}&\multicolumn{2}{c}{N}&\multicolumn{2}{c}{Mean}\\
\hline
\multicolumn{3}{l}{Female Investor}&\multicolumn{2}{c}{17,882}&\multicolumn{2}{c}{0.24}\\
\multicolumn{3}{l}{Senior Investor}&\multicolumn{2}{c}{17,882}&\multicolumn{2}{c}{0.84}\\
\multicolumn{3}{l}{Angel Investor}&\multicolumn{2}{c}{17,882}&\multicolumn{2}{c}{0.11}\\
\multicolumn{3}{l}{Top University}&\multicolumn{2}{c}{13,785}&\multicolumn{2}{c}{0.31}\\
\multicolumn{3}{l}{Graduate School}&\multicolumn{2}{c}{9,232}&\multicolumn{2}{c}{0.61}\\
\multicolumn{3}{l}{Not-for-profit Fund}&\multicolumn{2}{c}{13,156}&\multicolumn{2}{c}{0.02}\\
\hline
\end{tabular}}
\end{center}
\begin{tablenotes}
\item 	
\footnotesize \emph{Notes.} This table reports descriptive statistics for active venture capitalists, defined as those whose email addresses are verified by the testing email. These investors received recruitment emails in Experiment A and cold call pitch emails in Experiment B. Panel A reports the geographical distribution of the sample investors. ``Others" includes South Africa, Cayman Islands, Malaysia, etc. Panel B reports the industries that recruited investors are interested in. An investor can indicate multiple preferred industries. ``Others" includes special industries, such as packaging technology industry. 3.8\% of investors' industry preferences cannot be found online and I have assumed that they are interested in all the industries when sending out cold call pitch emails. Panel C reports investors' demographic information and investment philosophies. ``Female Investor" is an indicator variable that equals one if the investor is female, and zero otherwise. ``Senior Investor" is an indicator variable which equals one if the investor is senior (defined as C-level positions, principals, vice president, and partners), and zero otherwise. ``Angel Investor" is an indicator variable that equals one if the investor is an angel investor or belongs to an angel group, and zero otherwise. If an investor is both an angel investor and also an institutional investor, I treat her as an angel investor. ``Not-for-profit Fund" is an indicator variable that equals one if the investor works in a not-for-profit impact fund based on the ``primary investor type" in the Pitchbook Database. ``Top University" and ``Graduate School" are indicator variables that equal one if the investor has attended a top university (i.e., Ivy League colleges, MIT, Duke, Caltech, Amherst, Northwestern, Stanford, UC Berkeley, University of Chicago and Williams College) or has attended graduate school, respectively.
\end{tablenotes}
\end{table}


\begin{table}
 \caption{Summary Statistics of Recruited Investors in Experiment A}
\begin{center} 
\label{irr_investor_summary}
\scalebox{0.85}{ 
\begin{tabular}{l m{4cm} c m{2cm} c m{3cm} m{4cm}} 
\multicolumn{7}{c}{Panel A: Investor Stated Interest Across Sectors}\\
\hline
\multicolumn{3}{l}{Sector (Repeatable)}&\multicolumn{2}{l}{N}&Fraction (\%) & Fraction (\%) \\
\multicolumn{3}{l}{}&\multicolumn{2}{l}{}&& in Pitchbook\\
\hline
\multicolumn{3}{l}{Information Technology}&\multicolumn{2}{l}{39}&55.7\%& 58.3\%\\
\multicolumn{3}{l}{Consumers}&\multicolumn{2}{l}{10}&14.3\%& 28.4\%\\
\multicolumn{3}{l}{Healthcare}&\multicolumn{2}{l}{17}&24.3\%&22.1\%\\
\multicolumn{3}{l}{Clean Technology}&\multicolumn{2}{l}{3}&4.3\%&0.7\%\\
\multicolumn{3}{l}{Business-to-Business}&\multicolumn{2}{l}{7}&10.0\%&8.5\%\\
\multicolumn{3}{l}{Finance}&\multicolumn{2}{l}{11}&15.7\%&9.7\%\\ 
\multicolumn{3}{l}{Media}&\multicolumn{2}{l}{4}&5.8\%&8.0\%\\
\multicolumn{3}{l}{Energy}&\multicolumn{2}{l}{5}&7.1\%&15.9\%\\
\multicolumn{3}{l}{Education}&\multicolumn{2}{l}{3}&4.3\%&2.2\%\\
\multicolumn{3}{l}{Life Sciences}&\multicolumn{2}{l}{2}&2.9\%&9.9\%\\
\multicolumn{3}{l}{Transportation \& Logistics} &\multicolumn{2}{l}{4}&5.7\%&5.7\%\\
\multicolumn{3}{l}{Others} &\multicolumn{2}{l}{6}&8.6\%&12.8\%\\
\multicolumn{3}{l}{Industry Agnostic}&\multicolumn{2}{l}{6}&8.6\%&26.1\%\\
\hline
&&&&&&\\
&&&&&&\\
\multicolumn{7}{c}{Panel B: Investor Stated Interest Across Stages}\\
\hline
\multicolumn{3}{l}{Stage (Repeatable)}&\multicolumn{2}{l}{N}&Fraction (\%)&Fraction (\%)\\
\multicolumn{3}{l}{}&\multicolumn{2}{l}{}&& in Pitchbook\\
\hline
\multicolumn{3}{l}{Seed Stage}&\multicolumn{2}{l}{47}&67.1\%&41.9\%\\
\multicolumn{3}{l}{Series A}&\multicolumn{2}{l}{45}&64.3\%&31.8\%\\
\multicolumn{3}{l}{Series B}&\multicolumn{2}{l}{17}&24.3\%&15.0\%\\
\multicolumn{3}{l}{Series C or Later Stages}&\multicolumn{2}{l}{5}&7.1\%&11.2\%\\
\hline
&&&&&&\\
&&&&&&\\
\multicolumn{7}{c}{Panel C: Investor Stated Demographic Information}\\
\hline
\multicolumn{2}{l}{} & \multicolumn{2}{l}{N} & \multicolumn{2}{l}{Mean} & Mean\\
\multicolumn{3}{l}{}&\multicolumn{2}{l}{}&& in Pitchbook\\
\hline
\multicolumn{2}{l}{Female Investor} & \multicolumn{2}{l}{69} & \multicolumn{2}{l}{0.20} &0.24\\ 
\multicolumn{2}{l}{Minority Investor} & \multicolumn{2}{l}{64} & \multicolumn{2}{l}{0.42} &N/A\\ 
\multicolumn{2}{l}{Senior Investor} & \multicolumn{2}{l}{69} & \multicolumn{2}{l}{0.86} &0.80\\ 
\hline
&&&&&&\\
&&&&&&\\
\multicolumn{7}{c}{Panel D: Investor Stated Investment Philosophy}\\
\hline
\multicolumn{2}{l}{} & \multicolumn{2}{l}{N} & \multicolumn{2}{l}{Mean} & S.D\\
\hline
\multicolumn{2}{l}{Cold Email Acceptance} & \multicolumn{2}{l}{69} & \multicolumn{2}{l}{0.74} &0.44\\ 
\multicolumn{2}{l}{Prefer ESG} & \multicolumn{2}{l}{69} & \multicolumn{2}{l}{0.11} &0.32\\ \multicolumn{2}{l}{Direct Investment} & \multicolumn{2}{l}{69} & \multicolumn{2}{l}{0.94} &0.24\\
\hline
\end{tabular}}
 \end{center}
\end{table}

\begin{table}
\begin{center}
\scalebox{0.85}{ 
\begin{tabular}{l c c c c c c} 
\emph{Continued}&&&&&&\\
&&&&&&\\
\multicolumn{7}{c}{Panel E: Available Venture Capital Companies' Financial Performance}\\
\hline
\multicolumn{4}{c}{}&\multicolumn{3}{c}{Percentile}\\
\cline{5-7}\\
& N & Mean & S.D & 10 & 50 &90\\
\hline
\emph{Recruited Sample}&&&&&&\\
Total Active Portfolio& 54 & 41.40 & 44.51 & 10 & 24 &102\\
Total Exits & 46 & 32.74 & 48.39 & 1 & 9 &110\\
VC Company Age & 52 & 11.75 & 8.95 & 3 & 8.5 &25\\
AUM (Unit: \$1 Million)& 33 & 547.46 & 1029.10  & 30 & 111.7 &1700\\
Dry Power (Unit: \$1 Million)& 33 & 163.86 & 307.04 & 6.43 & 44.35 &313.59\\
Fraction of Female Founders&69&0.12&0.13&0.03&0.10&0.21\\
in Portfolio Companies&&&&&&\\
&&&&&&\\
\emph{Pitchbook Sample}&&&&&&\\
Total Active Portfolio&16,742&14.59&35.88&1&5&33\\
Total Exits &9,104&15.47&41.27&1&4&35\\
VC Company Age &9,115&9.63& 11.12&1&6&21\\
AUM (Unit: \$1 Million)&3,970&1840.83&21939.09&12.5&116.25&1500\\
Dry Power (Unit: \$1 Million)&3,674&124.18&691.72&0.02&19.98&220.29\\
Fraction of Female Founders&41,973&0.12&0.23&0&0&0.5\\
in Portfolio Companies&&&&&&\\
\hline
\end{tabular}}
\end{center}
\begin{tablenotes}
\item
\footnotesize \emph{Notes.} This table reports descriptive statistics for the investors who have participated in the lab-in-the-field experiment (i.e., Experiment A). In total, 69 different investors from 68 institutions, mostly venture funds, provided evaluations of 1216 randomly generated startup profiles. Panel A reports the sector distribution of investors. Each investor can indicate their interest in multiple industries. ``Others" includes HR tech, Property tech, infrastructure, etc. ``Industry Agnostic" means the investor does not have strong preferences based on sector. Panel B reports the stage distribution of investors, and each investor can invest in multiple stages. ``Seed Stage" includes pre-seed, angel investment, and late-seed stages. ``Series C or later stages" includes growth capital, series C, D, etc. Panel C reports the demographic information of these recruited investors. ``Female Investor" is an indicator variable which equals to one if the investor is female, and zero otherwise. ``Minority Investor" is an indicator variable which equals to one if the investor is Asian, Hispanic, or African American, and zero otherwise. Investors who prefer not to disclose their gender or race are not included in these variables. ``Senior Investor” is equal to one if the investor is in a C-level position, or is a director, partner, or vice president. It is zero if the investor is an analyst (intern) or associate investor. ``Cold Email Acceptance" is an indicator variable which equals one if the investor feels that sending cold call emails is acceptable as long as they are well-written, and zero if the investor feels that it depends. ``Prefer ESG" is an indicator variable which equals one if the investor prefers ESG-related startups, and zero otherwise. ``Direct Investment" is an indicator variable which equals to one if the investor can directly make the investment, and zero if their investment is through limited partners or other channels. Panel E provides the financial information of the 68 VC funds that these investors work for. However, we can only recover parts of their financial information from the Pitchbook Database.
\end{tablenotes}
\end{table}


\begin{center}
\begin{table}
 \caption{Randomization of Profile Components in Experiment A}
\label{irr_randomization}
\scalebox{0.85}{ 
\begin{tabular}{m{5.9cm} m{9.44cm} m{4.65cm} } 
\hline
Profile Component&Randomization Description&Analysis Variable\\
\hline
\emph{Startup Team Characteristics}&&\\
First and Last Names& Drawn from list of names that are indicative of& White Female (25\%)\\ 
&selected race and gender (See names in Online&Asian Female (25\%)\\
&Appendix Tables \ref{appendix_irr_full_name})&White Male (25\%)\\
&&Asian Male (25\%)\\
Number of Founders&The team can have 1 founder or 2 co-founders& 1 Founder (8/16)\\
Age&Founders' age is indicated by the graduation year&Age\\
&Young Founders: Old Founders=50\% : 50\%&\\
&Young Founders' graduation years: uniformly distributed between 2005 and 2019.&\\
&Old Founders' graduation years: uniformly distributed between 1980 and 2005.&\\
Educational Background&Drawn from top school list and common school list&Top School (8/16)\\
&(See school list in Online Appendix Table \ref{irr:schoolist})&\\
Entrepreneurial Experiences&The team can have serial founder(s) or only & Serial Founder (8/16)\\
&first-time founder(s)&\\
&&\\
\emph{Startup Project Characteristics}&&\\
Company Age&Founding dates are randomly drawn from & Company Age\\
&the following four years \{2016, 2017, 2018, 2019\}&\\
Comparative Advantages&Randomly drawn from a comparative advantage&\\
&list (See Online Appendix Section \ref{sec:appendix_irr}), the number  & 1 Advantage (4/16)\\
&of drawn advantages is between 1 to 4&2 Advantages (4/16)\\
&&3 Advantages (4/16)\\
&&4 Advantages (4/16)\\
Traction&Half randomly selected profiles generate no revenue &Positive traction (8/16)\\
&Half randomly selected profiles generate positive&\\
 & revenue. Previous monthly return: uniform &\\
& distribution [5K, 80K]; Growth rate: uniform  &\\
&distribution [5\%, 60\%]&\\
Company Category&Randomly assigned as either B2B or B2C& B2B (8/16)\\
Number of Employees&Randomly assigned with one of four categories &0-10 (8/16) \\
&&10-20 (8/16)\\
&&20-50 (8/16)\\
&&50+ (8/16)\\
Target Market&Randomly assigned as either domestic market or & Domestic (8/16)\\
&international market&\\
Mission&Randomly assigned with one of three categories&For profit (25\%)\\
&``For profit", ``For profit, consider IPO within 5&For profit, IPO (25\%)\\
&years", ``Besides financial gains, also care about the social and environmental impacts"&For profit, ESG (50\%)\\
Location&Randomly assigned as either U.S. or Outside & U.S. (70\%)\\
& the U.S.&\\
Number of Existing Investors&Randomly assigned as one of the four categories&Number of investors\\
&with equal probability \{0,1,2,3+\}&\\
\hline
\end{tabular}}
 \begin{tablenotes}
\item 
\footnotesize \emph{Notes.} This table provides the randomization of each startup profile's components and the corresponding analysis variables. Profile components are listed in the order that they appear on the hypothetical startup profiles. Weights of characteristics are shown as fractions when they are fixed across subjects and percentages when they represent a draw from a probability distribution. Variables in the right-hand column are randomized to test how investors respond to these analysis variables.
\end{tablenotes}
\end{table}
\end{center}

\begin{table}
 \caption{Implicit Gender and Racial Discrimination in Experiment A}
  \label{implicit bias gender and race}
\begin{center} 
\scalebox{0.9}{ 
\begin{tabular}{l c c c c c} 
\hline
Dependent Variable&Response Time&Q1&Q2&Q3&Q4\\
&(Unit: Second)&Profitability&Availability&Contact&Investment\\
&(1)&(2)&(3)&(4)&(5)\\
\hline
\emph{Panel A: Gender}&&&&&\\
&&&&&\\
Second Half of Study&-27.20***&2.42&2.27*&0.85&0.95**\\
&(2.48)&(1.66)&(1.62)&(2.02)&(0.31) \\
Female Founder&-1.34&1.56&1.27&0.89&0.56*\\
&(2.20)&(1.66)&(1.34)&(1.94)&(0.32)\\
Female Founder $\times$&&-4.26*&-1.67&-3.67&-1.03**\\
Second Half of Study&&(2.47)&(1.80)&(3.05)&(0.52)\\
&&&&&\\
Investor FE &Yes&Yes&Yes&Yes&Yes\\
Observations&1,216&	1,216&	1,184&	1,216&	1,176\\
R-squared&0.34&0.31&0.53&0.47&0.35\\
&&&&&\\
&&&&&\\
\emph{Panel B: Race}&&&&&\\
&&&&&\\
Second Half of Study&-27.20***&2.37&1.88&-0.28&	0.76**\\
&(2.48)&(1.92)&(1.37)&(2.24)&(0.29)\\
Asian Founder&0.54&2.26&-0.14&	0.41&	0.31\\
&(2.43)&(2.09)&(1.53)&(2.39)&(0.35)\\
&&&&&\\
Asian Founder $\times$&&-4.41&-0.93&-1.51&-0.69\\
Second Half of Study&&(3.00)&(1.88)&(3.16)&(0.53) \\
&&&&&\\
Investor FE &Yes&Yes&Yes&Yes&Yes\\
Observations&1,216&	1,216&	1,184&	1,216&	1,176\\
R-squared&0.34&	0.31&	0.53&	0.47&	0.35\\
\hline
\end{tabular}}
\end{center} 
\begin{tablenotes}
\footnotesize \item \emph{Notes.} This table reports regression results of how investors' response time and evaluation results respond to a startup founder's gender and race. Panel A tests investors' implicit discrimination based on a founder's gender. Panel B tests investors' implicit discrimination based on a founder's race. ``Female Founder" is equal to one if the startup founder has a female first name, and zero otherwise. ``Asian Founder" is equal to one if the startup founder has an Asian last name, and zero otherwise. ``Second Half of Study" is an indicator variable for startup profiles shown among the last half of resumes viewed by an investor. In column (1), the dependent variable is investors' response time, which is defined as the number of seconds before each page submission, winsorized at the 95th percentile (59.23 seconds on average). Columns (2)-(5) show the profitability ratings, availability ratings, contact interest ratings, and investment interest ratings, respectively. R-squared is indicated for each OLS regression. All the regressions add investor fixed effects. Standard errors in parentheses are clustered at the investor level. ***$p<0.01$, **$p <0.05$, *$p<0.1$\par
\end{tablenotes}
\end{table}


\begin{sidewaystable}
\begin{center} 
\caption{Quantile-Regression Estimates for Investors' Implicit Gender Discrimination}
\label{irr_investor_profile_gender_quantile}
\scalebox{0.9}{
\begin{tabular}{ m{5cm} c c c c c c c c c c}
&&&&&&&&&&\\
\toprule 
&10th&20th&30th&40th&50th&60th&70th&80th&90th&Mean\\
&[1]&[2]&[3]&[4]&[5]&[6]&[7]&[8]&[9]&[10]\\
\midrule
\emph{Panel A. Tech Sector}&&&&&&&&&&\\
&&&&&&&&&&\\
Female Founder&0.00&-4.00&-4.00&1.00&-7.00&-10.00**&-6.00&-5.00&-10.00**&-3.21\\
&(4.47)&(4.80)&(6.37)&(6.70)&(7.54)&(4.94)&(5.19)&(4.20)&(4.27)&(3.46)\\
Quantile of Dep. Var.&5&15&22&31&50&60&71&80&100&48.51\\
&&&&&&&&&&\\
Observations&392&392&392&392&392&392&392&392&392&392\\
&&&&&&&&&&\\
&&&&&&&&&&\\
\emph{Panel B. All Sectors}&&&&&&&&&&\\
&&&&&&&&&&\\
Female Founder&-2.00&-3.00&-1.00&-10.00*&-10.00**&-8.00**&-4.00&-7.00**&0.00&-2.30\\
&(3.64)&(4.47)&(4.52)&(5.43)&(4.34)&(4.06)&(4.80)&(3.21)&(0.90)&	(2.42)\\
Quantile of Dep. Var.&10&20&30&43&55&66&78&90&100&54.22\\
&&&&&&&&&&\\
Observations&608&608&608&608&608&608&608&608&608&608\\
\bottomrule
\end{tabular}}
\end{center}
\begin{tablenotes}
\item \footnotesize \emph{Notes.} This table reports the effects of a startup founder's gender on the quantiles and the mean of investors' contact interest ratings (i.e., $Q_3$) of the second half of the experiment. In each of Columns [1]–[9], the dependent variable is the $k$th percentile ($k\in{10,20,...,90}$) of the distribution of the startup's perceived attractiveness measured by investors' contact interest ratings (i.e., $Q_3$). In Column [10], the dependent variable is the average investor's contact interest ratings of the second half of the profiles. Panel A focuses on the evaluation results of investors working in the tech sector. Panel B uses the evaluation results of all recruited investors. Standard errors in parentheses are clustered at the investor level. $*p < 0.10, **p < 0.05, ***p < 0.01$\\
\end{tablenotes}
\end{sidewaystable}


\begin{sidewaystable}
\begin{center} 
\caption{Quantile-Regression Estimates for Investors' Implicit Racial Discrimination}
\label{irr_investor_profile_race_quantile}
\scalebox{0.9}{
\begin{tabular}{ m{5cm} c c c c c c c c c c}
&&&&&&&&&&\\
\toprule 
&10th&20th&30th&40th&50th&60th&70th&80th&90th&Mean\\
&[1]&[2]&[3]&[4]&[5]&[6]&[7]&[8]&[9]&[10]\\
\midrule
\emph{Panel A. Tech Sector}&&&&&&&&&&\\
&&&&&&&&&&\\
Asian Founder&0.00&2.00&-1.00&-6.00&-3.00&-2.00&-9.00*&-4.00&-9.00**&	-2.54\\
&(3.34)&(3.58)&(4.23)&(4.82)&(5.89)&(4.85)&(5.08)&(5.92)&(4.31)&(2.52)\\
Quantile of Dep. Var.&5&15&22&31&50&60&71&80&100&48.51\\
&&&&&&&&&&\\
Observations&392&392&392&392&392&392&392&392&392&392\\
&&&&&&&&&&\\
&&&&&&&&&&\\
\emph{Panel B. All Sectors}&&&&&&&&&&\\
&&&&&&&&&&\\
Asian Founder&1.00&1.00&-1.00&-5.00&-2.00&-9.00**&-6.00&-10.00***&0.00&	-2.00\\
&(2.67)&(3.26)&(3.54)&(4.42)&(4.46)&(3.79)&(4.40)&(3.62)&(0.94)&(1.86)\\
Quantile of Dep. Var.&10&20&30&43&55&66&78&90&100&54.22\\
&&&&&&&&&&\\
Observations&608&608&608&608&608&608&608&608&608&608\\
\bottomrule
\end{tabular}}
\end{center}
\begin{tablenotes}
\footnotesize \item \emph{Notes.} This table reports the effects of a startup founder's race on the quantiles and the mean of investors' contact interest ratings (i.e., $Q_3$) of the second half of the experiment. In each of Columns [1]–[9], the dependent variable is the $k$th percentile ($k\in{10,20,...,90}$) of the distribution of the startup's perceived attractiveness measured by investors' contact interest ratings (i.e., $Q_3$). In Column [10], the dependent variable is the average investor's contact interest ratings of the second half of the profiles. Panel A focuses on the evaluation results of investors working in the tech sector. Panel B uses the evaluation results of all recruited investors. Standard errors in parentheses are clustered at the investor level. $*p < 0.10, **p < 0.05, ***p < 0.01$\\
\end{tablenotes}
\end{sidewaystable}


\begin{sidewaystable}
\begin{center} 
\caption{Quantile-Regression Estimates for Investors' Implicit Discrimination (Investment Interest Ratings)}
\label{irr_investor_profile_Q4}
\scalebox{0.9}{
\begin{tabular}{ m{5cm} c c c c c c c c c c}
&&&&&&&&&&\\
\toprule 
&10th&20th&30th&40th&50th&60th&70th&80th&90th&Mean\\
&[1]&[2]&[3]&[4]&[5]&[6]&[7]&[8]&[9]&[10]\\
\midrule
\emph{Panel A. Gender}&&&&&&&&&&\\
&&&&&&&&&&\\
Female Founder&-1.00&-1.00&-1.00**&0.00&-1.00*&	-1.00*&	0.00&	0.00&	-1.00&-0.48\\
&(0.64)&(0.85)&(0.44)&(0.53)&(0.52)&(0.53)&(0.40)&(0.47)&(1.29)&(0.36)\\
Quantile of Dep. Var.&0&2&4&5&6&7&10&10&12&6.29\\
&&&&&&&&&&\\
Observations&591&591&591&591&591&591&591&591&591\\
&&&&&&&&&&\\
&&&&&&&&&&\\
\emph{Panel B. Race}&&&&&&&&&&\\
&&&&&&&&&&\\
Asian Founder&0.00&0.00&-1.00**&0.00&-1.00*&-1.00**&-1.00*&0.00&-2.00***&-0.47\\
&(0.55)&(0.77)&(0.44)&(0.57)&(0.54)&(0.47)&(0.59)&(0.46)&(0.60)&(0.33)\\
Quantile of Dep. Var.&0&2&4&5&6&7&10&10&12&6.29\\
&&&&&&&&&&\\
Observations&591&591&591&591&591&591&591&591&591\\
\bottomrule
\end{tabular}}
 \end{center}
\begin{tablenotes}
\item \footnotesize \emph{Notes.} This table reports the effects of a startup founder's gender and race on the quantiles and the mean of investors' investment interest ratings (i.e., $Q_4$) of the second half of the IRR experiment. In each of Columns [1]–[9], the dependent variable is the $k$th percentile ($k\in{10,20,...,90}$) of the distribution of the startup's perceived attractiveness measured by investors' investment interest ratings (i.e., $Q_4$). In Column [10], the dependent variable is the average investor's investment interest ratings of the second half of the profiles. Panel A focuses on implicit gender discrimination. Panel B focuses on implicit racial discrimination. Standard errors in parentheses are clustered at the investor level. $*p < 0.10, **p < 0.05, ***p < 0.01$\\
\end{tablenotes}
\end{sidewaystable}


\begin{table}
 \caption{Contact-Based Heterogeneous Effects of Founder's Gender, Race and Age in Experiment A}
\label{2nd_irr_hetgender}
\begin{center} 
\scalebox{0.85}{
\begin{tabular}{l l c c c c} 
\toprule
&&(1)&(2)&(3)&(4)\\
\multicolumn{2}{l}{Dependent Variable}&Profitability&Availability&Contact&Investment\\
\midrule
\multicolumn{2}{l}{\emph{Panel A: Gender}}&&&&\\
&&&&&\\
\multicolumn{2}{l}{\emph{$\beta^3<0$ (Not Contact Female Founders)}}&&&&\\
 &&&&&\\
Female Founder	&&-16.40***&	-2.85&	-21.81***&	-2.61***\\
&&(2.62)&	(1.79)&	(2.74)&	(0.47)\\
Ratios of Anti-Women&&0.42&0.43&0.42&0.41\\
&&&&&\\
\multicolumn{2}{l}{\emph{$\beta^3>0$ (Contact Female Founders)}}&&&&\\
&&&&&\\
Female Founder&& 7.93***&	1.54&	13.69***&	1.08**\\
&&(2.01)&(1.32)&(1.79)&(0.34)	\\
Ratios of Pro-Women&&0.58&0.57&0.58&0.59\\
&&&&&\\
&&&&&\\
\multicolumn{2}{l}{\emph{Panel B: Race}}&&&&\\
 &&&&&\\
\multicolumn{2}{l}{\emph{$\beta^3<0$ (Not Contact Asians)}}&&&&\\
 &&&&&\\
Asian Founder	&&-12.12***&	-1.43&	-17.60***&	-2.01***\\
&&(2.42)&	(1.83)&	(2.48)&	(0.46)\\
Ratios of Anti-Asian&&0.45&0.46&0.45&0.46\\
&&&&&\\
\multicolumn{2}{l}{\emph{$\beta^3>0$ (Contact Asians)}}&&&&\\
&&&&&\\
Asian Founder&& 6.34***&-0.78&	12.41***&	0.95***\\
&&(2.10)&(1.71)&(2.30)&(0.35)	\\
Ratios of Pro-Asian&&0.55&0.54&0.55&0.54\\
&&&&&\\
&&&&&\\
\multicolumn{2}{l}{\emph{Panel C: Age}}&&&&\\
&&&&&\\
\multicolumn{2}{l}{\emph{$\beta^3<0$ (Not Contact Older Founders)}}&&&&\\
 &&&&&\\
Older Founder	&&-13.17***&	-1.98&	-17.23***&	-2.03***\\
&&(2.54)&	(1.80)&	(2.60)&	(0.45)\\
Ratios of Anti-Older&&0.38&0.40&0.38&0.38\\
&&&&&\\
\multicolumn{2}{l}{\emph{$\beta^3>0$ (Contact Older Founders)}}&&&&\\
&&&&&\\
Older Founder&& 7.83***&2.06&	14.47***&	1.34***\\
&&(1.96)&(1.32)&(2.01)&(0.38)	\\
Ratios of Pro-Older&&0.62&0.60&0.62&0.62\\
&&&&\\
&&&&\\
Investor FE &&Yes&Yes&Yes&Yes\\
Observations &&608&592&608&591\\
\bottomrule
\end{tabular}}
\end{center} 
\end{table}

\clearpage
\begin{center}
\begin{table}
\begin{tablenotes}
\footnotesize \item \emph{Notes.} This table reports the contact decision-based heterogeneous effect of a startup founder's gender, race, and age by using profile evaluations in the second half of the study. Panels A, B, and C report the heterogeneous effect of investors who want to contact female, Asian, and older founders and those who want to contact male, white, and younger founders, respectively. ``Female Founder" is equal to one if the startup founder is female, and zero otherwise. Ratios of ``Anti-Women" is the number of profiles with $\beta^3 < 0$ divided by the number of profiles used. Ratios of ``Pro-Women" is the number of profiles with $\beta^3 > 0$ divided by the total number of profiles used. ``Asian Founder" is equal to one if the startup founder is Asian, and zero otherwise. Ratios of Anti-Asian is the number of profiles with $\beta^3 < 0$ divided by the total number of profiles used. Ratios of Pro-Asian is the number of profiles with $\beta^3 > 0$ divided by the number of profiles used. ``Older Founder" is equal to one if the startup founder graduated from college in 2005 or before, and zero otherwise. Ratios of ``Anti-Older" is the number of profiles with $\beta^3 < 0$ divided by the total number of profiles used. Ratios of ``Pro-Older" is the number of profiles with $\beta^3 > 0$ divided by the number of profiles used. All the regression results are estimated using the ``leave-one-out" estimator after adding investor fixed effects. Standard errors in parentheses are bootstrapped for the two-stage calculations. *** p$<$0.01, ** p$<$0.05, * p$<$0.1
\end{tablenotes}
\end{table}
\end{center}


\clearpage
\begingroup
\centering
\begin{table}
\caption{Discrimination in the Donation Game of Experiment A}
\label{irr_dictator}
\scalebox{0.85}{
\begin{tabular}{m{10cm} c c m{0.5cm} c c}
\hline
&\multicolumn{2}{c}{OLS}&&\multicolumn{2}{c}{Probit Model}\\
Dependent Variable&\multicolumn{2}{c}{Donated Amount (Unit:\$)}&&\multicolumn{2}{c}{1\{Donate All the \$15\}}\\
\cline{2-3} \cline{5-6}
&&&&&\\
&(1)&(2)&&(3)&(4)\\
\hline
\emph{Panel A: Gender}\\
&&&&&\\
Female Founder&-3.20*&-3.21*&&-0.77**&-0.77** \\
&(1.67)&(1.68)&&(0.37)&(0.37)\\
Female Founder $\times$ Female Investor&7.20&6.85&&1.45*&1.40* \\
& (4.49)&(4.68)&&(0.81)&(0.83) \\
Female Investor&-7.41**&-7.19**&&-1.33**&-1.30** \\
&(2.66)&(2.77)&&(0.52)&(0.53) \\
Control Profitability Evaluations&No&Yes&&No&Yes\\
&&&&&\\
Observations&70&70&&70&70 \\
R-squared&0.13&0.13&&0.09&0.09\\
&&&&&\\
\emph{Panel B: Race}\\
&&&&&\\
Asian Founder&1.04&1.02&&0.11&0.11 \\
&(1.87)&(1.86)&&(0.40)&(0.39)\\
Asian Founder $\times$ Asian Investor &1.05&1.13&&0.42&0.41 \\
&(3.47)&(3.51)&&(0.68)&(0.69) \\
Asian Investor&-4.71*&-4.70*&&-1.09**&-1.09**\\
&(2.42)&(2.45)&&(0.49)&(0.49)\\
Control Profitability Evaluations&No&Yes&&No&Yes\\
&&&&&\\
Observations&70&70&&70&70 \\
R-squared&0.09&0.09&&0.08&0.08\\
&&&&&\\
\emph{Panel C: White Male vs Non-white Male}\\
&&&&&\\
White Male Founder&-6.81**&-7.08**&&-1.34**&-1.42**  \\
&(2.28)&(2.30)&&(0.62)&(0.62)  \\
White Male Founder $\times$ White Male Investor&9.78**&10.29**&&2.26**&2.38**  \\
&(2.89)&(3.10)&&(0.86)&(0.89) \\
White Male Investor&1.56&1.42&&0.32&0.29  \\
&(1.94)&(1.97)&&(0.37)&(0.38)\\
Control Profitability Evaluations&No&Yes&&No&Yes\\
&&&&&\\
Observations&70&70&&70&70 \\
R-squared&0.21&0.21&&0.08&0.08\\
\hline
\end{tabular}}
\begin{tablenotes}\item \footnotesize \emph{Notes.} This table tests whether discrimination and homophily exist in the donation game of Experiment A. In Columns (1) and (2), the dependent variable is the donated amount measured in dollars. In Columns (3) and (4), the dependent variable is a dummy variable which equals one if the investor donates all the \$15, and zero otherwise. ``Female Founder", ``Asian Founder", and ``White Male Founder" are indicators which equal one if the displayed startup founders are female, Asian, and white male, respectively, and equal zero otherwise. ``Female Investor", ``Asian Investor", and ``White Male Investor" are indicators showing whether the venture capitalist is female, Asian, and white male, respectively. The ``Profitability Evaluations" are the coefficients $\beta_i$ of the regression $Q_{1ij}=\beta_0+\beta_i \text{Startup Characteristics}_{ij} +\epsilon_{ij}$ for each investor $i$. ``Startup Characteristics" is ``Female Founder" in Panel A, ``Asian Founder" in Panel B, and ``White Male Founder" in Panel C. One investor participates in this experiment twice for different funds, but results are similar when removing his responses. Robust standard errors are reported in parentheses. *** p$<$0.01, ** p$<$0.05, * p$<$0.1
\end{tablenotes}
\end{table}
\endgroup

\clearpage
\begin{table}
 \caption{Investor Responses to Randomized Emails in Experiment B}
\begin{center} 
\label{table_open}
\scalebox{0.85}{
\begin{tabular}{l c c c c c c} 

\multicolumn{2}{l}{\emph{Panel A: Summary Statistics of Email Responses}}&&&&&\\
\hline
&N&Mean&Median&S.D.&Min&Max\\
Open Rate& 3,720&12.03\%&0&0.33&0&1\\
Staying Time (Unit: s)&3,381&24.10&10.33&26.73&0.01&86.63\\
Click Rate&519&1.68\%&0&0.13&0&1\\
Email Replies&472&1.53\%&0&0.12&0&1\\
\hline
&&&&&&\\
&&&&&&\\
\multicolumn{2}{l}{\emph{Panel B: Email Opening Behaviors}}&&&&&\\
\hline
&&&&&&\\
&&\multicolumn{5}{c}{Dependent Variable: \textbf{$1$}(\emph{Opened})}\\
\cline{3-7}
&&(1)&(2)&(3)&(4)&(5)\\
&&Full Sample&Full Sample&Full Sample&``Pure Ivy"&Full Sample\\
\hline
Female Founder&&0.010***&&&&0.010***\\
&&(0.004)&&&&(0.004)\\
Asian Founder&&&0.007*&&&0.006\\
&&&(0.004)&&&(0.004)\\
Prestigious University&&&&0.007*&0.012**&0.007*\\
&&&&(0.004)&(0.005)&(0.004)\\
Project Advantage&&&&&&0.001\\
&&&&&&(0.004)\\
\multicolumn{2}{l}{Asian Founder $\times$ March Chinese Virus Period}&&-0.009&&&\\
\multicolumn{2}{c}{}&&(0.010)&&&\\
March Chinese Virus Period&&&-0.040**&&&\\
&&&(0.020)&&&\\
&&&&&&\\
Control&&Yes&Yes&Yes&Yes&Yes\\
Startup FE&&Yes&Yes&Yes&Yes&Yes\\
Observations&&30,909&30,909&30,909&16,578&30,909\\
Adjusted R-squared&&0.005&0.005&0.005&0.006&0.005\\
\hline
\end{tabular}}
 \end{center}
\end{table}

\clearpage
\begin{table}
\begin{center} 
\scalebox{0.85}{
\begin{tabular}{m{4cm} c c c} 
\emph{Continued}&&&\\
&&&\\
\multicolumn{2}{l}{\emph{Panel C: Staying Time}}&&\\
\hline
&&&\\
&\multicolumn{3}{c}{Dependent Variable: Staying Time (Unit: s)}\\
\cline{2-4}
&(1)&(2)&(3)\\
&Full Sample&Full Sample&Opened Emails\\
\hline
Female Founder&0.12&	0.25*&	0.31\\
&(0.19)&	(0.13)&	(0.88)\\
Asian Founder&0.28&	0.38**&	2.49*\\
&(0.13)&	(0.19)	&(1.34)\\
Prestigious University&0.11	&0.11&	-0.12\\
&(0.13)&	(0.13)&	(0.88)\\
Project Advantage&	0.12&	0.12&	0.92\\
&(0.13)&	(0.13)	&(0.88)\\
US Investor&-0.24&	-0.24&	1.30\\
&(0.20)	&(0.20)&	(1.20)\\
March&1.23&	1.68*&	6.11\\
&(0.93)&	(0.93)&	(4.98)\\

&&&\\
Female Founder $\times$&0.24&&\\
March&(0.26)&&\\
&&&\\
\multicolumn{2}{l}{Asian Founder $\times$}&-0.66**&	-5.48***\\
March&&(0.26)& (1.74)\\
&&&\\
Control&Yes&Yes&Yes\\
Pitch Email FE&Yes&Yes&Yes\\
Observations &30,909&30,909&	3,720\\
Adjusted R-squared&	0.002&	0.003&	0.002\\
\hline
\end{tabular}}
 \end{center}
 \begin{tablenotes}
\item \footnotesize
\emph{Notes.} This table summarizes investors' email responses in Experiment B and reports regression results of investors' email opening behaviors in response to randomized pitch emails. Panel A summarizes important investors' information acquisition behaviors in the pitch email setting. Panels B and C report regression results of how startup characteristics affect investors' email opening behaviors and investors' staying time on each pitch email, respectively. In Panel B, the dependent variable is a dummy variable, which is one when an investor opens the pitch email, and zero otherwise. ``Female Founder" equals one if the first name of the email sender is a female name, and zero otherwise. Similarly, ``Asian Founder" equals one if the last name of the email sender is an East Asian name, and zero otherwise. ``Prestigious University" equals one if the startup founder graduated from a prestigious university, and zero otherwise. ``Project Advantage"  is an indicator variable which is one when the email's subject line includes the corresponding comparative advantages. ``March Chinese Virus Period” is equal to one when the email was sent between 03/18/2020-03/24/2020 when President Trump used the wording ``Chinese Virus.", and zero otherwise. ``March" equals one if the pitch email was sent in 03/2022, and zero if the pitch email was sent in 04/2022. Control variables include ``US Investor" and ``Female Investor", which are indicator variables for being a US investor and being a female investor, respectively. Columns (1), (2), (3), and (5) use all the observations collected in the correspondence test. In Column (4), results are reported for the sub-sample where the startup team graduated from purely Ivy League colleges, Stanford and MIT. ``Pure Ivy" indicates cases like ``Team from Columbia University" while ``Mixed Ivy" indicates cases like ``Team from Columbia University and Juilliard Music School". For some startups in the music or medical industry, I combined an Ivy League college with a university well-known for having a top program in that specific area for the treatment group. In Panel C, the dependent variable is the time spent on each pitch email measured in seconds. In Columns (1) and (2), I include unopened emails and replace their email staying time with 0 seconds. Considering the potential truncation issue, I also report the sub-sample of opened emails in Column (3). Standard errors in parentheses are clustered at the investor level. *** p$<$0.01, ** p$<$0.05, * p$<$0.1\par
\end{tablenotes}
\end{table}

\clearpage
\begin{table}
 \caption{Interaction Effects Based on Email Opening Rate in Experiment B}
\label{table_interaction}
\begin{center}
\scalebox{0.85}{ 
\begin{tabular}{l c c c c c c } 
&&&&&&\\
\toprule
&&&&&&\\
&&\multicolumn{5}{c}{Dependent Variable: \textbf{$1$}(\emph{Opened}) }\\
\cline{3-7}
\multicolumn{2}{l}{} &(1)&(2)&(3)&(4)&(5)\\
\multicolumn{2}{l}{} &Full&``Mixed Ivy"&``Pure Ivy"&Full&``Pure Ivy"\\
\multicolumn{2}{l}{} &Sample&&&Sample&After 03/24\\

\midrule
\multicolumn{2}{l}{Female Founder}&0.006&	0.002&	0.009&&\\		
\multicolumn{2}{l}{}&(0.005)&	(0.008)&	(0.007)	&&\\
\multicolumn{2}{l}{Asian Founder}&&&&		0.009*&	0.026***\\
\multicolumn{2}{l}{}&&&&		(0.005)& (0.008)\\
\multicolumn{2}{l}{Prestigious University}&0.003&	-0.010&	0.013*&	0.010**&	0.030***\\
	&&(0.005)&	(0.008)&	(0.007)&	(0.005)&	(0.008)\\

\multicolumn{2}{l}{Prestigious University $\times$ Female Founder}&0.008&0.020*&-0.002&&\\
\multicolumn{2}{l}{}&(0.007)&(0.011)&(0.010)&&\\

\multicolumn{2}{l}{Prestigious University$\times$ Asian Founder}&&&&		-0.007&-0.032***\\
\multicolumn{2}{l}{}&&&&(0.007)&(0.011)\\
&&&&&&\\
\multicolumn{2}{l}{Control}&Yes&Yes&Yes&Yes&Yes\\
\multicolumn{2}{l}{Pitch Email FE}&Yes&Yes&Yes&Yes&Yes\\
\multicolumn{2}{l}{Observations}&	30,909&	14,331&16,578&	30,909&	13,006\\
\multicolumn{2}{l}{R-squared}&	0.005&	0.004&0.006&	0.005&	0.007\\
\bottomrule
\end{tabular}}
\end{center} 
\begin{tablenotes}
\item \footnotesize
\emph{Notes.} This table reports regression results of interaction effects between founders' educational backgrounds and founders' gender or race using investors' email opening rate as the outcome variable. The dependent variable is a dummy variable, which is one if an investor opens the pitch email, and zero otherwise. ``Female Founder" equals one if the first name of the email sender is a female name, and zero otherwise. Similarly, ``Asian Founder" equals one if the last name of the email sender is an East Asian name, and zero otherwise. ``Prestigious University" is an indicator variable for adding an Ivy League educational background in the email's subject line. Control variables include ``US Investor" and ``Female Investor", which are indicator variables for being a U.S. investor and being a female investor, respectively. To identify underlying dominant mechanisms, I include the interaction term of ``Prestigious University" and ``Female Founder" in Columns (1)-(3) and also the interaction term  of ``Prestigious University" and ``Asian Founder" in Columns (4)-(5). Column (1) reports the regression results using all the observations in the correspondence test. In column (2), results are reported for the ``Mixed Ivy" sub-sample, which indicates cases like ``Team from Columbia University and Juilliard Music School." For some startups in the music or medical industry, I combined an Ivy League college with a university well-known for having a top program in that specific area for the treatment group. In Column (3), results are reported for the ``Pure Ivy" sub-sample, which indicates cases like ``Team from Columbia University". The universities that founders have graduated from in the ``Pure Ivy” cases are the Ivy League colleges, Stanford, and MIT. In Column (5), results are reported for the sub-sample where pitch emails are sent after 03/24 and emails belong to the ``Pure Ivy" cases in order to increase the statistical power. Note that President Trump stopped using ``Chinese Virus" after 03/23/2020. $R^2$ is the adjusted $R^2$ for OLS regressions. Standard errors in parentheses are clustered at the investor level. *** p$<$0.01, ** p$<$0.05, * p$<$0.1.
\end{tablenotes}
\end{table}

\clearpage
\begin{table}
 \caption{Heterogeneous Effect Based on Investors' Investment Philosophies in Experiment B}
\label{table_ESG}
\begin{center}
\scalebox{0.85}{
\begin{tabular}{l c c c c c c } 
\toprule
&&&&&&\\
& \multicolumn{6}{c}{Dependent Variable: \textbf{$1$}(\emph{Opened}) }\\
\cline{2-7}
&(1)&(2)&(3)&(4)&(5)&(6)\\
&Full Sample&Impact Fund&Common Fund&Full Sample&Impact Funds&Common Fund\\
\midrule
Female Founder&0.012***&0.103***&0.011***&&&\\
&(0.004)&(0.033)&(0.004)&&&\\
Asian Founder&&&&0.004&0.008&0.004\\
&&&&(0.004)&(0.032)&(0.004)\\
Impact Fund&-0.048**&&&-0.010&&\\
&(0.020)&&&(0.024)&&\\
Female Founder $\times$&0.083**&&&&&\\
Impact Fund&(0.033)&&&&&\\
&&&&&&\\
Asian Founder $\times$&&&&0.011&&\\
Impact Fund&&&&(0.032)&&\\
&&&&&&\\
Control&Yes&Yes&Yes&Yes&Yes&Yes\\
Startup FE&Yes&Yes&Yes&Yes&Yes&Yes\\
Observations&23,649&368&23,281&23,649&	368&23,281\\
R-squared&0.006&0.075&0.006&0.006&0.049&0.005\\
\bottomrule
\end{tabular}}
\end{center}
\begin{tablenotes}
\item \footnotesize
\emph{Notes.} This table reports the heterogeneous effect of investors' email opening behaviors in response to randomized pitch emails based on their investment philosophies in Experiment B. I only include investors whose investment philosophies are available on Pitchbook, which accounts for 76.5\% of all the observations. The dependent variable is a dummy variable, which is one when an investor opened the pitch email, and zero otherwise. ``Female Founder" equals one if the first name of the email sender is a female name, and zero otherwise. Similarly, ``Asian Founder" equals one if the last name of the email sender is an East Asian name, and zero otherwise. ``Impact Fund" equals one if the investor works in a fund with ESG-related investment preferences based on Pitchbook Data, and zero otherwise. ESG-related preferences include supporting minority founders, caring about the environmental and social impact, etc. Control variables include ``US Investor" and ``Female Investor", which are indicator variables for being a U.S. investor and being a female investor, respectively. Columns (1) and (4) report the regression results for all observations with available investment philosophies. Columns (2) and (5) report the regression results for investors working in impact funds. Columns (3) and (6) report the regression results for investors working in profit-driven VC funds which do not pursue impact investing strategies. $R^2$ is the adjusted $R^2$ for OLS regressions. Standard errors in parentheses are clustered at the investor level. *** p$<$0.01, ** p$<$0.05, * p$<$0.1
\end{tablenotes}
\end{table}

\clearpage

\begin{figure}
    \begin{center}
    \includegraphics[scale=1.30]{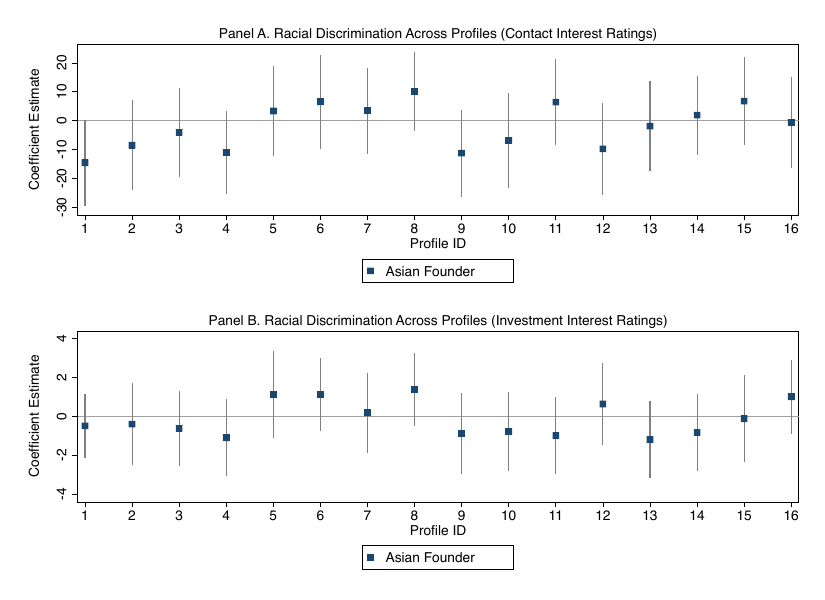}
    \caption{Race Discrimination Across Profiles}
    \label{fig:race_dynamic}
 \caption*{\footnotesize \emph{Notes.} This figure illustrates the impact of startup founders' race on investors' evaluations across profiles. The x-axis is the profile ID, which indicates the order of profiles displayed to each investor. The y-axis reports the coefficient of ``Asian Founder" and the 95\% confidence interval using the following regression $Q_{ij}^{k}=\alpha_j+\beta_j \text{Asian Founder}_{ij}+\epsilon_{ij}$ for each profile ID $j$. Robust standard errors are used in these cross-sectional regressions. Panel A focuses on investors' contact interest ratings (i.e., $Q_3$). Panel B focuses on investors' investment interest ratings (i.e., $Q_4$).}    
 \end{center}
\end{figure}

\clearpage
\begin{center}
\section*{Online Appendix}
\end{center}
\appendix
\onehalfspacing

\setcounter{table}{0}
\setcounter{figure}{0}
\renewcommand{\thetable}{\Alph{section}\arabic{table}}
\renewcommand{\thefigure}{\Alph{section}\arabic{figure}}

\section{Data Construction Process}\label{sec:data_construction}
To construct an individual-level global venture capitalist database, the paper mainly uses the following commercial databases as well as manually collected data: Pitchbook, CBInsight, ExactData, SDC New Issue Database and Rocketreach. The Pitchbook database contains extremely comprehensive information about venture capital and angel investors' demographic information and contact information. The following types of investors are selected from Pitchbook: Angel Group, Angel Individual Investor, Corporate Venture Capital, Family Office, and Venture Capital. CBInsight is used to complement the Pitchbook. ExactData is a U.S. professional data company that collects investors' information from online websites and various VC industry events. Rocketreach is one of the largest platforms that provides contact information for company employees. Given the company name list in the SDC New Issue Database, it is feasible to extract the employees' contact information on Rocketreach.\footnote{Zdatabase is provided by Zero2IPO Research Center and is currently one of the most comprehensive, accurate and timely databases covering the VC and PE industry in China. Considering that the research was implemented in English, I only included investors from Hong Kong and excluded investors from the Mainland.}\par

\vspace{2mm}
All key variables used in the analysis, including gender, location and industry, are manually verified through multiple social platforms including LinkedIn, company websites, personal websites and online news if such information is not available on Pitchbook. For VC funds' ESG preferences, the paper treats not-for-profit VC funds as impact funds and for-profit VC funds as common funds. An alternative way is to classify VC funds based on selected ESG-representative keywords in their company description \cite{barber2020impact}. Based on this keyword method, impact funds can account for roughly 7\% of the total observations. However, the heterogeneous effect analysis based on these two classification methods is similar.\par

\begin{figure}[h]
\centering
\includegraphics[scale=0.3]{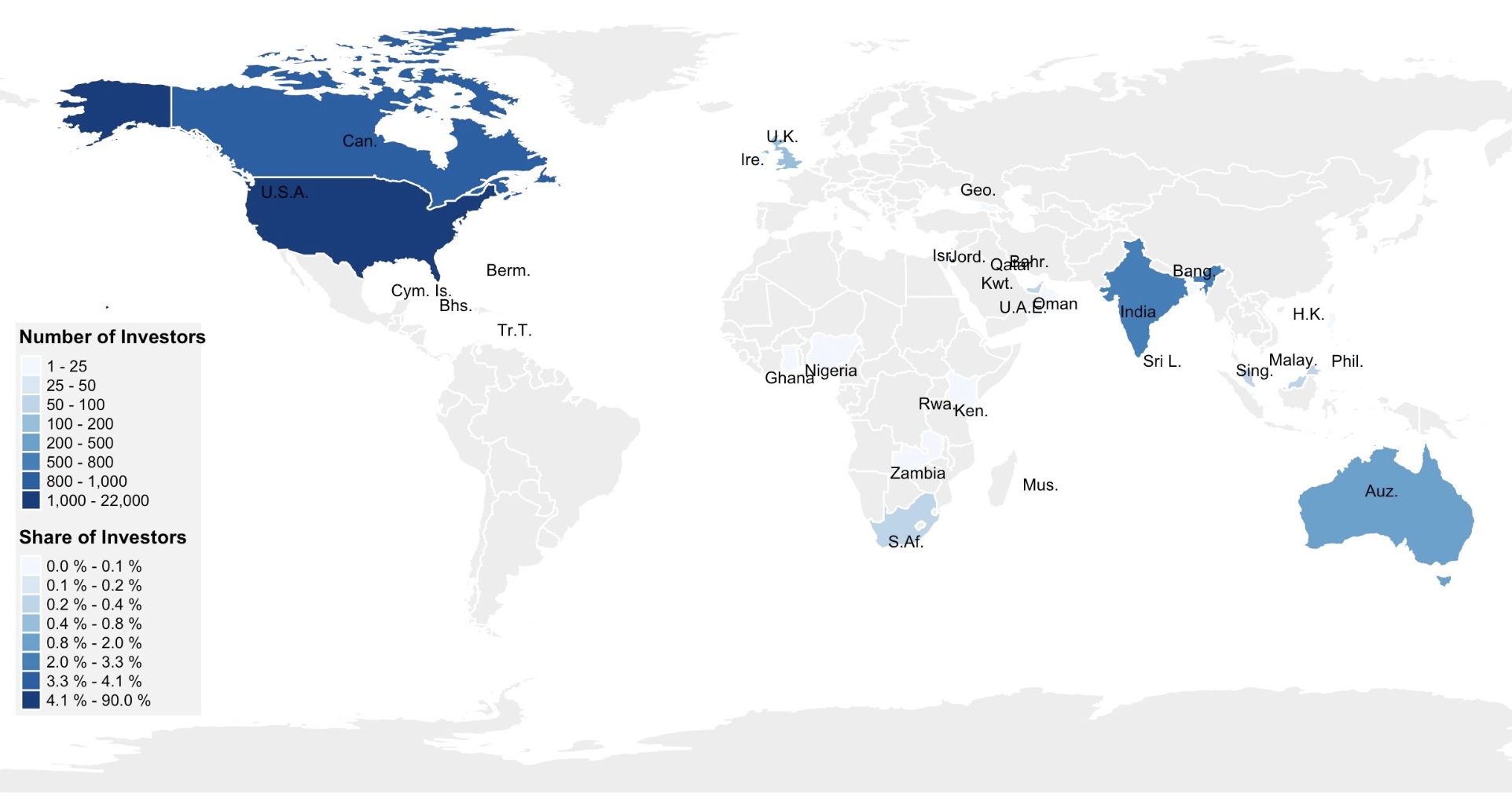}
\caption{Geographical Distribution of Global Investors}
\label{map_global}
\end{figure}

\begin{figure}[h]
\centering
\includegraphics[scale=0.25]{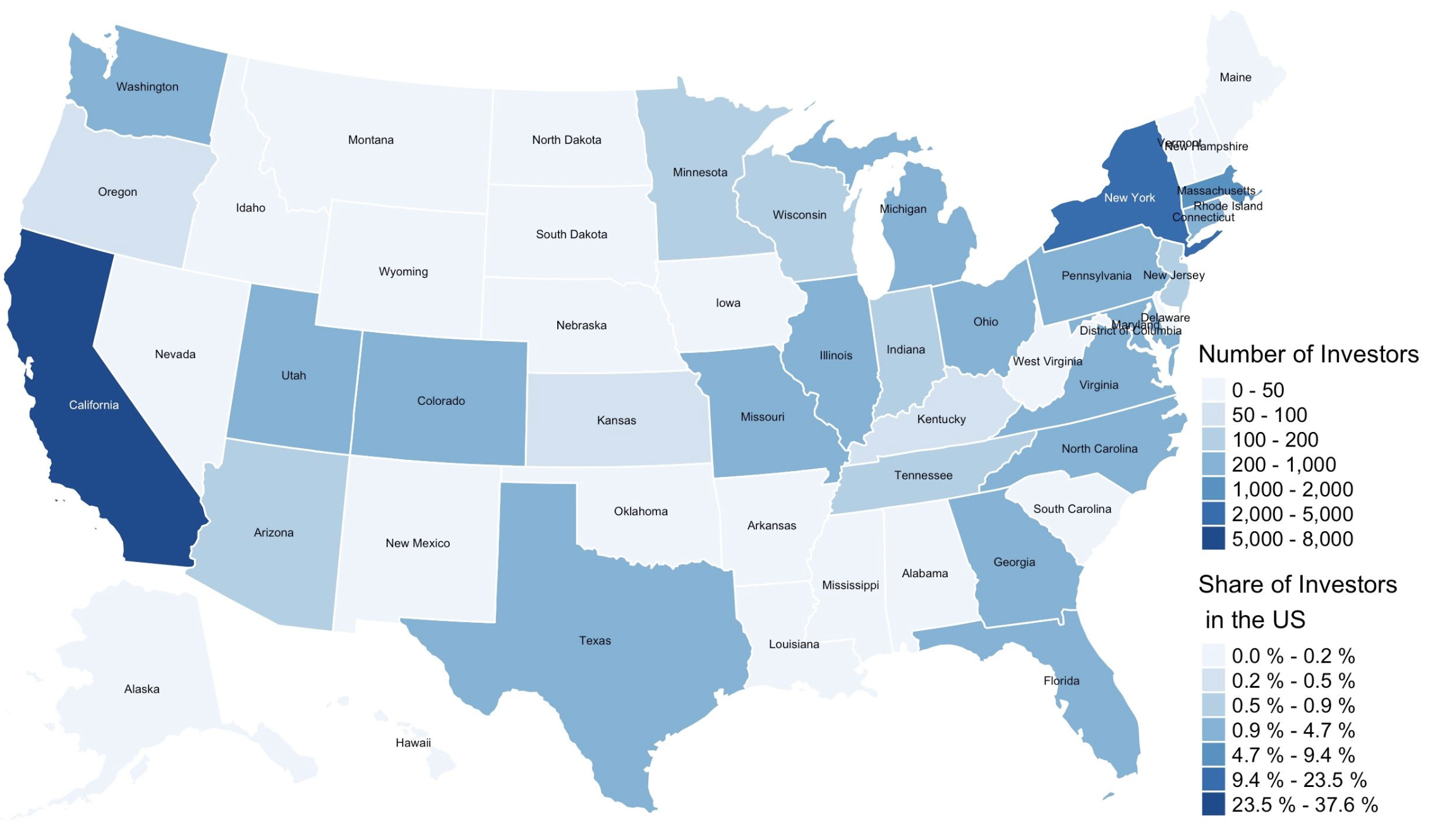}
\caption{Geographical Distribution of U.S. Investors}
\label{map_US}
\end{figure}

\setcounter{table}{0}
\setcounter{figure}{0}
\renewcommand{\thetable}{\Alph{section}\arabic{table}}
\renewcommand{\thefigure}{\Alph{section}\arabic{figure}}

\section{Experiment A}\label{sec:appendix_irr}

For each startup profile, a subset of comparative advantages is randomly drawn from the following: ``trade secrets/patents registered", ``celebrity endorsement", ``exclusive partnerships", ``accumulated many pilot consumers", ``adoption of the latest technology", ``pricing advantage", ``great product design", ``the 1st mover", ``lower cost", and ``economies of scale".\par

\begin{table}[h]
 \caption{Full Names Populating Startup Profiles in Experiment A}
\begin{center} 
\label{appendix_irr_full_name}
\scalebox{0.85}{
\begin{tabular}{l l l l } 
\hline
Asian Female&White Female&Asian Male&White Male\\
\hline
Cynthia	Huynh&Amber Morris&Evan Liu&Patrick	Kelly\\
Jennifer Tang&Erica Carpenter&Alan Wu&Stephen	Bennett\\
Amanda Cheung&Anna Hoffman&Bryan Liang&Steven	Martin\\
Christina Chang&Amanada	Gray&William Chung&Jeremy	White\\
Linda Chung&Tiffany Roberts&Nicholas Wang&Jason	Adams\\
Brittany Yi&Lisa Taylor&Charles	Luu&Donald	Schultz\\
Megan Ho&Karen Carroll&Zachary Ho&Jack Wright\\
Emily Xu&Danielle Collins&Marcus Yoon&Victor	Becker\\
Jacqueline Lin&Megan Bennett&George Thao&Michael	Hughes\\
Kayla Wang&Brenda Cox&Vincent Huynh&Keith Meyer\\
Cassandra Kwon&Kathleen Phillips&Luke Yang&Anthony	Roberts\\
Julie Chan&Amber Sullivan&Justin Dinh&Justin	Cooper\\
Monica Luong&Madeline Walsh&Matt Hoang&Benjamin Hill\\
Amber Hoang&Abigail	Kelly&Jacob	Xu&Mark	Myers\\
Sara Truong&Alicia Cook&Donald Choi&Phillip	Baker\\
Katrina Tsai&Amanda Jensen&Dennis Lin&Vincent	Peterson\\
Abigail Zhao&Angela	Larson&Victor Kwon&Dennis	Reed\\
Vanessa	Choi&Hayley	Thompson&Jason Pham&Frank	Phillips\\
\hline
\end{tabular}}
 \end{center} 
\end{table}

\begin{table}
\captionsetup{labelformat=empty}
\begin{center} 
\scalebox{0.85}{
\begin{tabular}{l l l l } 
\emph{Continued}&&&\\
\hline
Asian Female&White Female&Asian Male&White Male\\
\hline
Patricia Li&Christine Campbell&Eric	Duong&Shane	Taylor\\
Lisa Zhou&Caroline Parker&Stephen Hsu&William	Welch\\
Caroline Lu&Kristy Baker&Kevin Jiang&Bryan	Ward\\
Melissa	Hwang&Tina Reed&Jeffrey	Chen&Ian	Russell\\
Mary Pham&Sara Burke&Erik Luong&Brian Wilson\\
Amy	Hu&Victoria	Snyder&Philip Zhao&Seth	Schwartz\\
Jenna Nguyen&Molly Weaver&Jeremy	Yu&Jared Walsh\\
Margaret Liang&Melissa Stone&Seth Truong&Zachary Parker\\
 Danielle Liu&Melanie	Wilson&Ian Zhou&John Carpenter\\
Megan Dinh&Rachael Ward&Matthew	Chang&Jeffery Cook\\
Melanie	Yang&Elizabeth Miller&Scott	Lu&Nathan Nelson\\
Amanda Thao&Mary Hill&Sean Hwang&Matthew	Rogers\\
Sarah Yu&Amy Moore&Patrick Hu&George Barker\\
Nichole Liu&Vanessa Smith&Mark Chan&Sean Beck\\
Christine Cho&Teresa Anderson&Jack Zhu&David Hall\\
Victoria Xiong&Catherine Schultz&Timothy Cheng&Andrew Miller\\
Teresa Wong&Heather Martin&Benjamin Nguyen&Peter Keller\\
Kara Yoon&Kathryn Myers&Steven Tang&Luke Jensen\\
Kathleen Cheng&Katie Meyer&Travis Wong&Kevin Hansen\\
Angela Wu&Valerie Price&David Zheng&Dustin Sullivan\\
Catherine Zheng&Melinda Evans&Paul Ngo&Philip	Morris\\
Hayley Huang&Sandra	Wright&Anthony Yi&Evan Moore\\
Karen Ngo&Christina	Russell&Shane Huang&Paul	Burke\\
Elizabeth Duong&Kayla Allen&Robert Zhang&Matt	Price\\
Laura Luu&Jacqueline Schmidt&Kenneth Tsai&Marcus Collins\\
Rebecca	Hsu&Jennifer Welch&Richard Xiong&Richard Thompson\\
Melinda	Zhang&Michelle Nelson&Brian	Cho&Thomas Snyder\\
Katherine Le&Sarah Fisher&Joel Le&Christopher Larson\\
Tara Jiang&Brittany	Rogers&Michael Li&Travis Gray\\
Alicia Zhu&Grace Keller&Trevor Cheung&Charles Hoffman\\
Molly Huynh&Julie Beck&Adam Liu&Joel Stone\\
Samantha Tang&Monica Cooper&Peter Wu&Joseph Allen\\
\hline
\end{tabular}}

\begin{tablenotes}
\item \footnotesize \emph{Notes.} This table provides name lists of hypothetical startup founders used in the matching tool of Experiment A. 50 names were selected to be highly indicative of each combination of race and gender. Considering that white and Asian startup founders account for most of the highly innovative startups, there are only four combinations listed here: Asian Female, White Female, Asian Male, White Male. A name drawn from these lists is displayed at the beginning of each startup profile and in the evaluation questions. Given that Asian and white Americans have very similar naming patterns as documented by \cite{fryer2004causes}, the paper chose their first names from the same name pool. After a list of potential full name candidates is generated, the researcher further removed names owned by famous startup founders or CEOs. That's why there are slight differences between first names for Asian founders and first names for white founders. Names were selected uniformly and without replacement within the chosen column of the table.\\
 \end{tablenotes}
 \end{center} 
\end{table}

\clearpage

\clearpage
\begin{table}
 \caption{ Educational Background in Experiment A}
\begin{center} 
\label{irr:schoolist}
\scalebox{0.85}{ 
\begin{tabular}{l l l} 
\hline
School Category & Universities& Percentage \\
\hline
Top School Example&Brown University&50\%\\
&Columbia University &\\
&Cornell University&\\
&Dartmouth College&\\
&Harvard University& \\
&Princeton University& \\
&University of Pennsylvania&\\
&Yale University&\\
&California Institute of Technology&\\
&MIT&\\
&Northwestern University&\\
&Stanford University&\\
&University of Chicago&\\
&\\
Common School Example &Thomas Jefferson University& 50\% \\
&University of Arkansas&\\
&Hofstra University&\\
&University of Mississippi&\\
&Virginia Commonwealth University&\\
&Adelphi University&\\
&University of Maryland-Baltimore County&\\
&University of Rhode Island&\\
&St.John's University&\\
&University of Detroit Mercy&\\
&University of Idaho&\\
&Biola University&\\
&Chatham University&\\
&Bellarmine University&\\
&Bethel University&\\
&Loyola University New Orleans &\\
&Robert Morris University &\\
&Regis University&\\
&Widener University&\\
&Laurentian University&\\
& Auburn University &\\
& Rochester Institute of Technology&\\
&University of Tulsa &\\
&DePaul University&\\
\hline
\end{tabular}}

\begin{tablenotes}
\item \footnotesize \emph{Notes.} This table provides the school list used to generate the educational background of each hypothetical startup founder. The percentage of top school and common school is 50\% vs. 50\% to increase the statistical power. Representative top schools include the Ivy League schools as well as California Institute of Technology, MIT, Northwestern University, and Stanford University. Since the incubators that we collaborate with have more connections with Columbia University and Stanford University, we give more weight to these universities. Common Schools are universities those U.S. News ranks are below 150th in 2020. A Canadian school is added since one of the incubators is from Canada.\\ 
\end{tablenotes}
 \end{center} 
\end{table}


\begin{table}
 \caption{Aggregate-level Gender, Racial and Age Discrimination in Experiment A}
 \label{Group-level IRR ATE}
\begin{center} 
\scalebox{0.85}{ 
\begin{tabular}{l c c c c} 
\hline
Dependent Variable&Q1&Q2&Q3&Q4\\
&Profitability&Availability&Contact&Investment\\
&(1)&(2)&(3)&(4)\\
\hline
\emph{Panel A: Gender}&&&&\\
&&&&\\
Female Founder&-0.56&0.46&-0.94&0.04\\
&(1.26)&(0.92)&(1.46)&(0.19)\\
&&&&\\
Investor FE &Yes&Yes&Yes&Yes\\
Control Mean &44.30&63.84&55.01&6.02\\
Profile Observations&1,216&1,184&1,216&	1,176\\
R-squared&0.31&	0.53&0.47&0.34\\
&&&&\\
\emph{Panel B: Race}&&&&\\
&&&&\\
Asian Founder&0.05&	-0.61&	-0.34&	-0.04\\
&(1.19)&(1.04)&(1.40)&(0.20)\\
&&&&\\
Investor FE &Yes&Yes&Yes&Yes\\
Control Mean &44.31&65.51&55.51&6.12\\
Profile Observations&1,216&	1,184&	1,216&	1,176\\
R-squared&	0.31&	0.53&	0.47&	0.34\\
&&&&\\
\emph{Panel C: Age}&&&&\\
&&&&\\
Older Founder&-0.47&0.14&0.29&0.14\\ 
&(1.22)&(0.97)&(1.46)&(0.25)\\
&&&&\\
Investor FE &Yes&Yes&Yes&Yes\\
Control Mean &44.74&64.47&54.50&5.99\\
Observations&1,216&1,184&1,216&	1,176\\
R-squared&0.31&	0.53&	0.47&	0.34\\
\hline
\end{tabular}}
\end{center} 
\begin{tablenotes}
 \item \footnotesize \emph{Notes.} This table describes aggregate-level evaluation results using total profile evaluations, including profiles in the first half of the study and profiles in the second half of the study. Some investors skip the evaluation questions of availability or investment. Panel A tests gender discrimination. ``Female Founder" is equal to one if the startup founder has a female first name, and zero otherwise. Panel B tests racial discrimination. ``Asian Founder" is equal to one if the startup founder has an Asian last name, and zero otherwise. Panel C tests agism. ``Older Founder" is equal to one if the startup founder graduates before 2005, and zero otherwise. In Column (1), the dependent variable is the profitability evaluation, which indicates the percentile rank of each startup profile compared with an investor's previous invested startups in terms of its potential financial returns. In Column (2), the dependent variable is the availability evaluation, which indicates how likely the investors feels the startup team will accept his/her investment rather than other investors' offers. In Column (3), the dependent variable is the contact interest rating, which describes the probability that the investor contacts this startup. In Column (4), the dependent variable is the relative investment interest ranging from 1 to 20, which describes the relative investment amount compared with the investor's general investment amount. The unit is one-tenth of the relative investment compared with investors' average investment amount. For example, if the investor's average invested amount for each deal is \$1M and Q4 is equal to 5, then it means the investor only wants to invest \$1M $\times$ 5 $\times$ 10\% = \$500,000 in this startup. If Q4 is 20, then the investment amount is  \$1M $\times$ 20 $\times$ 10\% = \$2M. All the regressions add investor fixed effects. Standard errors in parentheses are clustered at the investor level. ***$p<0.01$, **$p <0.05$, *$p<0.1$ \par

\end{tablenotes}
\end{table}


\begin{sidewaystable}
\begin{center} 
\caption{Quantile-Regression Estimates for Investors' Implicit Discrimination (Remove Extreme Preferences)}
\label{irr_investor_profile_outlier_quantile}
\scalebox{0.9}{
\begin{tabular}{ m{7cm} c c c c c c c c c c}
&&&&&&&&&&\\
\toprule 
&10th&20th&30th&40th&50th&60th&70th&80th&90th&Mean\\
&[1]&[2]&[3]&[4]&[5]&[6]&[7]&[8]&[9]&[10]\\
\midrule
\multicolumn{4}{l}{\emph{Panel A. Implicit Gender Discrimination}}&&&&&&&\\
&&&&&&&&&&\\
Female Founder&-2.00&-3.00&-1.00&-8.00&-10.00**&-7.00*&-4.00&-6.00*&0.00&-1.65\\
&(3.60)&(4.38)&(4.43)&(5.57)&(4.29)&(3.94)&(4.86)&(3.28)&(0.87)&(2.34)\\
Quantile of Dep. Var.&10&20&30&43&55&66&78&89.5&100&54.29\\
&&&&&&&&&&\\
Observations&600&600&600&600&600&600&600&600&600&600\\
&&&&&&&&&&\\
&&&&&&&&&&\\
\multicolumn{4}{l}{\emph{Panel A. Implicit Racial Discrimination}}&&&&&&&\\
&&&&&&&&&&\\
Asian Founder&0.00&1.00&-1.00&-7.00&-4.00&-9.00**&	-7.00&-10.00***&0.00&-2.62\\
&(2.49)&(3.11)&(3.40)&(4.64)&(4.42)&(3.70)&(4.43)&(3.62)&(0.92)&(1.76)\\
Var.&10&20&30&43&55&66&78&89.5&100&54.29\\
&&&&&&&&&&\\
Observations&600&600&600&600&600&600&600&600&600&600\\
\bottomrule
\end{tabular}}
\end{center}
\begin{tablenotes}
\item \footnotesize \emph{Notes.} This table reports the effects of a startup founder's gender and race on the quantiles and the mean of investors' contact interest ratings (i.e., $Q_3$) in the second half of the experiment. The sample removes evaluations from the investor who has extreme preferences as shown in Figure \ref{fig:distribution_outlier_attitudes}. In each of Columns [1]–[9], the dependent variable is the $k$th percentile ($k\in{10,20,...,90}$) of the distribution of the startup's perceived attractiveness measured by investors' contact interest ratings (i.e., $Q_3$). In Column [10], the dependent variable is the average investor's contact interest ratings of the profiles displayed in the second half of the study. ``Female Founder" is equal to one if the startup founder is female, and zero otherwise. ``Asian Founder" is equal to one if the startup founder is Asian, and zero otherwise. Panel A focuses on implicit gender discrimination. Panel B focuses on implicit racial discrimination. Standard errors in parentheses are clustered at the investor level. $*p < 0.10, **p < 0.05, ***p < 0.01$\\
\end{tablenotes}
\end{sidewaystable}


\begin{sidewaystable}
\begin{center} 
\caption{Quantile-Regression Estimates for Investors' Implicit Age Discrimination}
\label{irr_investor_profile_age_quantile}
\scalebox{0.9}{
\begin{tabular}{ m{5cm} c c c c c c c c c c}
&&&&&&&&&&\\
\toprule 
&10th&20th&30th&40th&50th&60th&70th&80th&90th&Mean\\
&[1]&[2]&[3]&[4]&[5]&[6]&[7]&[8]&[9]&[10]\\
\midrule
\emph{Panel A. Tech Sector}&&&&&&&&&&\\
&&&&&&&&&&\\
Older Founder&0.00&10.00*&4.00&10.00**&5.00	&-1.00&	-5.00&	-5.00&	-8.00*&	2.61\\
&(4.27)&(5.55)&(4.05)&(4.81)&(5.51)&(6.27)&(6.20)&(4.15)&(4.70)&(2.57)\\
Quantile of Dep. Var.&5&15&22&31&50&60&71&80&100&48.51\\
&&&&&&&&&&\\
Observations&392&392&392&392&392&392&392&392&392&392\\
&&&&&&&&&&\\
&&&&&&&&&&\\
\emph{Panel B. Full Sample}&&&&&&&&&&\\
&&&&&&&&&&\\
Older Founder&2.00&3.00&2.00&7.00*&1.00&1.00&-4.00&-6.00*&0.00&1.41\\
&(3.15)&(2.93)&(3.33)&(4.14)&(4.33)&(4.16)&(5.43)&(3.34)&(0.75)&(1.98)\\
Quantile of Dep. Var.&10&20&30&43&55&66&78&90&100&54.22\\
&&&&&&&&&&\\
Observations&608&608&608&608&608&608&608&608&608&608\\
\bottomrule
\end{tabular}}
\end{center}
\begin{tablenotes}
\item \footnotesize \emph{Notes.} This table reports the effects of a startup founder's age on the quantiles and the mean of investors' contact interest ratings (i.e., $Q_3$) in the second half of the experiment. In each of Columns [1]–[9], the dependent variable is the $k$th percentile ($k\in{10,20,...,90}$) of the distribution of the startup's perceived attractiveness measured by investors' contact interest ratings (i.e., $Q_3$). In Column [10], the dependent variable is the average investor's contact interest ratings of the profiles displayed in the second half of the study. ``Older Founder" is equal to one if the startup founder graduated from college in 2005 or before, and zero otherwise. Panel A focuses on the evaluation results of investors working in the tech sector. Panel B uses the evaluation results of all recruited investors. Standard errors in parentheses are clustered at the investor level. $*p < 0.10, **p < 0.05, ***p < 0.01$\\
\end{tablenotes}
\end{sidewaystable}

\clearpage
\begin{table}
 \caption{Comparison of Results from Different Incentive Structures in Experiment A}
\begin{center} 
\label{irr:incentive_comparison}
\scalebox{0.85}{
\begin{tabular}{l c c c c} 
&&&&\\
\hline
&(1)&(2)&(3)&(4)\\
&Profitability&Availability&Contact&Investment\\
\hline
\emph{Panel A: Gender}&&&&\\
Female Founder&-0.60&	0.57&	-0.34&	0.02\\
&(1.35)&(1.02)&(1.56)& (0.20)\\  
Female Founder $\times$ Matching&0.30&-0.77&	-4.18&0.13\\
&(3.74)&(2.25)&(4.18)&(0.57) \\
Matching&-13.80***&	48.13***&15.28***&-0.87***\\
&(1.87)&(1.12)&(2.09)&(0.28)\\
&&&&\\
Investor FEs &Yes&Yes&Yes&Yes\\
Observations&1,216&	1,184&	1,216&	1,176\\
R-squared&0.31&	0.53&	0.47&	0.34\\
&&&&\\
\emph{Panel B: Race}&&&&\\
Asian Founder&-0.28&	-0.61&	-0.75&	-0.18\\
&(1.34)&(1.15)&(1.49)&(0.21)\\
Asian Founder $\times$ Matching&2.26&	0.03&	2.81&	0.93\\
&(2.48)& (2.62)& (4.22)&(0.55)\\
Matching&-14.78***&	47.73***&	11.78***&-1.26***\\
&(1.24)&(1.31)&(2.11)&(0.27)\\
Investor FEs &Yes&Yes&Yes&Yes\\
Observations&1,216&	1,184&	1,216&	1,176\\
R-squared&0.31&	0.53&	0.47&	0.34\\
&&&&\\
\emph{Panel C: Age}&&&&\\
Older Founder&-0.77&-0.29&0.20&0.09\\
&(1.31)&(1.08)&(1.60)&(0.29)\\
Older Founder $\times$ Matching& 2.11&2.86&0.65&0.33 \\
&(3.52)&(2.29)&(3.84)&(0.49)\\
Matching&-14.79***&46.16***&12.81*** &-0.99***\\
&(1.95)&(1.26)&(2.12)&(0.26)\\
Investor FEs &Yes&Yes&Yes&Yes\\
Observations&1,216&	1,184&	1,216&	1,176\\
R-squared&0.31&0.53&0.47&0.34\\
\hline
\end{tabular}}
 \end{center} 
 \begin{tablenotes}
\item \footnotesize \emph{Notes.} This table compares the evaluation results of investors who are recruited by the following two incentive structures: ``matching incentive $+$ monetary incentive" and the ``matching incentive" only. ``Matching" equals to one if only the matching incentive is provided in the recruitment process, and zero otherwise. Panels A, B, and C show the comparison of evaluation results related to gender, racial, and age discrimination, respectively. The dependent variable is profitability evaluation (i.e., $Q_1$) in Column (1), availability evaluation (i.e., $Q_2$) in Column (2), contact interest rating (i.e., $Q_3$) in Column (3), and investment interest rating (i.e., $Q_4$) in Column (4), separately. All regression specifications add investor fixed effects. Standard errors in parentheses are clustered at the investor level.  *** p$<$0.01, ** p$<$0.05, * p$<$0.1.\smallskip
\end{tablenotes}
\end{table}

\clearpage
\begin{table}
 \caption{Correlations between Startup Characteristics ($1^{st}$ Half) \& Investors' Evaluations ($2^{nd}$ Half)}
\begin{center} 
\label{irr:robustnesscheck}
\scalebox{0.85}{
\begin{tabular}{l c c c c} 
&&&&\\
\hline
&(1)&(2)&(3)&(4)\\
&Profitability&Availability&Contact&Investment\\
\hline
Fraction of Female Founders&20.52&27.04&36.58&1.17\\
In the first Half&(15.80)&(15.76)&(21.59)&(2.59) \\
&&&&\\

Fraction of Asian Founders&0.34&2.66&-29.04&-0.38 \\
In the first Half&(10.48)&(14.19)&(16.21)&(1.90)\\
&&&&\\

Fraction of Older Founders&-8.11&24.44&21.34& 2.07 \\
In the first Half&(12.14)&(16.76)&(28.32)&(2.37)\\
&&&&\\
Observations&70&69&70&69\\
\hline
\end{tabular}}
 \end{center} 
 \begin{tablenotes}
\item \footnotesize \emph{Notes.} This table tests whether investors' evaluation ratings of the minority founders decrease in the second half of the study when they evaluate more minority founders' profiles in the first half of the study. The dependent variable is profitability evaluation (i.e., $Q_1$) in the second half of Experiment A in Column (1), availability evaluation (i.e., $Q_2$) in the second half of Experiment A in Column (2), contact interest rating (i.e., $Q_3$) in the second half of Experiment A in Column (3), and investment interest rating (i.e., $Q_4$) in the second half of Experiment A in Column (4), separately. ``Fraction of Female Founders In the first Half", ``Fraction of Asian Founders In the first Half" and ``Fraction of Older Founders In the first Half" stand for the fraction of female founders, Asian founders and older founders in the first half profiles, respectively. These cross-sectional regressions use robust standard errors. Significance has been adjusted for multiple hypothesis testing. One investor participated in the experiment twice. However, results are still robust after removing his responses. \smallskip
\end{tablenotes}
\end{table}


\begin{sidewaystable}
\begin{center} 
\caption{Quantile-Regression Estimates for Investors' Profitability Evaluations (Gender, $Q_1$)}
\label{irr_investor_profile_gender_quantile_Q1}
\scalebox{0.9}{ 
\begin{tabular}{ m{5cm} c c c c c c c c c c}
&&&&&&&&&&\\
\toprule 
&10th&20th&30th&40th&50th&60th&70th&80th&90th&Mean\\
&[1]&[2]&[3]&[4]&[5]&[6]&[7]&[8]&[9]&[10]\\
\midrule
\emph{Panel A. Tech Sector}&&&&&&&&&&\\
&&&&&&&&&&\\
Female Founder&-1.00&-2.00&-3.00&-4.00&	-3.00&	-1.00&	0.00&	0.00&-5.00&-3.88\\
&(2.06)&(3.33)&(3.81)&(4.46)&(4.40)&(4.42)&(3.94)&(3.45)&(3.50)&(2.70)\\
Quantile of Dep. Var.&10&20&30&36&42&50&60&70&79&44.29\\
&&&&&&&&&&\\
Observations&392&392&392&392&392&392&392&392&392&392\\
&&&&&&&&&&\\
&&&&&&&&&&\\
\emph{Panel B. All Sectors}&&&&&&&&&&\\
&&&&&&&&&&\\
Female Founder&-4.00*&-1.00&-4.00&-7.00**&-4.00&0.00&0.00&0.00&-1.00&-2.80\\
&(2.36)&(3.51)&(2.90)&(2.90)&(2.98)&(2.62)&(2.63)&(3.66)&(3.16)&(1.99)\\
Quantile of Dep. Var.&10&20&30&36&41.5&50&60&70&80&44.44\\
&&&&&&&&&&\\
Observations&608&608&608&608&608&608&608&608&608&608\\
\bottomrule
\end{tabular}}
 \end{center}
\begin{tablenotes}
\item \footnotesize \emph{Notes.} This table reports the effects of a startup founder's gender on the quantiles and the mean of investors' profitability evaluations (i.e., $Q_1$) in the second half of the study. In each of Columns [1]–[9], the dependent variable is the $k$th percentile ($k\in{10,20,...,90}$) of the distribution of the startup's perceived profitability measured by $Q_1$. In Column [10], the dependent variable is the average investor's profitability evaluations. Panel A focuses on the profile evaluations provided by tech investors. Panel B uses all recruited investors' evaluations. Standard errors in parentheses are clustered at the investor level. $*p < 0.10, **p < 0.05, ***p < 0.01$\\
\end{tablenotes}
\end{sidewaystable}


\begin{sidewaystable}
\begin{center} 
\caption{Quantile-Regression Estimates for Investors' Evaluations (Race, $Q_1$)}
\label{irr_investor_profile_race_quantile_Q1}
\scalebox{0.85}{
\begin{tabular}{ m{5cm} c c c c c c c c c c}
&&&&&&&&&&\\
\toprule 
&10th&20th&30th&40th&50th&60th&70th&80th&90th&Mean\\
&[1]&[2]&[3]&[4]&[5]&[6]&[7]&[8]&[9]&[10]\\
\midrule
\emph{Panel A. Tech Sector}&&&&&&&&&&\\
&&&&&&&&&&\\
Asian Founder&0.00&-3.00&3.00&-2.00&0.00&-1.00&0.00&-6.00&	-8.00**&-2.85\\
&(2.52)&(3.03)&(2.53)&(3.64)&(4.92)&(4.18)&(3.44)&(3.79)&(3.39)&(2.40)\\
Quantile of Dep. Var.&10&20&30&36&42&50&60&70&79&44.29\\
&&&&&&&&&&\\
Observations&392&392&392&392&392&392&392&392&392&392\\
&&&&&&&&&&\\
&&&&&&&&&&\\
\emph{Panel B. All Sectors}&&&&&&&&&&\\
&&&&&&&&&&\\
Asian Founder&2.00&0.00&0.00&-3.00&-4.00&-1.00&0.00&-7.00**&-8.00**&-2.44\\
&(2.55)&(2.76)&(2.59)&(3.22)&(3.34)&(2.72)&(2.43)&(3.25)&(3.21)&(1.83)\\
Quantile of Dep. Var.&10&20&30&36&41.5&50&60&70&80&44.44\\
&&&&&&&&&&\\
Observations&608&608&608&608&608&608&608&608&608&608\\
\bottomrule
\end{tabular}}
 \end{center}
\begin{tablenotes}
\item \footnotesize \emph{Notes.} This table reports the effects of a startup founder's race on the quantiles and the mean of investors' profitability evaluations (i.e., $Q_1$) in the second half of the study. In each of Columns [1]–[9], the dependent variable is the $k$th percentile ($k\in{10,20,...,90}$) of the distribution of the startup's perceived profitability measured by $Q_1$. In Column [10], the dependent variable is the average investor's profitability evaluations. Panel A focuses on the profile evaluations provided by tech investors. Panel B uses all recruited investors' evaluations. Standard errors in parentheses are clustered at the investor level. $*p < 0.10, **p < 0.05, ***p < 0.01$\\
\end{tablenotes}
\end{sidewaystable}


\begin{sidewaystable}
\begin{center} 
\caption{Quantile-Regression Estimates for Investors' Evaluations (Educational Background, $Q_1$)}
\label{valid_signal}
\scalebox{0.9}{
\begin{tabular}{m{5cm} c c c c c c c c c c}
&&&&&&&&&&\\
\toprule 
&10th&20th&30th&40th&50th&60th&70th&80th&90th&Mean\\
&[1]&[2]&[3]&[4]&[5]&[6]&[7]&[8]&[9]&[10]\\
\midrule
\emph{Panel A. Women-led Startups}&&&&&&&&&&\\
&&&&&&&&&&\\
Prestigious University&5.00*&9.00***&9.00***&10.00***&13.00***&	10.00***&10.00***&3.00&5.00**&6.89***\\
&(2.70)&(2.63)&(2.64)&(2.16)&(2.69)&(3.26)&(3.45)&(3.44)&(2.39)&(1.82)\\
Quantile of Dep. Var.&10&20&30&35&40&50&60&70&79&44.28\\
&&&&&&&&&&\\
Observations&605&605&605&605&605&605&605&605&605&605\\
&&&&&&&&&&\\
&&&&&&&&&&\\
\emph{Panel B. Asian-led Startups}&&&&&&&&&&\\
&&&&&&&&&&\\
Prestigious University&8.00***&8.00***&9.00***&10.00***&10.00***&	12.00***&6.00**&7.00***&8.00***&6.83***\\
&(2.71)&(2.75)&(2.66)&(2.24)&(2.24)&(2.22)&(2.50)&(2.45)&(2.19)&(1.62)\\
Quantile of Dep. Var.&10&20.5&30&41&54.5&65&75.5&90&100&53.81\\
&&&&&&&&&&\\
Observations&570&570&570&570&570&570&570&570&570&570\\
\bottomrule
\end{tabular}}
 \end{center}
\begin{tablenotes}
\item \footnotesize \emph{Notes.} This table reports the effects of a startup founder's educational background on the quantiles and the mean of investors' profitability evaluations (i.e., $Q_1$). The sample includes all investors' evaluations in both the first half and the second half of the study. In each of Columns [1]–[9], the dependent variable is the $k$th percentile ($k\in{10,20,...,90}$) of the distribution of the startup's perceived profitability measured by $Q_1$. In Column [10], the dependent variable is the average investor's profitability evaluations. Panel A focuses on the profile evaluations of women-led startups. Panel B focuses on the profile evaluations of Asian-led startups. ``Prestigious University" is an indicator variable that equals one if the startup founder graduated from a prestigious university, and zero otherwise. Standard errors in parentheses are clustered at the investor level. $*p < 0.10, **p < 0.05, ***p < 0.01$\\
\end{tablenotes}
\end{sidewaystable}

\clearpage
\begin{table}
 \caption{Compare the Performance of Women-led and Men-led Startups}
  \label{compare_performance_gender}
\begin{center} 
\scalebox{0.85}{
\begin{tabular}{l c c m{0.5cm} c c m{0.5cm} c c } 
\hline
&&&&&&&&\\
&\multicolumn{2}{c}{Raised New Funding}&&\multicolumn{2}{c}{Out of Business}&&\multicolumn{2}{c}{IPO/Acquisition}\\
\cline{2-3} \cline{5-6} \cline{8-9}\\
&(1)&(2)&&(3)&(4)&&(5)&(6)\\
\hline
&&&&&&&&\\
\multicolumn{4}{l}{\textit{Panel A. Global Startups}}&&&&\\
&&&&&&&&\\
All Female Founders&-0.049***&-0.009&&0.009**&0.003&&-0.005***&-0.002\\
&(0.008)&(0.008)&&(0.004)&(0.004)&&(0.001)&(0.001)\\
Mixed Gender Founders&0.003&0.017***&&-0.005*&-0.006**&&-0.002&-0.001\\
&(0.005)&(0.005)&&(0.003)&(0.003)&&(0.001)&(0.001)\\
Observations&44,215&44,215&&44,215&44,215&&44,215&44,215\\
R-squared&0.06&0.17&&0.06&0.07&&0.05&0.06\\
Control&No&Yes&&No&Yes&&No&Yes\\
Location FE&Yes&Yes&&Yes&Yes&&Yes&Yes\\
Industry FE&Yes&Yes&&Yes&Yes&&Yes&Yes\\
Stage FE&Yes&Yes&&Yes&Yes&&Yes&Yes\\
&&&&&&&&\\
\multicolumn{4}{l}{\textit{Panel B. U.S. Startups}}&&&&\\
&&&&&&&&\\
All Female Founders&-0.031***&0.010&&0.012*&0.004&&-0.008***&-0.003\\
&(0.012)&(0.011)&&(0.007)&(0.007)&&(0.002)&(0.003)\\
Mixed Gender Founders&0.002&0.024***&&-0.005&-0.008**&&-0.003&-0.001\\
&(0.008)&(0.007)&&(0.004)&(0.004)&&(0.003)&(0.003)\\
Observations&17,852&17,852&&17,852&17,852&&17,852&17,852\\
R-squared&0.05&0.19&&0.05&0.07&&0.04&0.06\\
Control&No&Yes&&No&Yes&&No&Yes\\
Location FE&Yes&Yes&&Yes&Yes&&Yes&Yes\\
Industry FE&Yes&Yes&&Yes&Yes&&Yes&Yes\\
Stage FE&Yes&Yes&&Yes&Yes&&Yes&Yes\\

&&&&&&&&\\
\multicolumn{4}{l}{\textit{Panel C. IT Industry}}&&&&\\
&&&&&&&&\\
All Female Founders&-0.056***&-0.015&&0.031***&0.024***&&-0.001&0.001\\
&(0.015)&(0.015)&&(0.009)&(0.009)&&(0.004)&(0.004)\\
Mixed Gender Founders&-0.006&0.015**&&-0.006&-0.008&&-0.002&-0.002\\
&(0.008)&(0.008)&&(0.005)&(0.005)&&(0.002)&(0.002)\\
Observations&18,539&18,539&&18,539&18,539&&18,539&18,539\\
R-squared&0.06&0.19&&0.08&0.09&&0.05&0.06\\
Control&No&Yes&&No&Yes&&No&Yes\\
Location FE&Yes&Yes&&Yes&Yes&&Yes&Yes\\
Stage FE&Yes&Yes&&Yes&Yes&&Yes&Yes\\
\hline
\end{tabular}}
\end{center} 
\end{table}

\clearpage
\begin{table}
\begin{center} 
\scalebox{0.85}{
\begin{tabular}{l c c m{0.5cm} c c m{0.5cm} c c } 
\hline
&&&&&&&&\\
&\multicolumn{2}{c}{Raised New Funding}&&\multicolumn{2}{c}{Out of Business}&&\multicolumn{2}{c}{IPO/Acquisition}\\
\cline{2-3} \cline{5-6} \cline{8-9}\\
&(1)&(2)&&(3)&(4)&&(5)&(6)\\
\hline
&&&&&&&&\\
\multicolumn{4}{l}{\textit{Panel D. Early Stage}}&&&&\\
&&&&&&&&\\
All Female Founders&-0.045***&-0.007&&0.011**&0.003&&-0.003**&-0.002\\
&(0.009)&(0.008)&&(0.005)&(0.005)&&(0.001)&(0.001)\\
Mixed Gender Founders&0.010*&0.023***&&-0.006*&-0.006*&&0.000&0.000\\
&(0.006)&(0.005)&&(0.003)&(0.003)&&(0.001)&(0.002)\\
Observations&31,962&31,962&&31,962&31,962&&31,962&31,962\\
R-squared&0.06&0.19&&0.06&0.08&&0.05&0.05\\
Control&No&Yes&&No&Yes&&No&Yes\\
Location FE&Yes&Yes&&Yes&Yes&&Yes&Yes\\
Industry FE&Yes&Yes&&Yes&Yes&&Yes&Yes\\
\hline
\end{tabular}}
\end{center} 
\begin{tablenotes}
\item \footnotesize \emph{Notes.} This table tests whether women-led ventures underperform men-led ventures during the 1 year period after the experiment (i.e., 2020/07/31-2021/07/31). The sample contains all the startups which have received funding between 2017/01/01 and 2020/07/31 and whose founders' gender information is observable in the Pitchbook data. Panel A examines the performance of global startups. Panel B focuses on the performance of only U.S. startups, defined as startups whose headquarters are located in the U.S. Panel C zooms into the IT-related startups. Panel D discusses startups whose latest financing round is still in the early stage or seed stage. In columns (1) and (2) of each panel, the dependent variable is equal to one if the startup either successfully raised new funding from the venture capital industry or the deal is in progress during 2020/07/31-2021/07/31. In columns (3) and (4), the dependent variable is equal to one if the startup's business status is ``Out of Business" in 2021/10. Ideally I should use the business status in 2021/07/31, however, this information is not available. ``Out of Business" is defined as either ``File Bankruptcy" or ``Out of Business" in Pitchbook. Results are still robust when including cases where the startup's website does not function anymore, such as reporting a 404 error. In columns (5) and (6), the dependent variable is equal to one if the startup filed an IPO or was acquired between 2020/07/31 and 2021/07/31. Columns (2), (4), and (6) include the following control variables which describe the last updated startup characteristics before 2020/07/31: number of deals, founding years, log (1+raised amount of the latest deal). Robust standard errors clustered at the headquarter location level are reported in parentheses. Significance: *$p < 0.10$, **$p < 0.05$, ***$p < 0.01$. \par
\end{tablenotes}
\end{table}

\clearpage
\begin{table}
\caption{Rule Out Attention Discrimination In the Profile Selection Process}
\label{table_attention_irr}
\begin{center}
\scalebox{0.85}{
\begin{tabular}{l c c c} 
\toprule
Dependent Variable&Response Time&Response Time&Response Time\\
&(1)&(2)&(3)\\
\midrule
Second Half of Study&-25.65***&-28.11***&-26.53***\\
&(3.35)&(3.23)&(3.63)\\
Female Founder&0.20&&\\
&(3.74)&&\\
Female Founder $\times$ Second Half of Study&-3.10&&\\
&(5.08)&&\\
Asian Founder&&-0.42&\\
&&(3.99)&\\
Asian Founder $\times$ Second Half of Study&&1.93&\\
&&(5.19)&\\
Older Founder&&&3.11\\
&&&(3.96)\\
Older Founder $\times$Second Half of Study&&&-1.37\\
&&&(5.16)\\
&&&\\
Investor FE &Yes&Yes&Yes\\
Observations&1216&1216&1216\\
R-squared&0.34&0.34&0.34\\
\bottomrule
\end{tabular}}
\end{center}
 \begin{tablenotes}
\item\footnotesize \emph{Notes.} This table tests whether investors' response time decreases for minorities in the second half of the study (i.e., ``attention discrimination"). The dependent variable is investors' response time, which is defined as the number of seconds before each page submission, winsorized at the 95th percentile (59.23 seconds on average). ``Female Founder" is equal to one if the startup founder has a female first name, and zero otherwise. ``Asian Founder" is equal to one if the startup founder has an Asian last name, and zero otherwise. ``Older Founder" is equal to one if the startup founder graduated from college in 2005 or before, and zero otherwise. ``Second Half of Study" is an indicator variable for startup profiles shown among the last half of resumes viewed by an investor. Standard errors in parentheses are clustered at the investor level. ***$p<0.01$, **$p <0.05$, *$p<0.1$ \par
\end{tablenotes}
\end{table}


\clearpage
\begin{figure}
    \centering
    \includegraphics[scale=0.63]{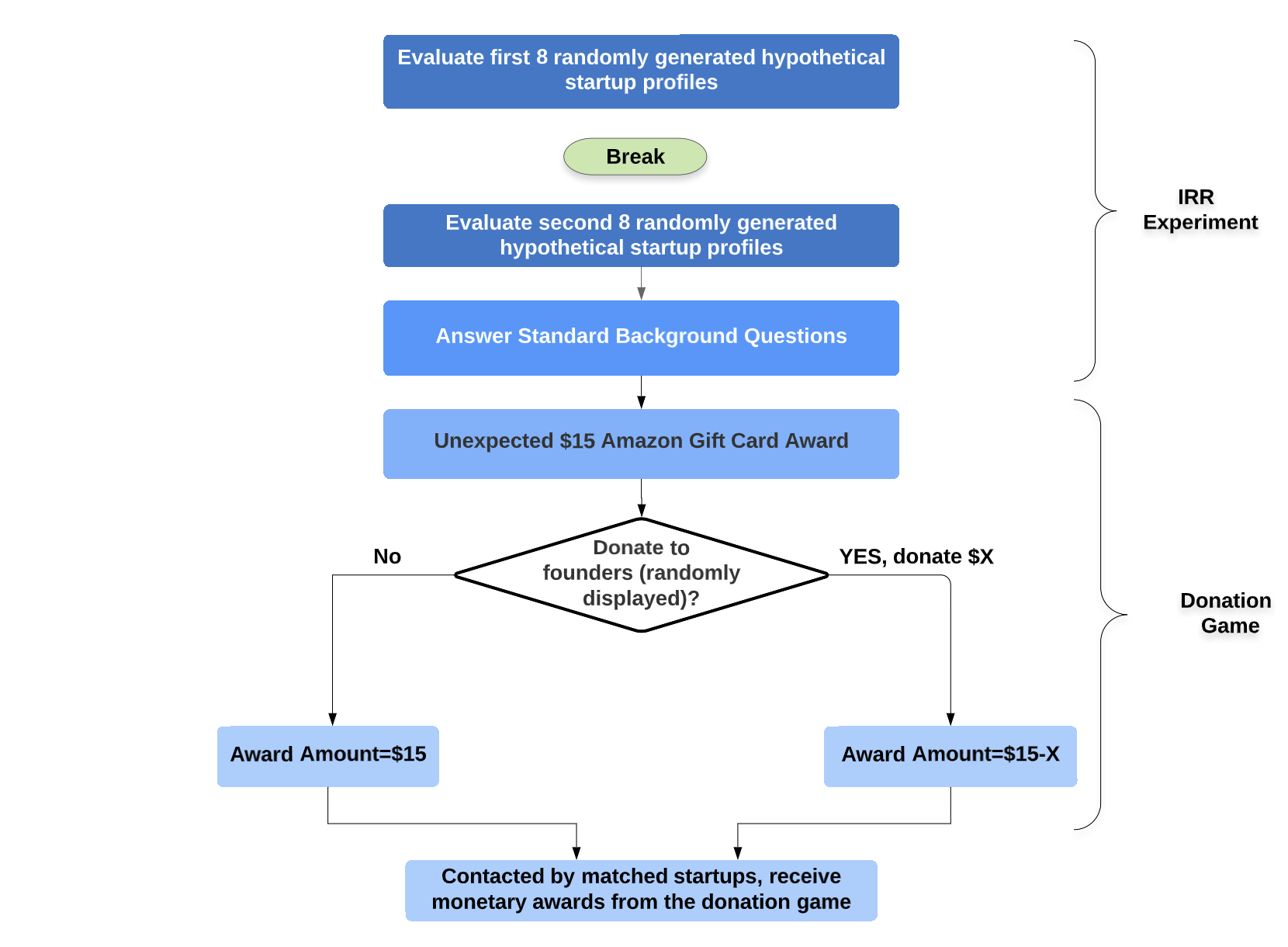}
    \caption{Experimental Flowchart for Experiment A}
    \label{fig:irr_flow_chart}
\end{figure}

\clearpage
\begin{figure}
    \centering
    \includegraphics[scale=0.66]{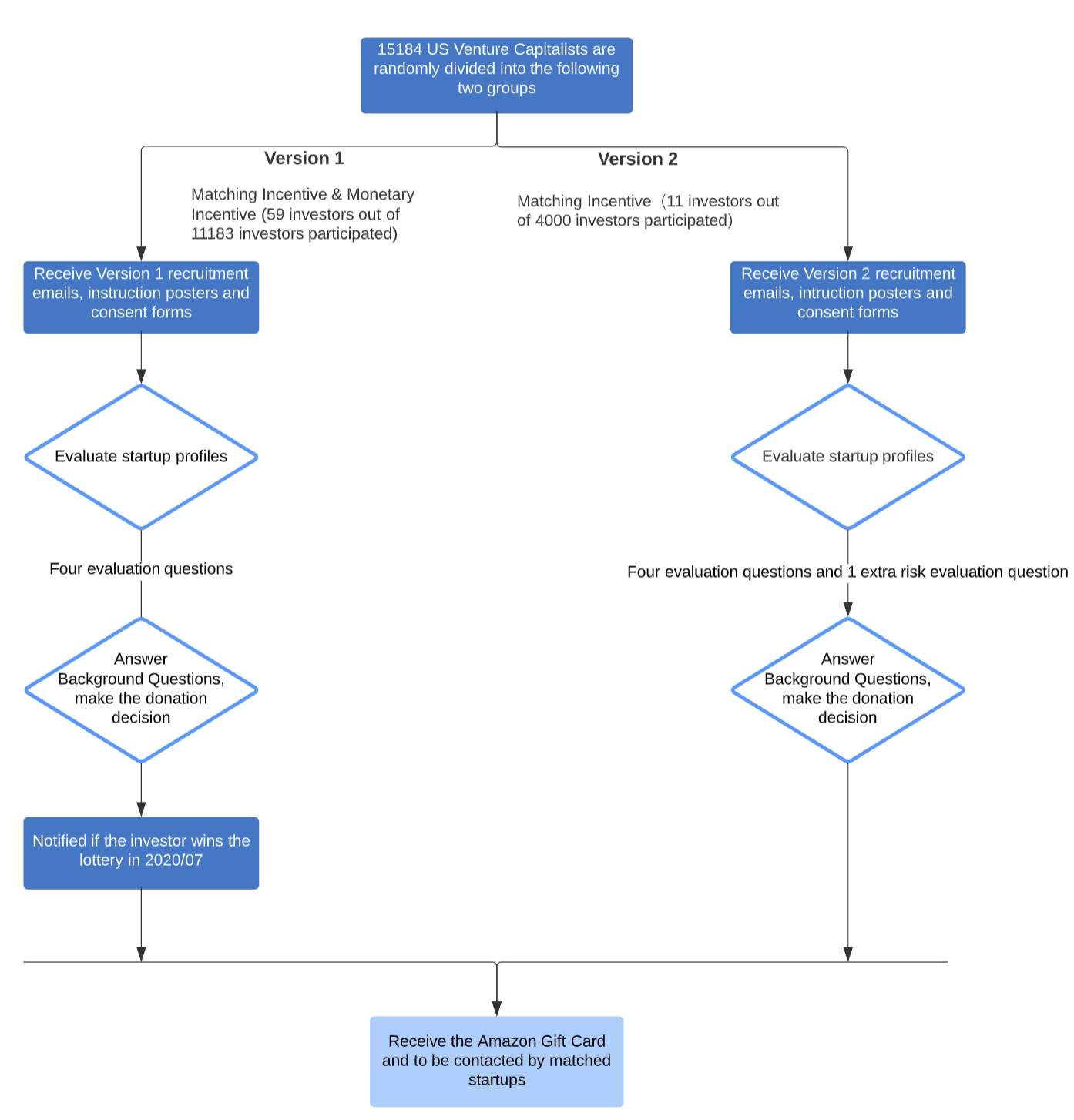}
    \caption{Incentive Structure in Experiment A}
    \label{fig:irr_flow_chart2}
\end{figure}

\clearpage
\begin{figure}
    \centering
    \includegraphics[scale=0.6]{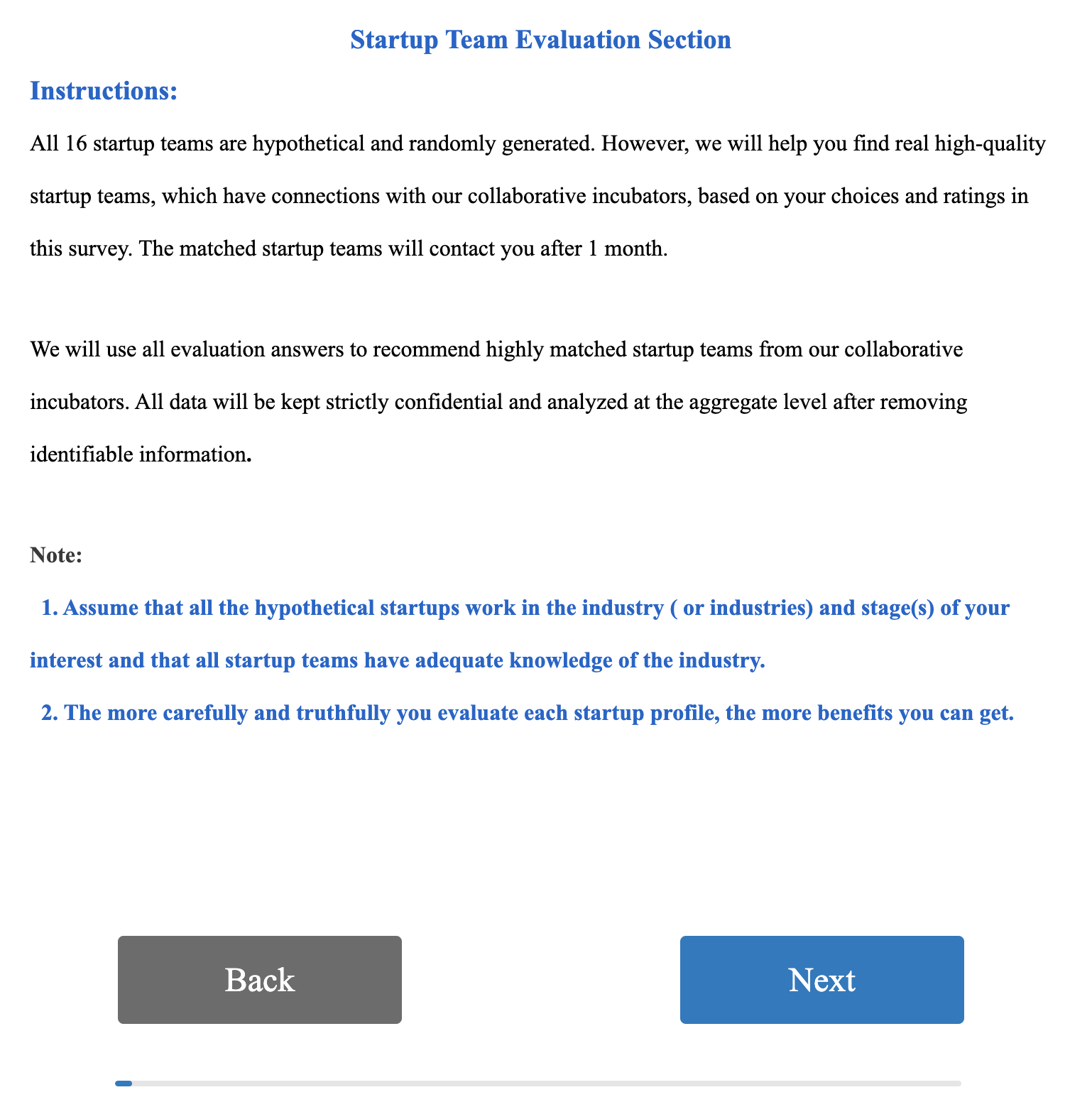}
    \caption{Instruction Page of Experiment A}
    \label{fig:instruction_version2}
\end{figure}

\clearpage
\begin{figure}
    \centering
    \includegraphics[scale=0.48]{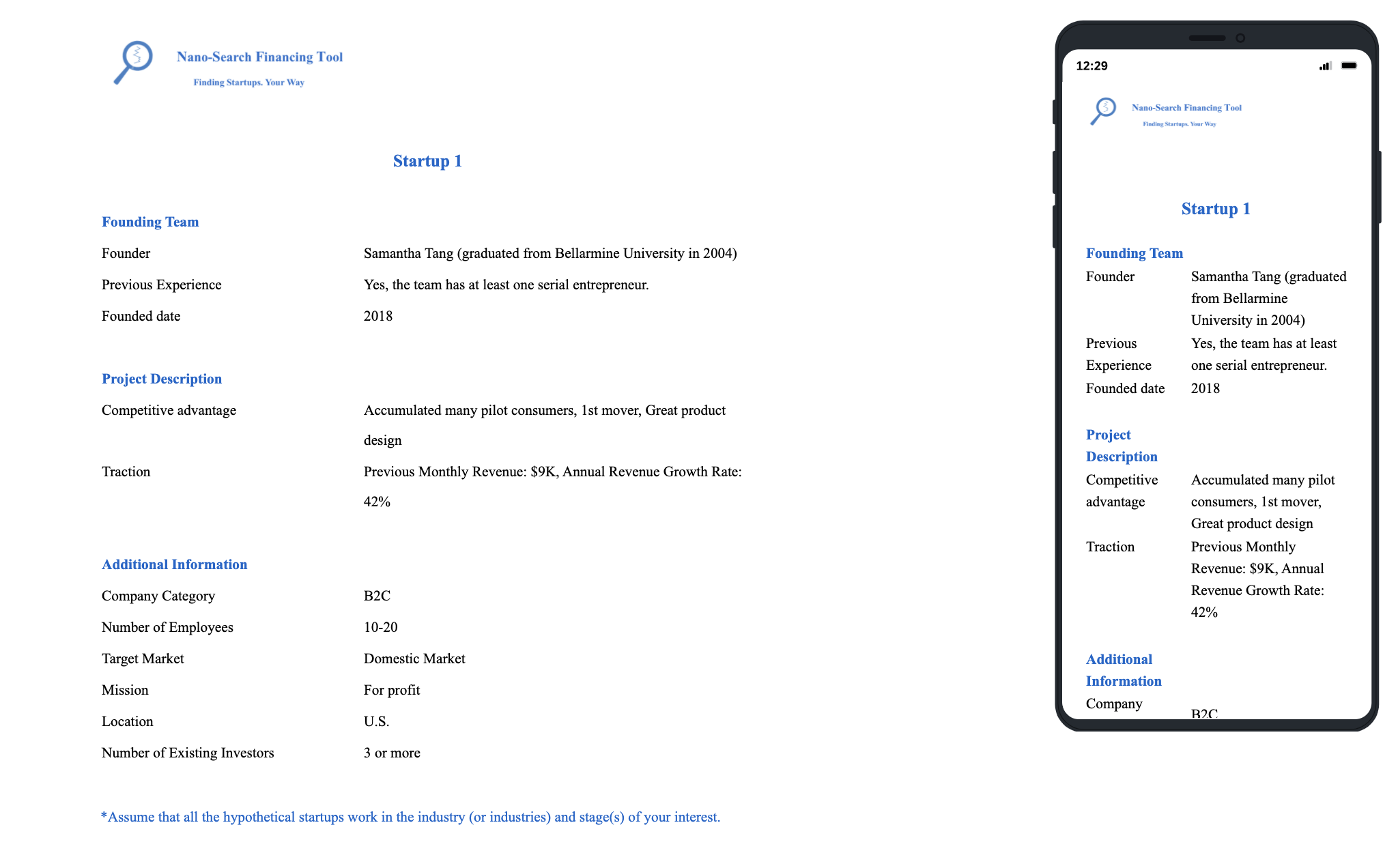}
    \caption{Startup Profile Example for Experiment A}
    \label{fig:startup_profile}
\end{figure}

\clearpage
\begin{figure}
    \centering
    \includegraphics[scale=0.7]{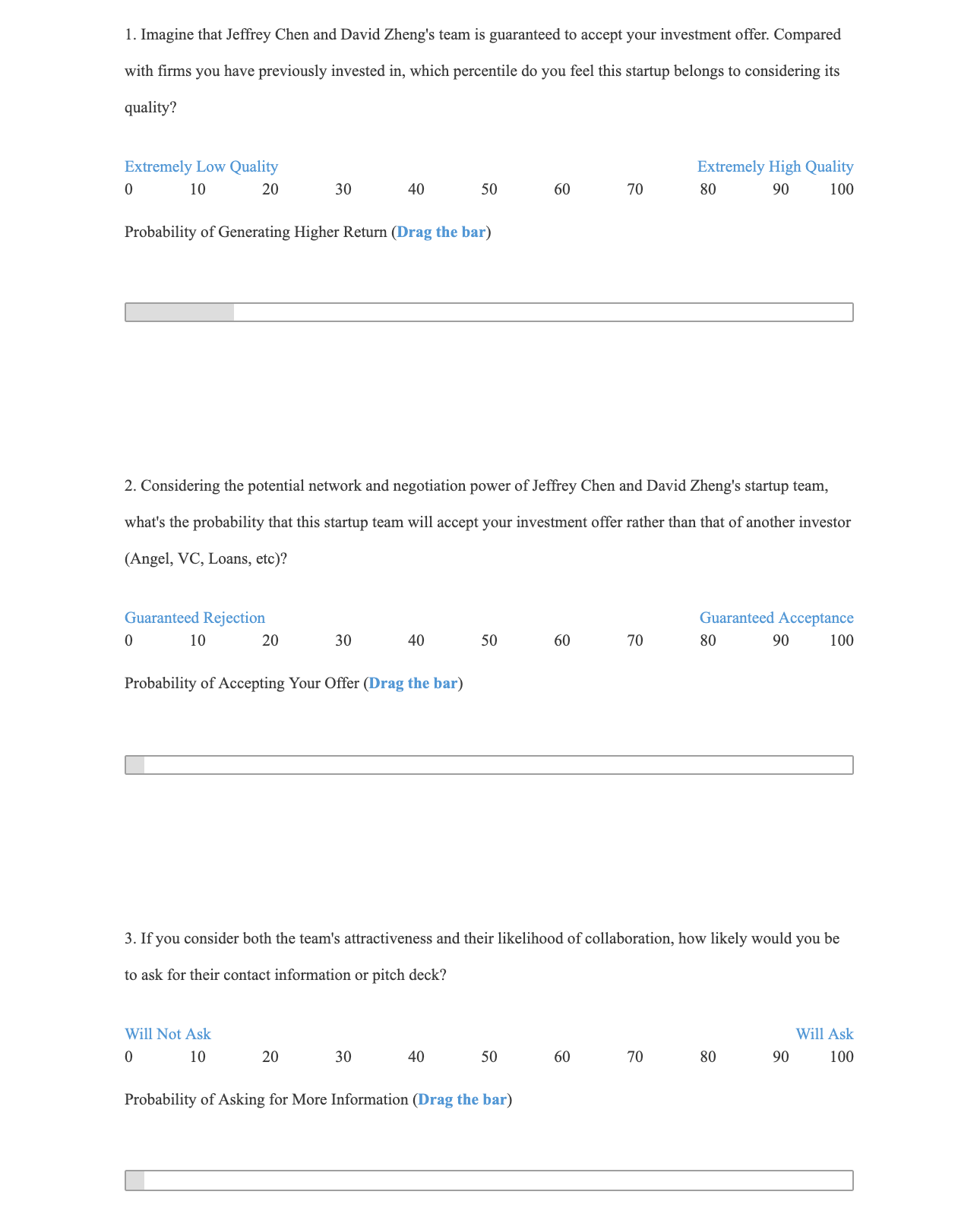}
    \caption{Evaluation Questions of Experiment A (Part 1)}
    \label{fig:Q1}
\end{figure}

\clearpage
\begin{figure}
    \centering
    \includegraphics[scale=0.7]{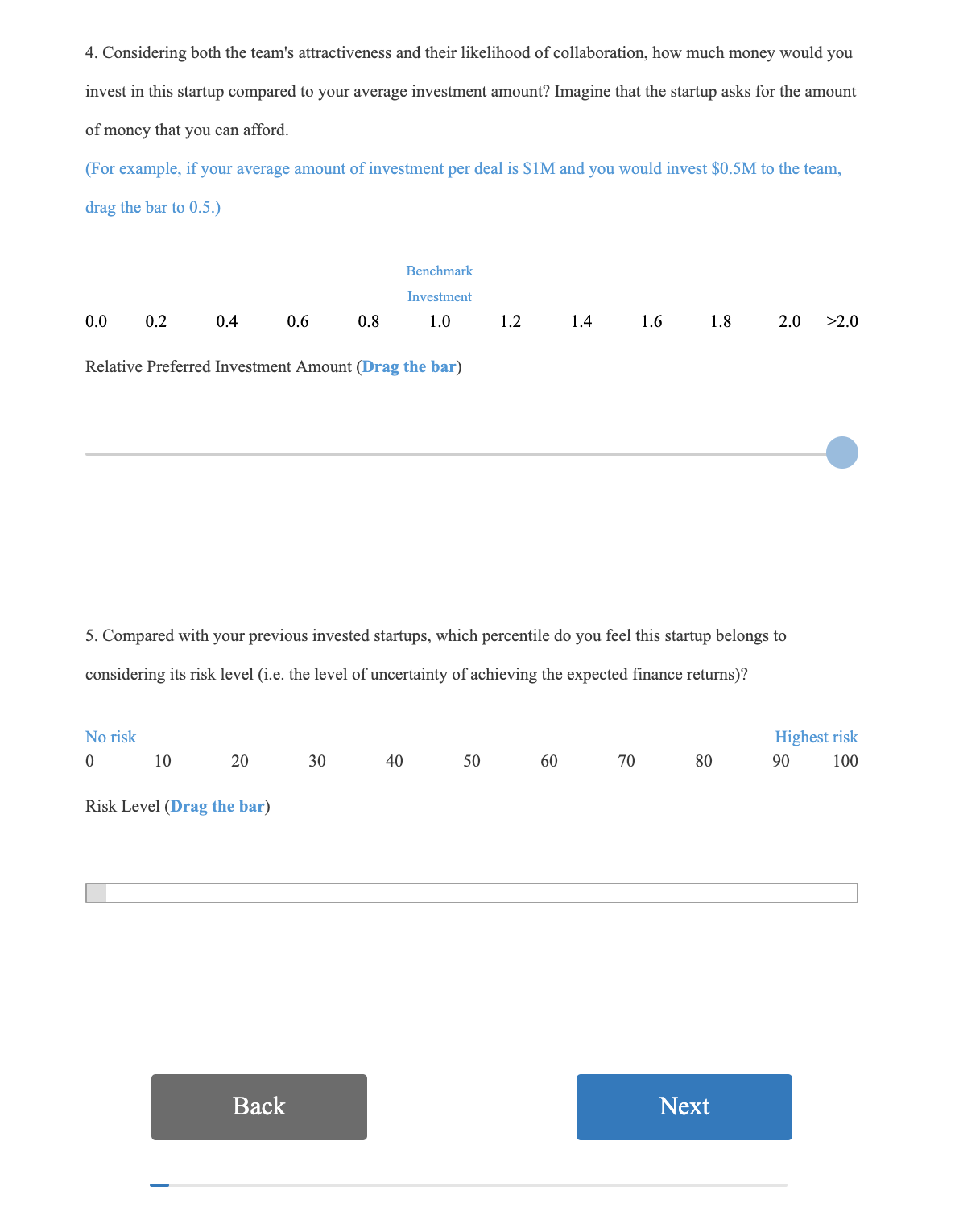}
    \caption{Evaluation Questions of Experiment A (Part 2)}
    \label{fig:Q5}
\end{figure}

\clearpage
\begin{figure}
    \centering
    \includegraphics[scale=0.7]{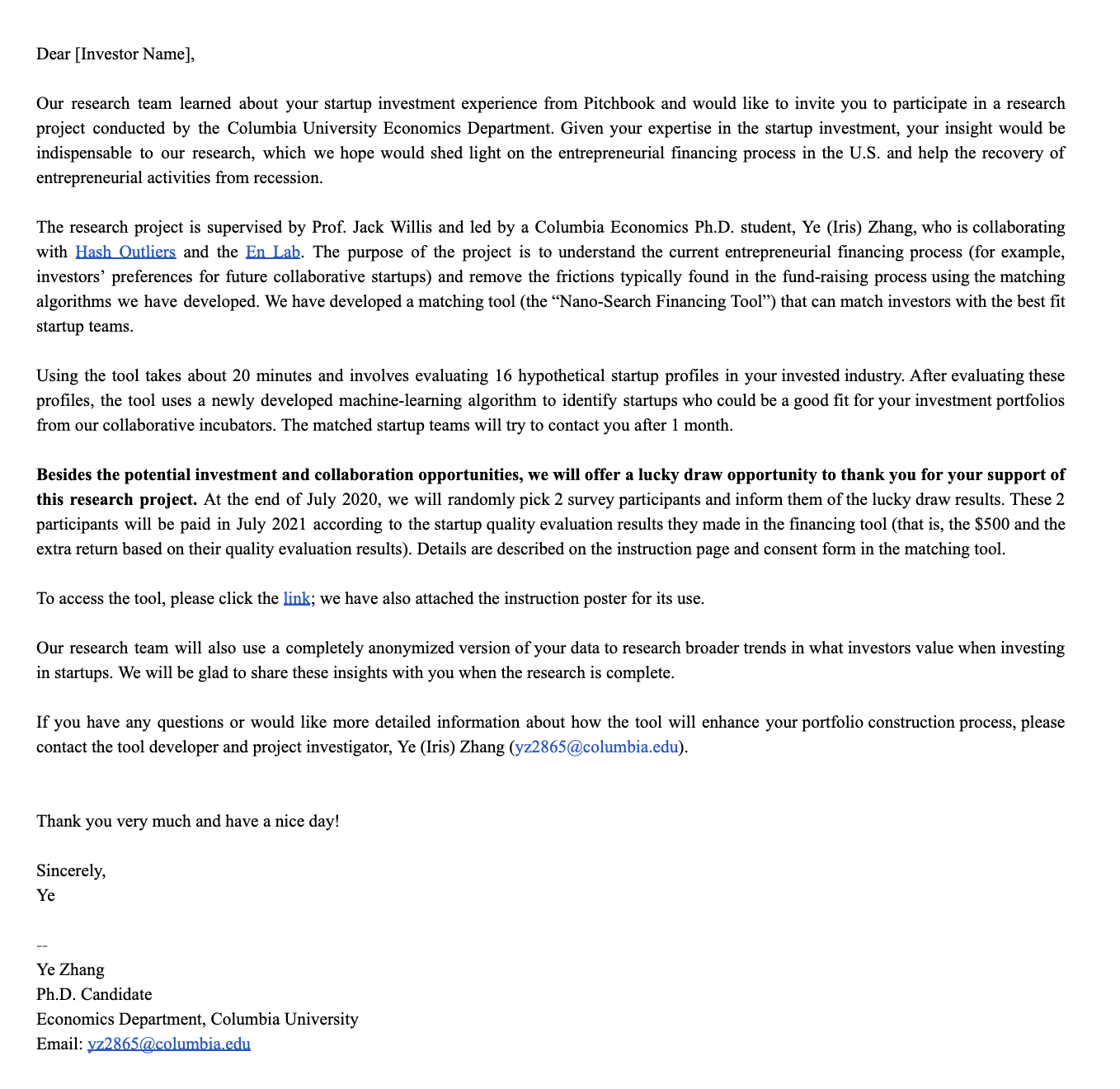}
    \caption{Recruitment Email of Experiment A (Version 1)}
    \label{fig:recruitment1}
      \caption*{\footnotesize{\emph{Notes.} Version 1 provides both the matching incentive and the monetary incentive to randomly selected U.S. venture capitalists.}}
\end{figure}

\clearpage
\begin{figure}
    \centering
    \includegraphics[scale=0.7]{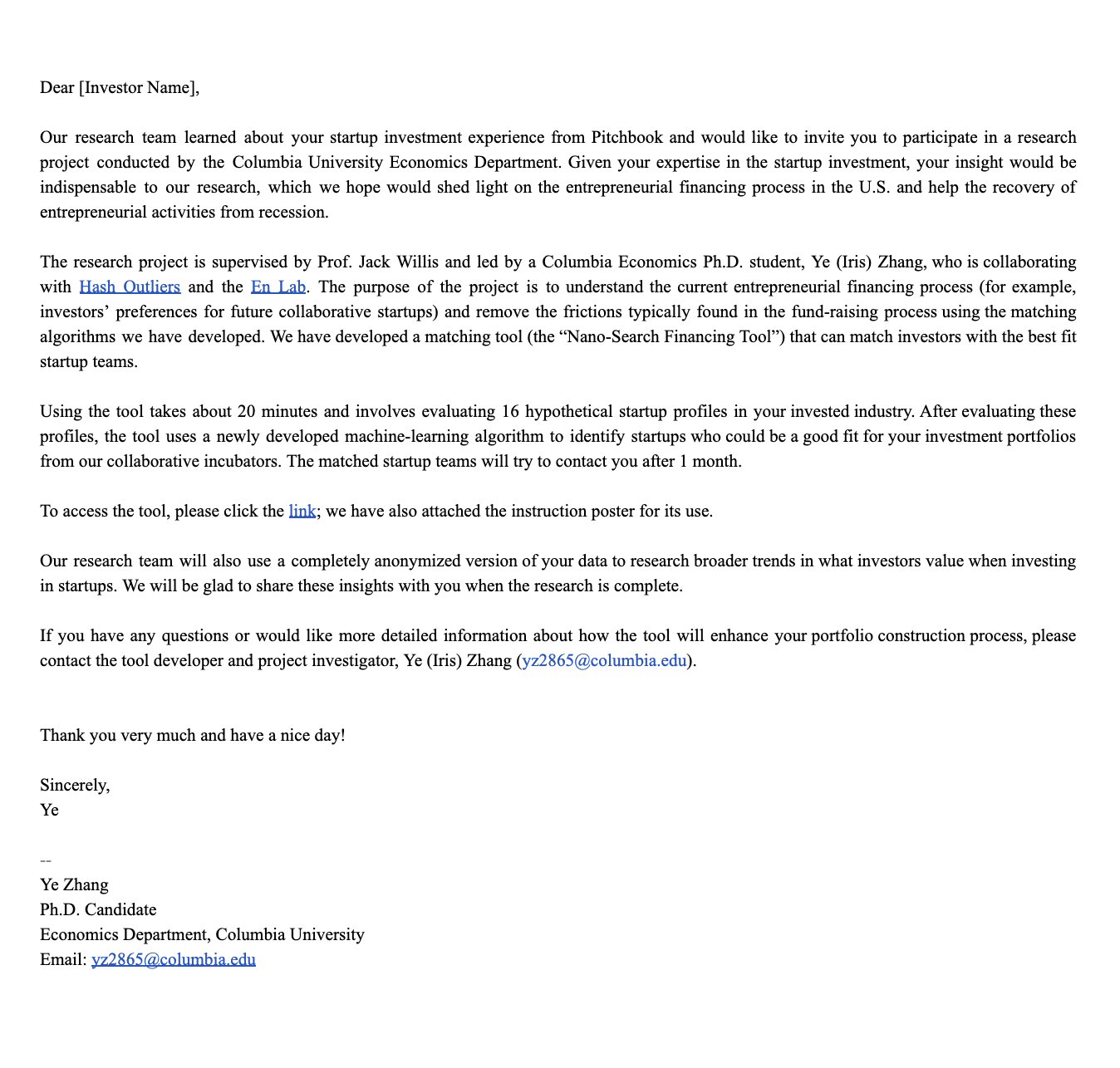}
    \caption{Recruitment Email of Experiment A (Version 2)}
    \label{fig:recruitment2}
    \caption*{\footnotesize{\emph{Notes.} Version 2 provides only the matching incentive to randomly selected U.S. venture capitalists.}}
\end{figure}

\begin{figure}
    \centering
    \includegraphics[scale=0.7]{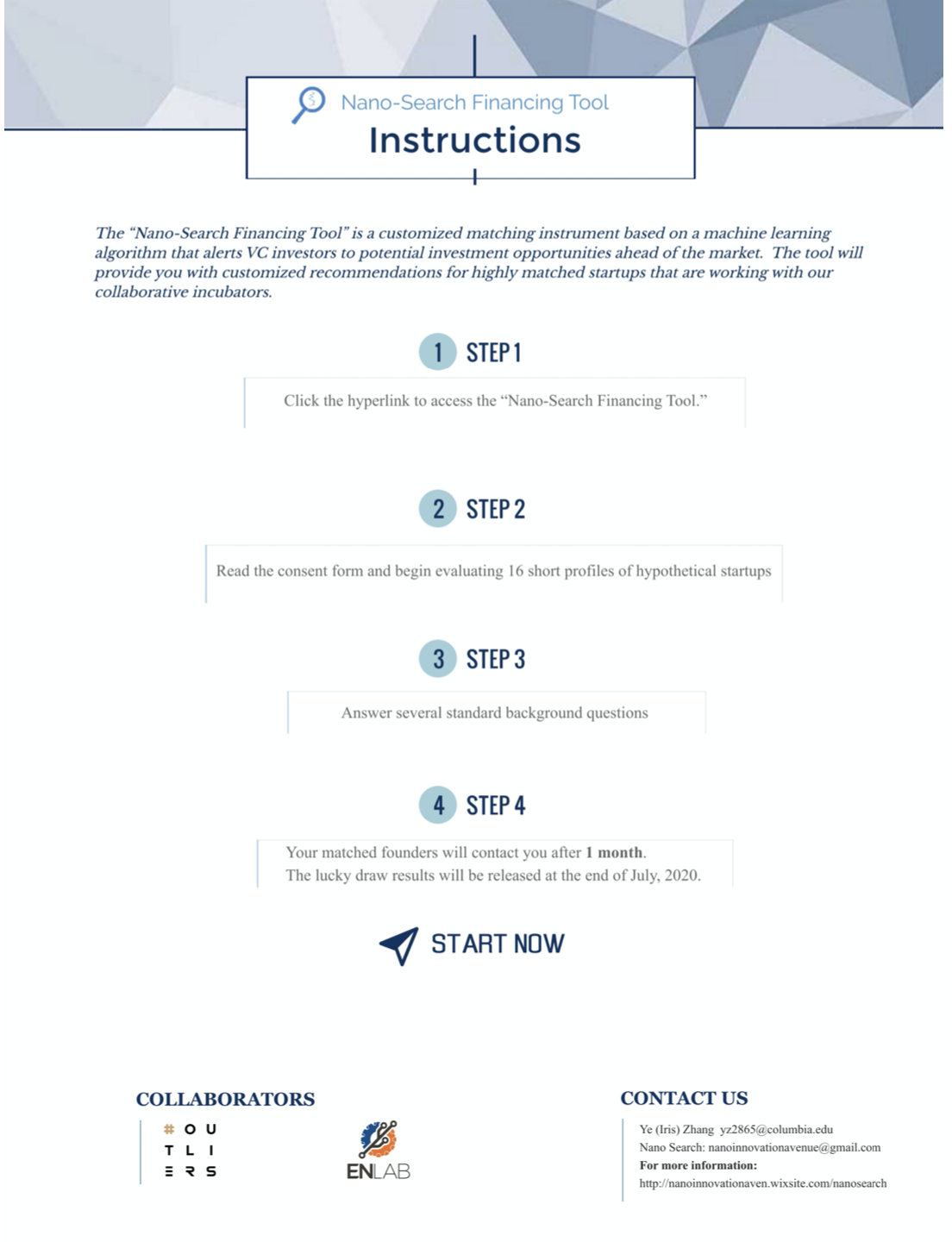}
    \caption{Recruitment Poster of Experiment A (Version 1)}
    \label{fig:poster1}
    \begin{flushleft}
    \caption*{\footnotesize{\emph{Notes.} Version 1 provides both the matching incentive and the monetary incentive to randomly selected U.S. venture capitalists.}}
    \end{flushleft}
\end{figure}

\clearpage
\begin{figure}
    \centering
    \includegraphics[scale=0.7]{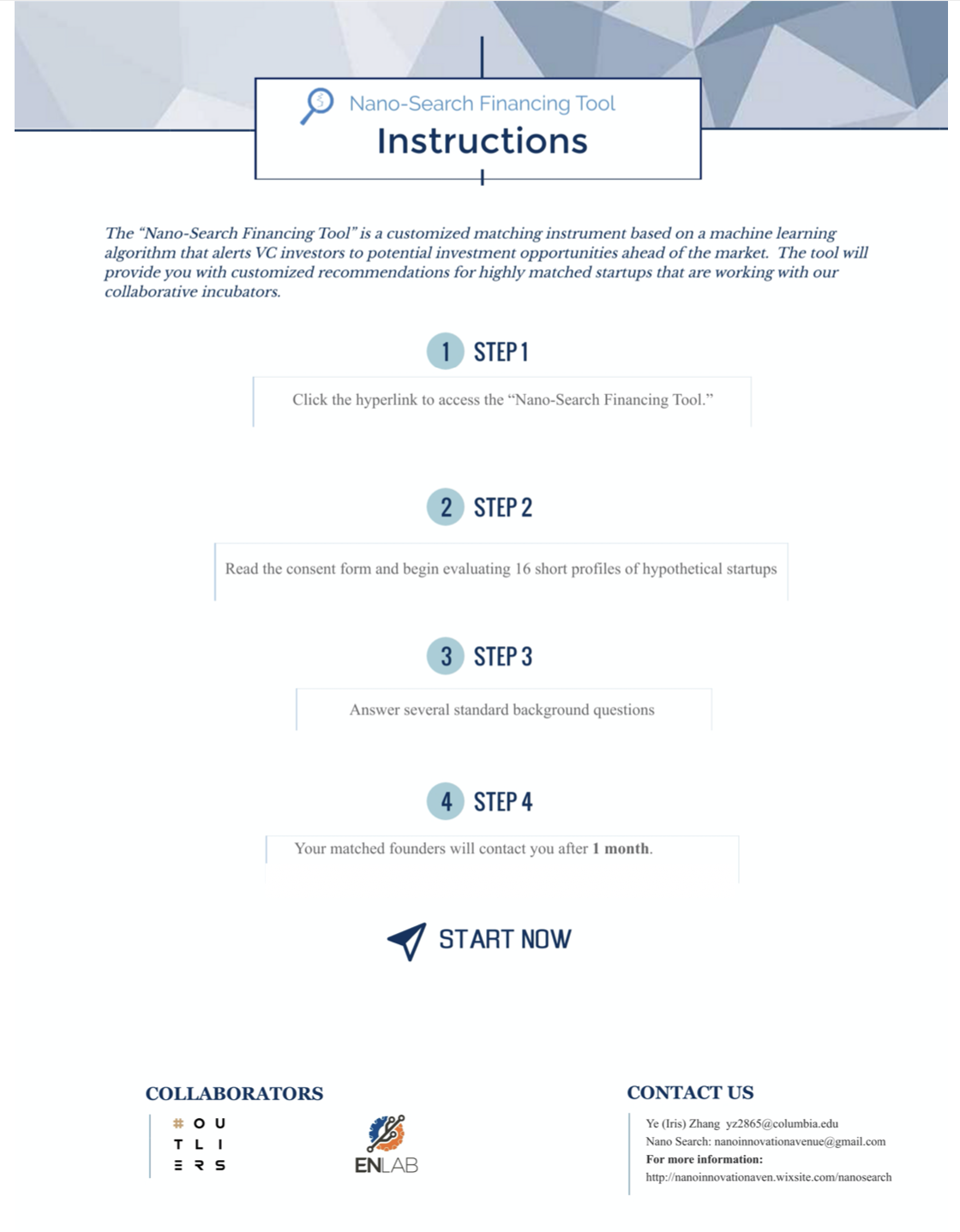}
    \caption{Recruitment Poster of Experiment A (Version 2)}
    \label{fig:poster2}
    \begin{flushleft}
    \caption*{\footnotesize{\emph{Notes.} Version 2 provides only the matching incentive to randomly selected U.S. venture capitalists.}}
    \end{flushleft}
\end{figure}

\clearpage
\begin{figure}
    \centering
    \includegraphics[scale=0.5]{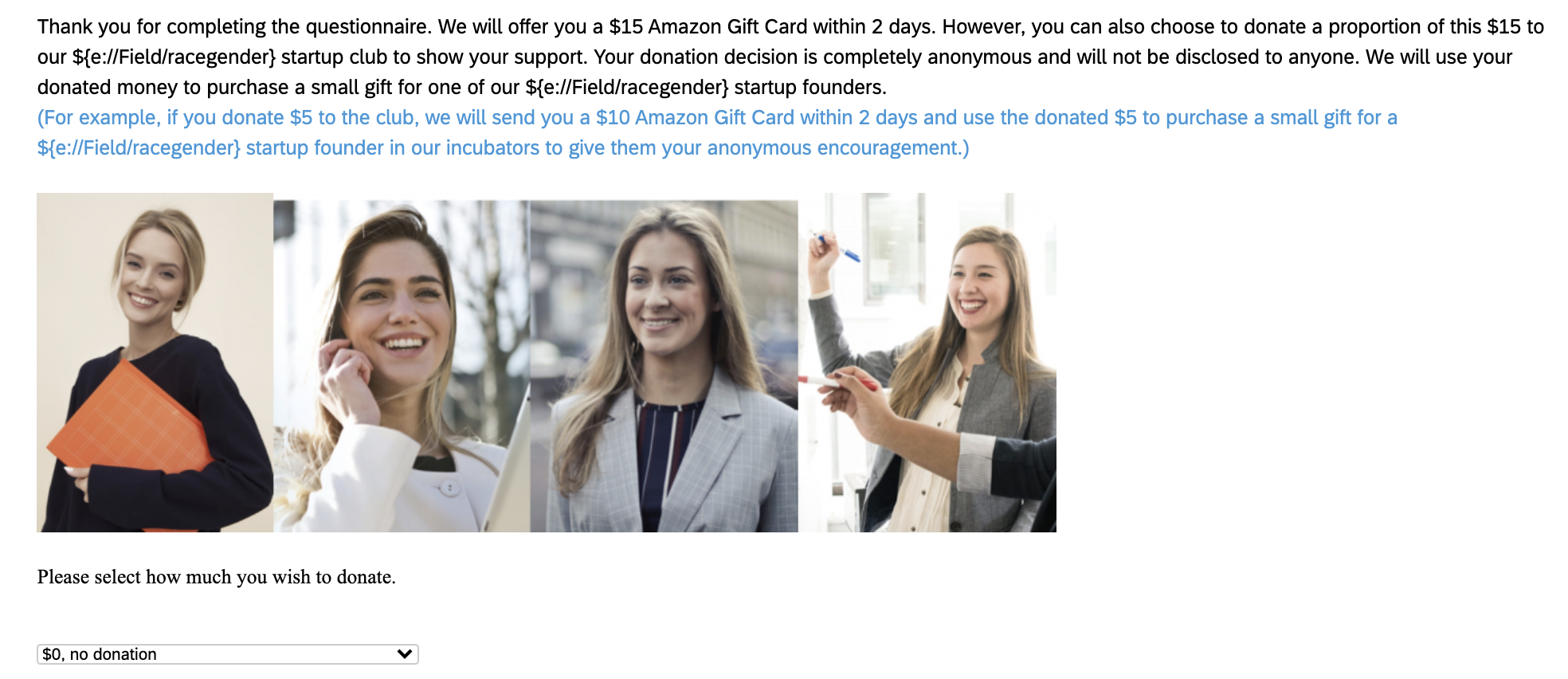}
    \caption{Founder Picture Example in the Donation Game of Experiment A}
    \label{donation_founder}
\end{figure}

\clearpage
\begin{figure}
    \centering
    \includegraphics[scale=0.45]{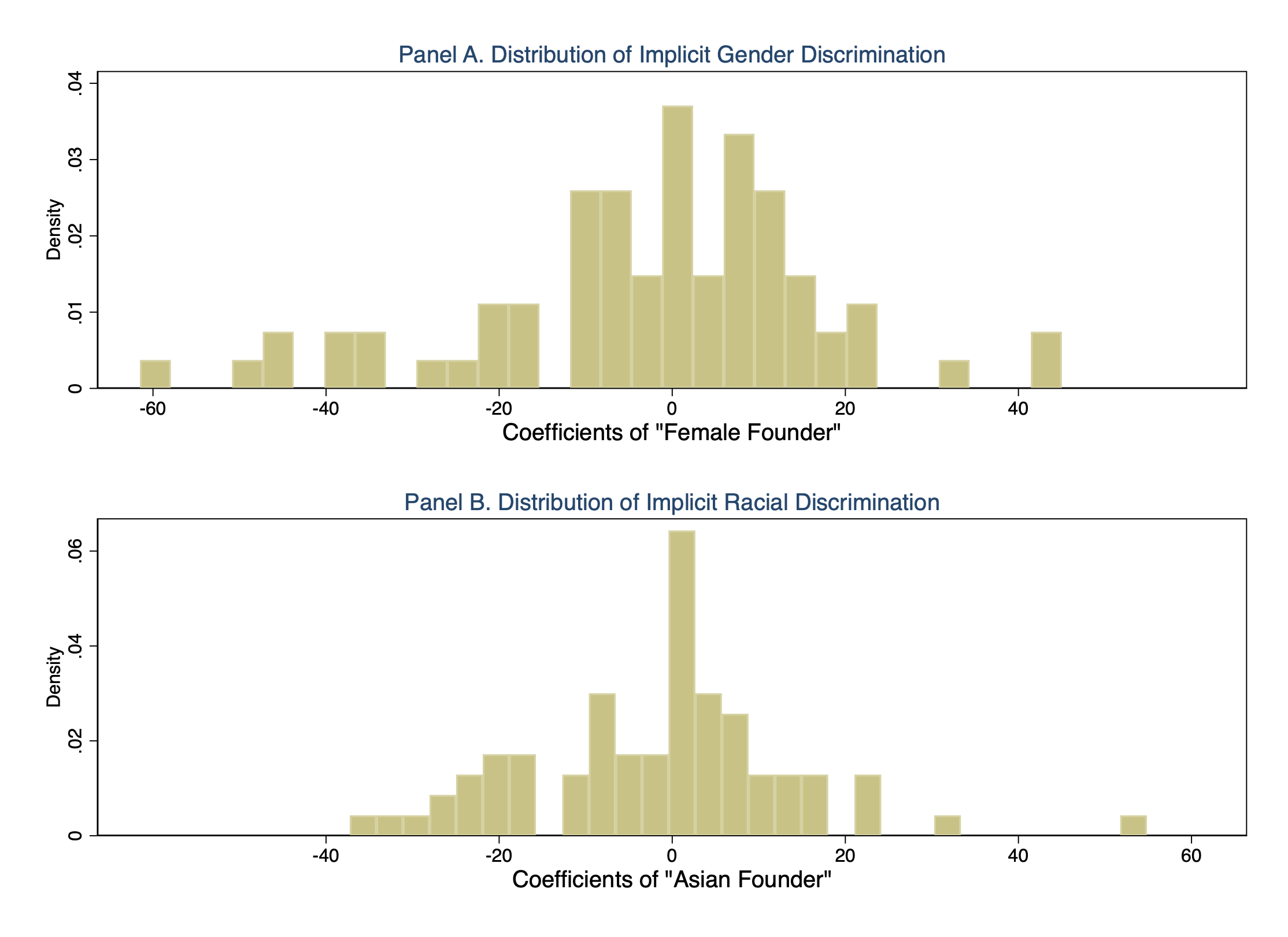}
    \caption{Distribution of Recruited Investors' Implicit Gender and Racial Discrimination}
    \justify
    \caption*{\emph{Notes.} This figure plots the distribution of recruited investors' attitudes towards female and Asian founders. The attitude of each investor $i$ is the coefficient $\beta_i$ of the following regression that uses the profile evaluations in the second half of the IRR experiment: $Q_{3ij}=\alpha_i+\beta_i\text{Startup Characteristics}_{ij}+\epsilon_{ij}$. Panel A demonstrates the distribution of investors' implicit gender discrimination. ``Startup Characteristics" is ``Female Founder" in Panel A. Panel B demonstrates the distribution of investors' implicit racial discrimination. ``Startup Characteristics" is ``Asian Founder" in Panel B. }
    \label{fig:distribution_outlier_attitudes}
\end{figure}

\setcounter{table}{0}
\setcounter{figure}{0}
\renewcommand{\thetable}{\Alph{section}\arabic{table}}
\renewcommand{\thefigure}{\Alph{section}\arabic{figure}}

\section{Correspondence Test}\label{sec:appendix_cor}

\subsection{Name Generation Process} \label{sec:namegeneration}

To generate a list of names that are highly indicative of race (Asian or white) and gender (male or female), this paper combines the approaches of \cite{fryer2004causes} and \cite{gornall_gender_2020}. First names highly indicative of gender are selected based on birth records in Social Security Administration (SSA) dataset.\footnote{The SSA dataset is available at \href{https://www.ssa.gov/OACT/babynames/limits.html}{https://www.ssa.gov/OACT/babynames/limits.html}, accessed on July 27, 2019. Birth Statistical Master File is available at \hyperlink{https://www.cdc.gov/nchs/data_access/vitalstatsonline.htm}{https://www.cdc.gov/nchs/data\_access/vitalstatsonline.htm}, accessed on July 27, 2019.} Last names highly indicative of race are generated based on 2010 U.S. Census data.\footnote{2010 Census surnames product is available at: \hyperlink{https://www.census.gov/topics/population/genealogy/data/2010_surnames.html}{https://www.census.gov/topics/population/genealogy/data/2010\_surnames.html} accessed on July 27, 2019.} The full lists of names are provided in Table \ref{namelist}. The following describes the detailed steps for generating these names.\par

\vspace{2mm}

\textbf{\emph{First Names}} --- Starting with U.S. female and male babies' first names in the SSA dataset, the paper has chosen common names to mitigate the concern that a distinctively ethnic first name can convey other information besides gender.\footnote{For example, such confounding information may be social status and economic background of a person (\cite{bertrand2004emily}). Considering that the naming pattern for Asians and white is very similar (\cite{fryer2004causes}), the paper selects indicative first names from the same name pool.} To avoid gender ambiguity, the paper removes ambiguous names, which are defined as names that were in both the top 1,000 male and top 1,000 female lists with a difference in frequency of less than 200,000 times. Then I pick the most frequent 100 names for each gender for further checks. To remove names that might be perceived as Hispanic or Jewish, the research team manually checked each potential candidate's name and its origin, keeping all the popular Christian names and removing names whose origins are mainly Jewish. The paper further removes names that are strongly indicative of religion, such as Moshe.\par

\vspace{2mm}

\textbf{\emph{Last Names}} --- The paper follows exactly the method of \cite{gornall_gender_2020} by starting with the most common 1,000 last names in the 2010 U.S. Census data. The white-sounding last names are the 50 most common last names, which are more than 85\% likely to be white and less than 3\% likely to be Hispanic. The Asian-sounding last names are all 26 last names on the most common list, which are more than 85\% likely to be Asian. The research team deleted surnames which did not show up in venture capital investors' names recorded by major VC data platforms.\footnote{For each selected last name, we searched the keyword ``last name venture capital investor" or ``last name angel investor" on Google and LinkedIn. If there was no investor which showed up with this last name, we deleted it from the name list. We also removed certain religious last names and some last names like ``Kaur" or ``Vang".} Asian Americans and white Americans have similar first name naming patterns as documented by \cite{fryer2004causes}. Therefore, this paper uses the last name to indicate the ethnicity status of each created fictitious startup founder. To prevent names from signaling extra information such as a founder's social status, the paper only selects commonly used names that do not have any systematic association with founders' social backgrounds.\par

\vspace{2mm}
\textbf{\emph{Additional Check}} --- We also hire 107 Amazon Mechanical Turk (MTurk) users in the U.S. to confirm that the perception of gender and race elicited by these names is in line with demographic data. For both first names and last names, we exclude any names that are not correctly classified by more than 90\% of MTurks. To prevent the generated founder names from being associated with famous startup founders' names, we search LinkedIn to ensure that there are no real famous founders or investors who have the same name and match the key details in the profile. If a conflict is found, we delete the full name and add a new name from the waiting list. 50 names for each race-gender combination for randomization are selected. Selected names are shown in Table \ref{namelist} and Table \ref{lastnamelist}. Gender and race are randomized independently. The corresponding names used for each hypothetical startup for Experiment B are provided in Table \ref{cornamelist}.\par

\subsection{Emailing Process and Preparation Work} \label{sec:emailsending}

\textbf{\emph{Emailing Process}} --- The paper mainly implements the following two steps to solve the technical difficulties of sending a large number of cold call emails to investors' email inboxes and to pass the existing spam filters.\footnote{Different email providers usually use different spam filtering algorithms. However, there are some common patterns for detecting spam emails. First, if there are many invalid email addresses sent out from the same domain at an extremely high frequency (for example, 10 emails sent out per second), then the emails sent are more likely to be labeled as spam emails. Second, if the email contains unverified website links or common words used in spam emails like ``Dear,” these emails are likely to fail the spam filter. However, none of these spam filtering algorithms are correlated with email senders' gender and race.} First, before sending large-scale pitch emails in 03/2020, I send out a testing email (see Figure \ref{cor:testingemail} in Appendix \ref{sec:appendix_cor}) which introduces public information about COVID-19 in 02/2020. The testing email is meant to identify which email addresses are invalid and to check the opening rate of cold emails irrelevant to investment opportunities.\footnote{Invalid email addresses are those that no longer exist or are no longer frequently checked by investors based on the bounced back email notifications. The investor database was constructed between 04/2018-12/2019. Therefore, more than 20\% of the collected email addresses are no longer valid due to the high turnover rate.} The opening rate of the testing email after 2 weeks was 2.8\%, while the average opening rate of the investment-related pitch emails in this experiment is 11.8\%. This indicates that investors only open the emails that they are interested in based on the email subject line and senders.\par

\vspace{2mm}
Second, I use Mailgun’s Managed Service, a third-party commercial email API delivery service provider, for sending a large number of emails.\footnote{\href{https://www.mailgun.com/}{https://www.mailgun.com/} Mailgun has more than 150,000 customers in 2020. It was founded in 2010 and was a part of the Y Combinator Winter 2011 cohort.} Compared with the traditional method of using multiple web hosts to combat spam policies, Mailgun is designed for developers and businesses, with an extremely powerful functionality to track the status of each email sent and achieve a high delivery rate through its emailing infrastructure. It also provides developers with complete freedom to customize email sender names, set the back-end database structure, and develop new email tracking functionalities with a user-friendly price compared with Gsuite, which is an email provider from Google.\footnote{If researchers have abundant research funding, they can also create multiple Gsuite accounts to combat spam policies. Gsuite is a “company-version” of gmail and is user-friendly to people without strong coding skills. The only drawback is its relatively expensive price, costing \$6 per account per month starting in 2020.} Before automatically sending pitch emails, I use GlockApp, a spam filter testing service provider, to test and improve my pitch email templates. \par

\vspace{2mm} 
Following the two-step email sending procedures mentioned above, the response rate is very stable along the whole recruitment process. \cite{gornall_gender_2020} use standard methods of sending out a large number of cold call pitch emails and the email response rate declined from 9.0\% for the first 4,000 emails to 5.3\% for the last 4,000 emails. This situation did not occur in this experiment. Moreover, the email sending procedures in this experiment allow for monitoring multiple investors’ information acquisition behaviors without hurting the email delivery rate too much.\par

\vspace{2mm} 
\textbf{\emph{Preparation Work}} --- To make sure that the i.i.d assumption holds for the experiment,\footnote{Abbreviation for ``independent and identically distributed".} the preparation work for this experiment is implemented in the following steps. First, to increase the response rate, I match investors with pitched startup ideas based on their industry/vehicle preferences so that healthcare-related pitch emails are sent to investors who are interested in the healthcare industry.\footnote{For investors recorded in the Pitchbook Database, I use the recorded industry preference for the matching purpose. For investors from other databases, I manually collected their industry preferences from information on their company websites, LinkedIn, and CBInsight. If the manually collected industry information is not accurate, this will increase the noise of the experiment's results and reduce the email response rate. However, it does not affect investors' email opening behaviors.} Second, considering the potential spillover effect within each VC fund, investors receiving the same pitch email ideas come from different VC funds.\footnote{For some VC funds, they usually have a weekly meeting to discuss promising investment opportunities before replying to cold call pitch emails. If investors receiving the same startup idea come from the same fund, their responses are likely to be correlated. However, this situation will not affect email opening behaviors and email reading time when they just receive pitch emails.} Each startup pitch email is sent to roughly 1000 investors who all work in different funds. Among these 1000 investors, they are randomly divided into 16 groups. Based on the factorial experimental design, the startup founder's gender, race, education, and project advantages should be randomized independently. Hence, we have $2\times 2\times 2\times 2=16$ groups. Third, it usually takes more than 2 weeks to send two sequential pitch emails to the same investor to avoid unnecessary attention and keep the i.i.d. assumption.\footnote{\cite{gornall_gender_2020} wait at least five days to send a sequential email, which raises the attention of some investors who draw attention to these cold emails on twitter in the middle of the experiment. Their experiment was finished between 11/2018-12/2018. To avoid such a situation, I slow down the pace of sending cold emails and extend the experiment's implementation period.} Each investor received 3 to 5 pitch emails between 03/2020-09/2020.\par

\begin{table}
 \caption{First Names Used in Experiment B}
\label{namelist}
\begin{center}
\scalebox{0.85}{
\begin{tabular}{l l l l l l l} 
\hline
\emph{Panel A: Female} &&&&&&\\
Jennifer&Elizabeth&Lisa&Laura&Megan&Emily&Erica\\
Natalie&Jacqueline&Victoria&Melanie&Tina&Kayla&Kristy\\
Melinda&Linda&Theresa&Kara&Amanda&Sarah&Amy\\
Angela&Christina&Rebecca&Tiffany&Mary&Brittany&Samantha\\
Katherine&Alicia&Monica&Kathryn&Patricia&Anna&Catherine\\
Veronica&Kathleen&Sandra&Cassandra&Valerie&Amber&Teresa\\
Allison&Amber&Katrina&Jenna&Megan&Jessica&Melissa\\
Nicole&Sara&Julie&Christine&Tara&Katie&\\
&&&&&&\\
(Extra)&&&&&&\\
Abigail&Danielle&Michelle&Rachael&Brenda&Margaret&Amanada\\
Hayley&Madeline&Molly&Vanessa&Rachael&Grace&Heather\\
Cynthia&Caroline&Karen&&&&\\
&&&&&&\\
\emph{Panel B: Male} &&&&&&\\
Robert&Brian&Kevin&Steven&Thomas&Adam&Patrick\\
Bryan&Keith&Donald&Peter&Jared&Phillip&Jeffery\\
Victor&Seth&Alan&Matt&David&Jason&John\\
William&Andrew&Justin&Anthony&Jonathan&Timothy&Nicholas\\
Jeremy&Richard&Jeffrey&Benjamin&Paul&Stephen&Nathan\\
Jacob&Gregory&Travis&Kenneth&Samuel&Edward&Derek\\
Ronald&Joel&Frank&Dennis&Erik&Philip&Christopher\\
James&Mark&Scott&Dustin&Zachary&Marcus&Gary\\
&&&&&&\\
(Extra)&&&&&&\\
Vincent&Jack&Luke&Michael&Evan&Joseph&Eric\\
Shane&Sean&Matthew&Ian&George&Trevor&Charles\\
\hline
\end{tabular}}
\end{center}
\begin{tablenotes}
\item \footnotesize \emph{Notes.} All listed first names which are indicative of gender are used for both Experiment A and Experiment B. It covers the popular first names of people who are between 24 years old and 45 years old. To make sure all the names are only indicative of gender, 107 Amazon Mechanical Turks are hired to classify potential names into different genders and provide their feedback on whether these names remind them of other information besides gender (e.g., economic background, race, immigration status, etc). For all the selected names listed above, more than 98\% of Amazon Mechanical Turks correctly classify the names into the corresponding gender. Names which are indicative of other information are also deleted. For example, ``Chelsea" is deleted because some M-turks feel it is associated with the upper-class; ``Luis," ``Carlos," or ``Antonio" are deleted because they are perceived as more likely to be Hispanic. The additional first names and last names used in \cite{gornall_gender_2020} are added in the ``extra" part.
 \end{tablenotes}
\end{table}

\begin{table}
 \caption{Last Names Used in Experiment B}
\label{lastnamelist}
\begin{center} 
\scalebox{0.85}{
\begin{tabular}{l l l l l} 
\hline
\emph{Panel A: Asian} &&&&\\
Yu&Zhao&Zhang&Jiang&Hwang\\
Huynh&Luong&Cheung&Hsu&Liang\\
Li&Hu&Xu&Zhu&Huang\\
Yang&Kwon&Choi&Nguyen&Pham\\
Hoang&Luu&Liu&Lu&Chen\\
Lin&Chang&Chung&Zheng&Xiong\\
Zhou&Ngo&Truong&Wu&Duong\\
Cho&Cheng&Yi&Dinh&Tang\\
Wong&Chan&Ho&Thao&Tsai\\
Le&Yoon&Wang&&\\
&&&&\\
\emph{Panel B: White} &&&&\\
Nelson&Russell&Roberts&Rogers&Adams\\
Cooper&Wright&Cox&Kelly&Phillips\\
Bennett&Bailey&Collins&Thompson&Stewart\\
Parker&Evans&Allen&Martin&Anderson\\
Clark&Campbell&Morris&Reed&Wilson\\
White&Taylor&Sullivan&Myers&Peterson\\
Murphy&Fisher&Cook&Hughes&Price\\
Gray&Moore&Hill&Baker&Hall\\
Smith&Miller&Ward&&\\
&&&&\\
(Extra)&&&&\\
Hansen&Welch&Hoffman&Meyer&Schmidt\\
Burke&Beck&Walsh&Carpenter&Schultz\\
Jensen&Keller&Snyder&Stone&Cohen\\
Barker&Becker&Schwartz&Larson&Weaver\\
Carroll&&&&\\
\hline
\end{tabular}}
 \end{center} 
\begin{tablenotes}
\item \footnotesize \emph{Notes.} The table contains selected last names indicating ethnic identity for hypothetical startup founders. To make sure all the names are only indicative of race and perceived correctly by people, 107 Amazon Mechanical Turks are hired to classify potential names into different races and provide their feedback on whether these names remind them of other information besides race (e.g., economic background, immigration status, etc.). For all the selected last names listed above, more than 95\% of the Amazon Mechanical Turks correctly classify the Asian last names into the corresponding race and more than 92\% of the Amazon Mechanical Turks correctly classify the white last names. All the ambiguous last names are removed. For example, ``Shah" is deleted because many M-turks feel it can also be a middle-eastern name; ``Long" is deleted because it can serve as both a white name and also an Asian name. Some last names are also removed if they are related to religion or very rare in the venture capital industry, like ``Kaur" and ``Vang." The additional first names and last names used in \cite{gornall_gender_2020} are added in the ``extra" part.\\
 \end{tablenotes}
\end{table}

\begin{table}
 \caption{Full Names Used in Experiment B}
\begin{center} 
\label{Startup and Entrepreneur Names List}
\label{cornamelist}
\scalebox{0.85}{ 
\begin{tabular}{l l l l l } 

\hline
Startup Names & White Female&Asian Female& White Male& Asian Male\\
\hline
\emph{Panel A: 1st round}&&&&\\
VoiceFocus&Kathleen Jensen&Kathleen Yi&Joseph Adams&Kevin Truong\\
Light Run&Lisa Thompson&Stephanie Lu&Vincent Snyder&Jeffrey Luong\\
Instrument Tell&Molly Weaver&Jennifer Dinh&Sean Miller&Justin Huang\\
Sign Reader&Megan Schwartz&Valerie Yu&Evan Meyer&Shane Chan\\
Bross&Catherine Welch&Rachael Pham&Eric Burke&Ryan Le\\
Chicky&Rachael Smith&Vanessa Zhu&Robert Reed&Trevor Thao\\
LoopuDeck&Mary Meyer&Melissa Liu&George Price&Vincent Xu\\
EasySample&Melissa Larson&Catherine Yang&Matthew Russell&Ian Zheng\\
YouTubys&Grace Clark&Christine Tang&Justin Hansen&Bryan Hu\\
OSS&Veronica Russell&Emily Thao&Shane Snyder&Luke Zhao\\
CPRX&Danielle Cook&Margaret Dinh&Scott Parker&Eric Pham\\
All-in&Julie Barker&Karen Wong&Marcus Becker&Derek Yoon\\
SkatED&Kathryn Beck&Abigail Chang&Andrew Moore&George Cheng\\
GeniusPlot&Christina Parker&Katie Kwon&David Sullivan&Marcus Wang\\
EasyTry-On&Katherine Snyder&Angela Ho&Richard Cook&Mark Chung\\
Krysco&Valerie Baker&Amanda Jiang&Patrick Ward&Kevin Hoang\\
Lens Bioimage Technology&Emily Bennett&Erica Zhou&Adam Hoffman&Peter Cheung\\
Medprint&Jacqueline Hughes&Patricia Yoon&Ian Cooper&Brian Dinh\\
BM International&Vanessa Phillips&Mary Luu&Edward Keller&Jack Luu\\
Vet Technology&Michelle Gray&Natalie Hwang&Jeremy Carroll&Michael Wu\\
Freight Future&Amanda Meyer&Danielle Cheng&Christopher Cohen&Edward Lin\\
AfroLab&Madeline Hill&Nicole Xu&Steven Collins&Stephen Liu\\
SmartTeacher&Jessica Evans&Melanie Ngo&William Welch&Jason Chung\\
CleanPlanet&Christine Fisher&Megan Liang&Jeffrey Barker&Nicholas Lu\\
FancyTravel&Melanie Schultz&Rebecca Zhao&Ryan Schwartz&Sean Xiong\\
MeSafeMicro&Cynthia Keller&Allison Duong&Samuel Kelly&Samuel Ngo\\
Talently&Caroline Stone&Heather Zhang&Jack Moore&Richard Thao\\
AgriSoft&Rebecca Miller&Katherine Truong&Gregory Morris&Jonathan Duong\\
EduPar&Erica White&Caroline Chung&Derek Jensen&Jeremy Jiang\\
Milkless&Hayley Becker&Christina Hsu&Luke Thompson&William Hwang\\
Durabuddy&Brenda Bailey&Madeline Tsai&Brian Reed&James Le\\
Constructech&Samantha Peterson&Samantha Le&Michael Myers&Patrick Nguyen\\
SolarWat&Patricia Stewart&Brenda Hoang&Thomas Beck&Christopher Huynh\\
\hline

\end{tabular}}
 \end{center}
\end{table}

\clearpage
\begin{table}
\begin{center} 
\scalebox{0.85}{ 
\begin{tabular}{l l l l l } 
&&&&\\
&&&&\\
\emph{Continued}&&&&\\
\hline
Startup Names & White Female& Asian Female& White Male& Asian Male\\
\hline
\emph{Panel B: 2nd round}&&&&\\
Highlight&Melanie Cohen&Cynthia Zhao&Scott Hughes&Steven Pham \\
AutoTrend&Jessica Hughes&Christina Hoang&Adam Ward&Anthony Ngo\\
PackingFirst&Rachael Welch&Abigail Wu&Andrew Phillips&Bryan Nguyen\\
ApexInfluence&Abigail Jensen&Madeline Chen&Richard Weaver&Jack Thao\\
CSandBet Co&Michelle Gray&Margaret Dinh&Michael Hoffman&Vincent Thao\\
Alyx Room Inc& Michelle Keller&Danielle Zhu&Nicholas White&Patrick Jiang\\
Laundrobot&Valerie Price&Mary Le&Anthony Russell&Scott Chen\\
Green Scan&Patricia Hill&Vanessa Tsai&Ian Larson&Peter Le\\
SmartBell&Brenda Myers&Nicole Luu&Derek Nelson&Justin Cheung\\
WarmHugs&Cynthia Reed&Brenda Thao&Edward Fisher&Brian Zhu\\
Athleticism&Erica Bennett&Hayley Zhou&Marcus Bennett&Bryan Pham\\
Life Orama Systems&Katie Cooper&Jessica Cho&Jason Cook&Labs Zhang\\
Quanta Meeting&Lisa Barker&Molly Hu&Vincent Collins&Trevor Ho\\
Indoor Health Monitor&Caroline Stewart&Jacqueline Yi&Jonathan Snyder&Joseph Truong\\
FinFollow&Grace Baker&Veronica Wong&Justin Meyer&Evan Li\\
Pillow Dream&Amanda Moore&Karen Liu&Bryan Murphy&Mark Yi\\
Fragrance Fresh&Christina Nelson&Heather Chan&Samuel Sullivan&Edward Hwang\\
SolarPlug&Margaret Walsh&Natalie Kwon&Steven Carpenter&Edward Li\\
FoodFormula&Danielle Snyder&Angela Yoon&Trevor Price&Richard Huynh\\
SmartClothes&Valerie Cox&Kathryn Liang&Luke Stewart&Matthew Yu\\
TourVirtual&Julie Russell&Emily Duong&Paul Becker&Jeremy Hu\\
Dyslexia+&Nicole Morris&Hayley Le&Patrick Carroll&Gregory Zhao\\
Wrinkless&Megan Hall&Samantha Tang&Matthew Burke&Ryan Yang\\
BioPack&Hayley Ward&Megan Ho&Jeremy Wilson&David Wu\\
Breathe Glove&Katherine Anderson&Catherine Wang&William Jensen&Adam Hsu\\
Foglessness&Madeline Sullivan&Grace Ngo&Anthony Schmidt&George Hoang\\
Momfit&Angela Thompson&Heather Hu&George Hall&Eric Lu\\
InsurMe&Stephanie Beck&Cynthia Truong&Stephen Anderson&Derek Duong\\
All-in-one&Vanessa Larson&Melissa Jiang&Labs Miller&Stephen Dinh\\
TalkThrough&Veronica Allen&Rachael Cheung&Nicholas Parker&Christopher Tang\\
A-BodyBank&Samantha Burke&Jennifer Choi&Sean Allen&William Kwon\\
StartSoon&Rebecca Hoffman&Valerie Nguyen&Ryan Cox&Luke Xiong\\
XManager&Molly Phillips&Lisa Lin&Andrew Adams&Scott Chen\\
OutGuard&Allison Cook&Caroline Huang&Eric Cohen&Ian Xu\\
\hline
\end{tabular}}
 \end{center}
  \begin{tablenotes}
\item \footnotesize \emph{Notes.} 33 startups are created for the first round experiment, which was implemented between 03/2020-04/2020. 34 startups are created for the second rounds of experiments, which were implemented between 2020/10-2020/11. All the startup founders' names are randomly generated using the commonly used first names and last names in the U.S. To prevent the fictitious startup founders from being associated with real people, I search LinkedIn, Google, and available university directories to make sure that no real students from the corresponding universities have the same names. If a conflict is discovered, I replace the conflicting names with other randomly generated names to avoid such a situation. Information of startups used in the later round correspondence test will be updated in the next version of draft.\medskip
\end{tablenotes}
\end{table}

\clearpage
\begin{center} 
\begin{table}
 \caption{Summary Statistics of Hypothetical Startups in Experiment B}
 \label{Hypothetical Startup Summary Statistics}
 \scalebox{0.85}{
\begin{tabular}{l l l} 
&&\\
\toprule
&N&Industry Covered\\
\midrule
\emph{Panel A: 1st round}&&\\
B2B&13&  Media, Music, Fashion, Advertisement, Real Estate, Construction, SAAS, Education,\\
&&Logistics, Energy, Agriculture \\
B2C&12& Media, Fashion, Sports, Food, SAAS, Traveling, Pets, Chemical Products, Education\\
Healthcare&8& Healthcare\\
Total&33&\\
&&\\
\emph{Panel B: 2nd round}&&\\
B2B&13&Entertainment, Media, Packaging, Advertisement, Finance, Management, Education \\
&&SAAS\\
B2C&14&Entertainment, Media, Energy, SAAS, Sports, Chemical Products, Food \\
Healthcare&7&Healthcare\\
Total&34&\\
&&\\
\emph{Panel C: Total} &&\\
B2B&26&Media, Music, Fashion, Advertisement, Real Estate, Construction, SAAS, Education,\\
&&Logistics, Energy, Agriculture, Entertainment, Packaging, Finance, Management\\
B2C&26&Media, Fashion, Sports, Food, SAAS, Traveling, Pets, Chemical Products, Education,\\
&&Entertainment, Energy, \\
Healthcare&15& Healthcare\\
Total&67&\\
\bottomrule
\end{tabular}}
\begin{tablenotes}
\item \footnotesize \emph{Notes.} This table reports descriptive statistics for the 67 startups used in the first-round and second-round correspondence tests. All the startups are classified into B2B (Business to Business), B2C (Business to Consumer), and Healthcare following the classification categories of \cite{gornall_gender_2020}. I also provide more granular industry information about the created startups in the table. Panel A reports the startup category distribution of the first-round correspondence test, which was implemented between 03/2020 and 04/2020 during the outbreak of COVID-19. Main results of Experiment B in this paper only use the first-round experiment’s results because many investors have realized the existence of this experiment when I implemented the second-round correspondence test. This makes the second-round experiments' results very noisy and less credible. Panel B reports the startup category distribution of the second-round correspondence test, which was implemented in 10/2020. Panel C reports the startup category distribution of all 67 startups used in the two rounds of correspondence tests. If a startup belongs to both B2B and B2C, I have labeled it as ``B2B."\par
\end{tablenotes}
\end{table}
\end{center}

\clearpage
\begin{center}
\begin{table}
 \caption{Trace Investors' Email Behaviors in Experiment B}
\label{cor:tracebehavior}
\scalebox{0.85}{ 
\small
\begin{tabular}{ m{3.3cm} m{5.67cm} m{2.95cm} m{4.96cm} m{1.77cm}} 
\toprule
Email Behaviors& Behavior Tracking Mechanisms& Merits &Limitations&Literature\\
\midrule
1. Email Opening Rate (Time Stamp)&Write each pitch email using HTML with a unique ID and insert a one-pixel invisible transparent picture into the email. If the picture is downloaded from the server, I assume the investor opened the pitch email when the picture was downloaded &Increases the experiment's power (high opening rate); only affected by the email's subject line rather than the email's contents&Noisy measurements (Some remote servers prevent users from downloading a picture while others automatically download a picture for their users. However, such server properties are unrelated to the experimental treatment.)&\\
2. Email Reading Time (Time Stamp)&Write each pitch email using HTML with a unique ID and insert a large invisible transparent picture (i.e. 500 MB) into the email. Set the speed of downloading the picture from our server to 10KB/s. If only 200KB is downloaded from the server, then the email staying time is 20s.&A continuous variable which measures attention; Increases the experiment's power;&Noisy measurements (Researchers cannot observe directly whether inventors are reading the email or simply leaving the email open while having lunch.)&\\
&&&&\\

3. Multiple Email Opening Rate \par(Not Enough Statistical Power)&If the one-pixel transparent picture inserted in the pitch email is downloaded multiple times as recorded in the server, then I assume the email is opened multiple times. This happens if the same investor opens the email multiple times or the email is forwarded to others who open it later.&Increases the experiment's power; a stronger indicator of investors' interest&Noisy measurement. Researchers cannot differentiate whether the email is opened multiple times by the same investor, or the email is forwarded to others.&\\
&&&&\\
4. Sentimental Analysis of Email Replies \par(Not Enough Statistical Power)&Use LIWC to analyze the sentiment of the content of each email reply. I used the following website which automatically generates analyzed results: \href{http://liwc.wpengine.com/}{http://liwc.wpengine.com/} &Relatively objective measurement of the investors' attitudes towards each pitch email&Low response rate during the recession, hence low experimental power&\cite{hong2015crime}\\
&&&&\\
5. Website Click Rate \par(Not Enough Statistical Power)&The Mailgun platform developed this function, and researchers can use it directly. Click \href{https://help.mailgun.com/hc/en-us/articles/360011566033-How-to-Enable-HTTPS-Tracking-Links}{here} for mechanism explanations provided by Mailgun. &Can be used when investors do not reply to the email&Low website click rate in the entrepreneurial financing setting&\cite{bartos_attention_2016}; \cite{bernstein_attracting_2017}\\
&&&&\\
6. Email Response Rate \& Reply's Contents \par(Not Enough Statistical Power)&Collected directly from the inbox and spam box&Commonly used callback measurements&Low response rate; The reply's contents may not represent true interest if investors try to be politically correct.&\cite{gornall_gender_2020}, etc.\\
\bottomrule
\end{tabular}}
\begin{tablenotes}
\item \footnotesize \emph{Notes.} This table provides detailed mechanisms of recording different email behaviors, the merits, the limitations of each tracked behavior measurement, and the previous correspondence tests in the literature that have used similar participants' behaviors. To realize these functions, I use the Mailgun platform, which is a professionally designed platform for large email campaign activities founded in 2010. Except for the first two email behavior measurements, the other measurements all suffer from the ``low-response-rate" problem.
\end{tablenotes}
\end{table}
\end{center}


\clearpage
\begin{table}
 \caption{Heteroscedastic Probit Estimates in Experiment B}
\label{probithet}
\begin{center} 
\scalebox{0.85}{
\begin{tabular}{l c c c} 
\hline
&&&\\
& \multicolumn{3}{c}{Dependent Variable: \textbf{$1$}(\emph{Opened}) }\\
\cline{2-4}
&(1)&(2)&(3)\\
&&&After 03/24\\
\hline
\emph{Panel A. Probit estimates} &&&\\
&&&\\
Female Founder (marginal)&0.010***&&\\
&(0.004)&&\\
Asian Founder &&0.006&0.007*\\
&&(0.004)&(0.004)\\
&&&\\
\emph{Panel B. Heteroscedastic probit estimates} &&&\\
&&&\\
Female Founder (marginal)&0.009***&&\\
&(0.004)&&\\
Asian Founder &&0.006&0.008*\\
&&(0.004)&(0.004)\\
&&&\\
Standard deviation of&&&\\
unobservables, female/male&0.81&&\\
&&&\\
Standard deviation of&&&\\
unobservables, Asian/white&&1.12&1.09\\
&&&\\
Test: ratio of standard&&&\\
deviations = 1 ( p-value)&0.27&0.55&0.701\\
&&&\\
Observations&30,909&30,909&25,525\\
\hline
\end{tabular}}
 \end{center}
\begin{tablenotes}
\item \footnotesize
\emph{Notes.} This table reports regression results from the heteroscedastic probit estimates for email opening rate after correcting potential biases from the difference in variance of unobservables following \cite{neumark_detecting_2012}. Marginal effects are computed as the change in the probability associated with being a ``female” founder using the continuous approximation, evaluating other variables at their means; the continuous approximation yields an unambiguous decomposition of the heteroscedastic probit estimates. The dependent variable is one if an investor opens the pitch email, and zero otherwise. Columns (1) and (2) use all the observation. Column (3) uses the observations from pitch emails sent after 03/24/2020. Standard errors are in parentheses. P-values are based on Wald tests. *** p$<$0.01, ** p$<$0.05, * p$<$0.1\\

\end{tablenotes}

\end{table}

\clearpage
\begin{table}
 \caption{Gender Homophily in Experiment B}
\label{table_homophily}
\begin{center}
\scalebox{0.85}{
\begin{tabular}{l c c c m{0.5cm} c c c } 
\toprule
&&&&&&&\\
& \multicolumn{3}{c}{Dependent Variable: \textbf{$1$}(\emph{Opened}) }&& \multicolumn{3}{c}{Dependent Variable: \emph{Staying Time}}\\
\cline{2-4} \cline{6-8}
&(1)&(2)&(3)&&(4)&(5)&(6)\\
&Full&Female&Male&&Full&Female&Male\\
&Sample&Investors&Investors&&Sample&Investors&Investors\\
\midrule
Female Founder&0.011**&0.008&0.011**&&0.147&0.301&0.076\\
	&(0.004)&(0.007)&(0.004)&&(0.996)&(1.884)&(0.997)\\
&&&&&&&\\
Female  Founder $\times$&-0.003	&&&&0.590&&\\
Female Investor& (0.008)&&&&(2.108)&&\\
&&&&&&&\\
Female Investor&-0.018***&&&&-2.752*&&\\
&(0.006)&&&&(1.561)	&&\\
&&&&&&\\
Control&Yes&Yes&Yes&&Yes&Yes&Yes\\
Startup FE&Yes&Yes&Yes&&Yes&Yes&Yes\\
Observations&30,909&7,277&	23,632&&3,720&767&2,953\\
R-squared&0.005&0.002&0.005&&0.000&0.001&0.000\\
\bottomrule
\end{tabular}}
\end{center}
\begin{tablenotes}
\item \footnotesize
\emph{Notes.} This table reports the heterogeneous effect of investors' email opening behaviors based on investors' gender in Experiment B, which tests the gender homophily mechanism. In Columns (1)-(3), the dependent variable is a dummy variable, which is one if an investor opens the pitch email, and zero otherwise. In Columns (4)-(6), the dependent variable is the time spent on each pitch email measured in seconds. In order to mitigate the truncation issue, I only include the opened emails in Columns (4)-(6). ``Female Founder" equals one if the first name of the email sender is a female name, and zero otherwise. Similarly, ``Asian Founder" equals one if the last name of the email sender is an East Asian name, and zero otherwise. ``Female Investor" is an indicator variable for being a female investor. Control variable includes ``US Investor", which is an indicator variable for being a U.S. investor. $R^2$ is the adjusted $R^2$ for OLS regressions. Standard errors in parentheses are clustered at the investor level. *** p$<$0.01, ** p$<$0.05, * p$<$0.1 
\end{tablenotes}
\end{table}

\subsection{Extension of Neumark's Model by Adding Strategic Channel}
In \cite{neumark_detecting_2012}, the higher the startup's perceived quality is, the more likely the investor will open the startup's email. However, if some emails are too good (i.e., ``overqualified"), investors may not want to spend time on them. Although this mechanism does not play an important role in Experiment B, this strategic channel can be added to the model of \cite{neumark_detecting_2012} by assuming the following non-monotonic hiring rule:\footnote{This subsection uses the notations in \cite{neumark_detecting_2012}.}

\begin{eqnarray*}
c_2'>\beta_1^{'}X^{I*}+X_F^{II}+\gamma^{'}+F>c_1'
\end{eqnarray*}
It is feasible to use MLE method to estimate the model:
\begin{center}
\begin{eqnarray*}
T_{ij}=&1\{c_1'<\beta X_1^{I*}+X_2^{II}+\gamma' G +\epsilon_{ij}<c_2'\}\\
T_{ij}=&1\{(c_1'-X_1^{I*}-\gamma' G)/\sigma_B<X_2^{II}+\epsilon_{ij}<(c_2'-X_1^{I*}-\gamma' G)/\sigma_B\}\\
\prod_{i=1}^{n}(\Phi(\frac{(c_2'-X_1^{I*}-\gamma')}{\sigma_B^F})-&\Phi(\frac{(c_1'-X_1^{I*}-\gamma')}{\sigma_B^F}))^{T_{i\in F,j}=1}(\Phi(\frac{(c_2'-X_1^{I*})}{\sigma_B^M})-\Phi(\frac{(c_1'-X_1^{I*})}{\sigma_B^M}))^{T_{i\in M,j}=1}\\
\end{eqnarray*}
\end{center}
Since it becomes a non-monotonic crossing threshold model, it is technically hard to non-parametrically estimate this model (see \cite{lee2018identifying}).

\vspace{3cm}

\begin{figure}[h]
    \centering
    \includegraphics[scale=0.35]{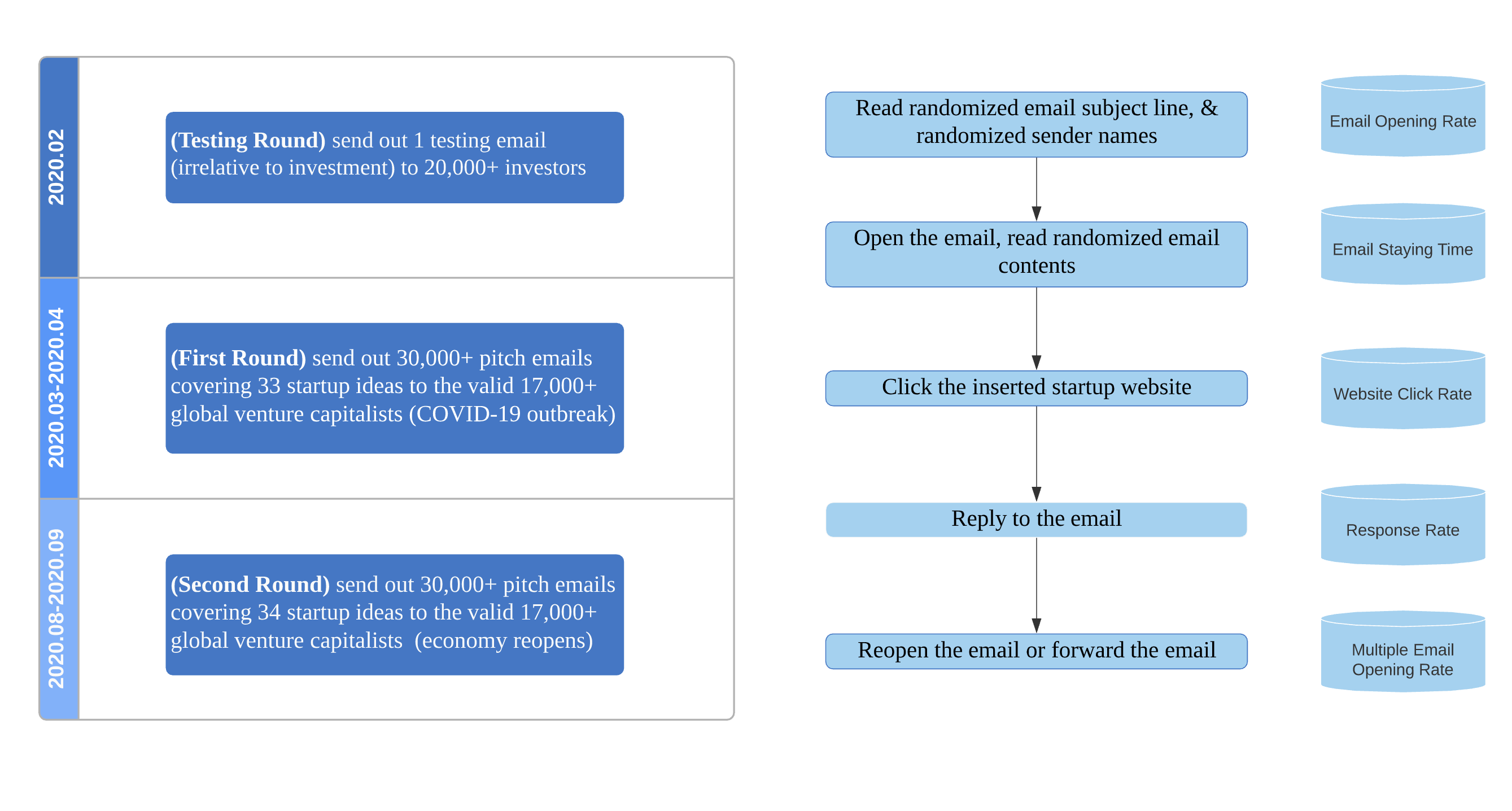}
    \caption{Experimental Flowchart for Experiment B}
    \captionsetup{labelformat=empty}
    \caption*{\footnotesize\emph{Notes.} This figure describes the experimental timeline, experimental design, and the traced email behaviors of investors.}
    \label{fig:ct_flow_chart}
\end{figure}

\clearpage
\begin{figure}
    \centering
    \includegraphics[scale=0.75]{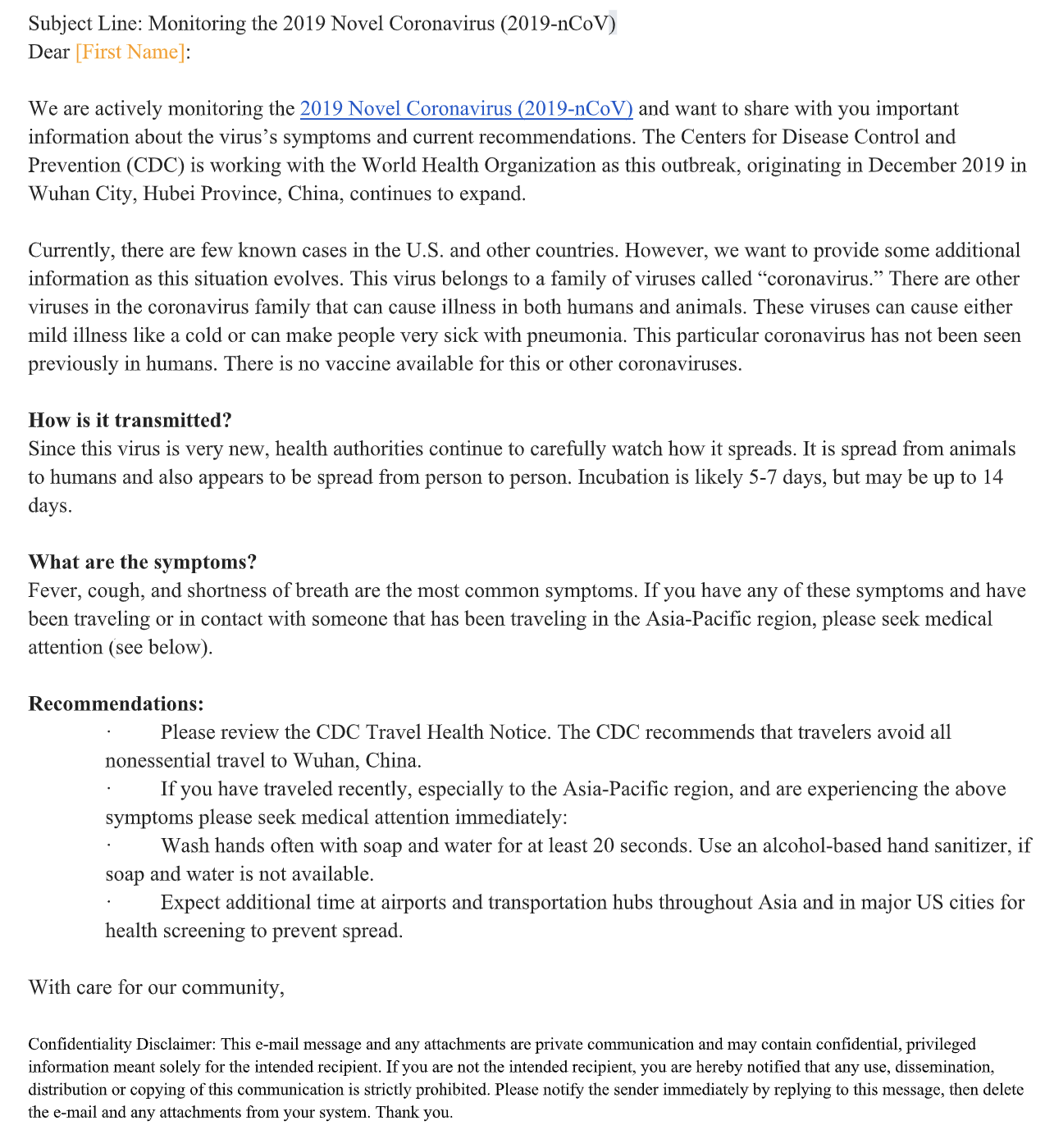}
    \caption{Example of the Testing Email in Experiment B}
    \label{cor:testingemail}
\end{figure}

\begin{figure}
    \centering
    \includegraphics[scale=0.6]{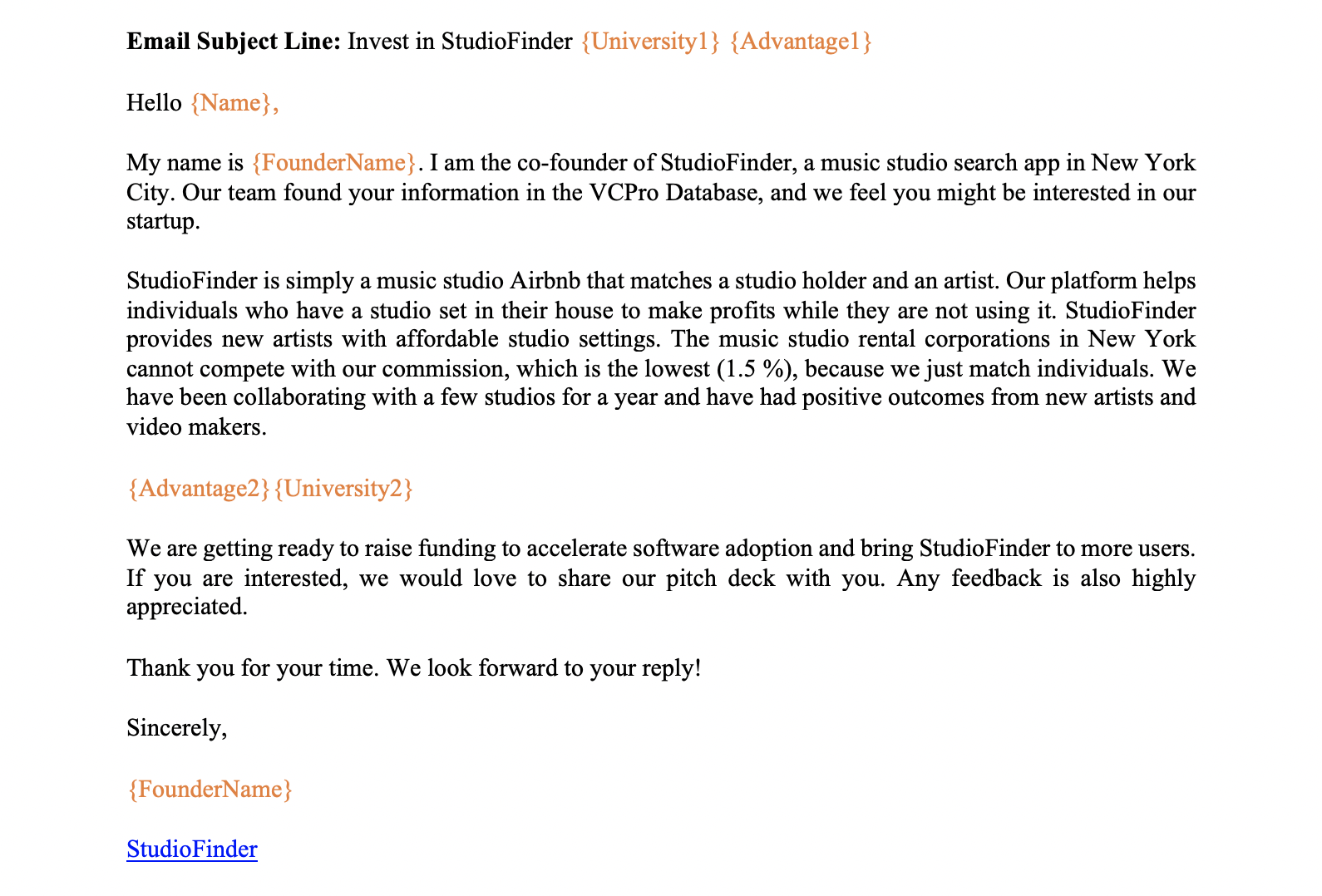}
    \caption{Example of a Pitch Email in Experiment B}
    \label{cor:pitchemail}
\end{figure}

\begin{figure}
    \centering
    \includegraphics[scale=0.3]{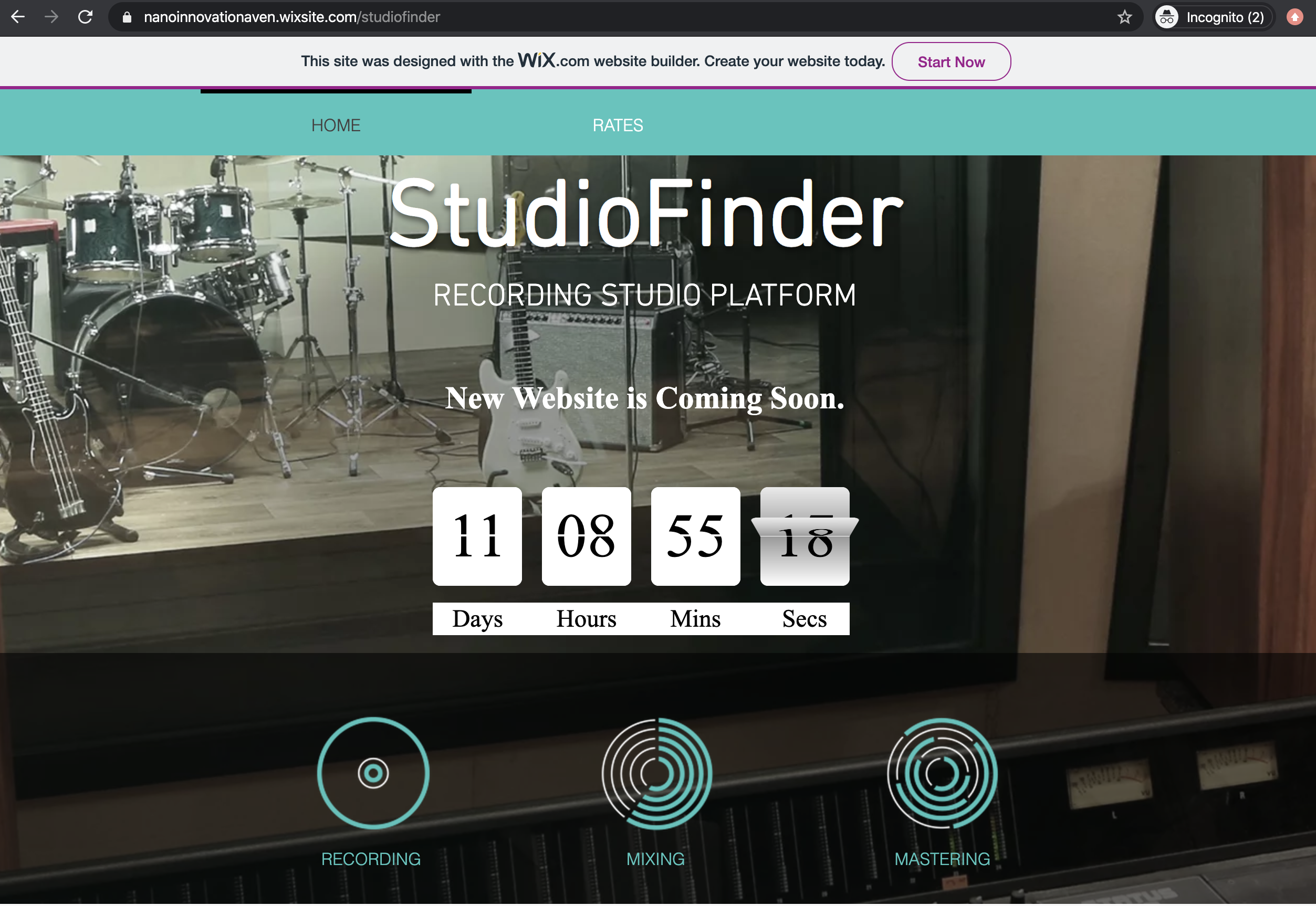}
    \caption{Example of a Startup Website in Experiment B}
    \label{cor:website}
\end{figure}

\setcounter{table}{0}
\setcounter{figure}{0}
\renewcommand{\thetable}{\Alph{section}\arabic{table}}
\renewcommand{\thefigure}{\Alph{section}\arabic{figure}}



\end{document}